\documentclass[aps,prb,twocolumn,showkeys,superscriptaddress,floatfix]{revtex4}
\usepackage{amsmath}
\usepackage{epsfig}
\usepackage{bm}
\usepackage[dvips]{color}
\usepackage{multirow}
\usepackage{rotating}
\usepackage{longtable}
\definecolor{grigio}{rgb}{0.8,0.8,0.8}

\begin{document}
\title{
Full configuration interaction approach to
the few-electron problem in artificial atoms}

\author{Massimo Rontani}
\email{rontani@unimore.it}
\homepage{http://www.nanoscience.unimo.it/max_index.html}
\affiliation{CNR-INFM National Research Center on nanoStructures
and bioSystems at Surfaces (S3), Modena, Italy}

\author{Carlo Cavazzoni}

\affiliation{CNR-INFM National Research Center on nanoStructures
and bioSystems at Surfaces (S3), Modena, Italy}

\affiliation{CINECA, Via Magnanelli 6/3, 40033 Casalecchio di Reno BO, Italy}

\author{Devis Bellucci}

\affiliation{CNR-INFM National Research Center on nanoStructures
and bioSystems at Surfaces (S3), Modena, Italy}

\affiliation{Dipartimento di Fisica, Universit\`a degli Studi di Modena
e Reggio Emilia, Via Campi 213/A, 41100 Modena MO, Italy}

\author{Guido Goldoni}

\affiliation{CNR-INFM National Research Center on nanoStructures
and bioSystems at Surfaces (S3), Modena, Italy}

\affiliation{Dipartimento di Fisica, Universit\`a degli Studi di Modena
e Reggio Emilia, Via Campi 213/A, 41100 Modena MO, Italy}

\date{\today}
\begin{abstract}
We present a new high-performance configuration interaction code
optimally designed for the calculation of the lowest energy
eigenstates of a few electrons in semiconductor quantum dots (also
called artificial atoms) in the strong interaction regime. The
implementation relies on a single-particle representation, but it is
independent of the choice of the single-particle basis and,
therefore, of the details of the device and configuration of
external fields. Assuming no truncation of the Fock space of Slater
determinants generated from the chosen single-particle basis, the
code may tackle regimes where Coulomb interaction very effectively
mixes many determinants. Typical strongly correlated systems lead to
very large diagonalization problems; in our implementation, the
secular equation is reduced to its minimal rank by exploiting the
symmetry of the effective-mass interacting Hamiltonian, including
square total spin. The resulting Hamiltonian is diagonalized via
parallel implementation of the Lanczos algorithm. The code gives
access to both wave functions and energies of first excited states.
Excellent code scalability in a parallel environment is
demonstrated; accuracy is tested for the case of up to eight
electrons confined in a two-dimensional harmonic trap as the density
is progressively diluted up to the Wigner regime, where 
correlations become dominant.
Comparison with previous Quantum Monte Carlo simulations in the
Wigner regime demonstrates power and flexibility of the method.
\end{abstract}

\maketitle

\section{Introduction}

Semiconductor quantum dots \cite{Jacak,Bimberg,Woggon,Tapah} (QDs)
are nanometer sized regions of space where free carriers are
confined by electrostatic fields. Typically, the confining field may
be provided by compositional design (e.g., the QD is formed by a
small gap material embedded in a larger-gap matrix) or by gating an
underlying two-dimensional (2D) electron gas. Different techniques
lead to nanometer size QDs with different shapes and strengths of
the confinement. As the typical de Broglie wavelength in
semiconductors is of the order of 10 nm, nanometer confinement leads
to a discrete energy spectrum, with energy splittings ranging from
fractions to several tens of meV.

The similarity between semiconductor QDs and natural atoms, ensuing from
the discreteness of the energy spectrum, is often pointed out.
\cite{MaksymPRL,Kastner,Ashoori96,Tarucha96,RontaniNato} Shell
structure \cite{Ashoori96,Tarucha96,Rontani98} and correlation effects
\cite{RontaniNato,BryantPRL,Garcia05}
are among the most striking
experimental demonstrations. More recently, the interest in new
classes of devices has fuelled investigations of QDs which are
coupled by quantum tunneling. \cite{Rontani04}
In addition to their potential for promising applications,
these systems extend the analogy between
natural and artificial atoms to the molecular realm (for a review
of both theoretical and experimental work see
Ref.~\onlinecite{Rontani01}).

Maybe the most prominent feature of {\em artificial} atoms is the
possibility of a fine control of a variety of parameters in the
laboratory. The nature of ground and excited few-electron states has
been shown to vary with artificially tunable quantities such as
confinement potential, density, magnetic field, inter-dot coupling.
\cite{Jacak,Tapah,RontaniNato,Rontani01,Rontani02,ReimannRMP,GoldoniNato}
Such flexibility allows for envisioning a vast range of applications
in optoelectronics (single-electron transistors,
\cite{Grabert92} lasers, \cite{Arakawa82,Kirstaedter94}
micro-heaters and micro-refrigerators based on thermoelectric
effects \cite{Edwards95}), life
sciences,\cite{Michalet05,Bailey04} as well as in several quantum
information processing schemes in the solid-state environment.
\cite{Loss98,FilippoQIP,Biolatti00}

Almost all QD-based applications rely on (or are influenced by)
electronic correlation effects, which are prominent in these systems.
The dominance of interaction in artificial atoms and molecules is
evident from the multitude of strongly correlated few-electron
states measured or predicted under different regimes: Fermi liquid,
Wigner molecule (the precursor of Wigner crystal in 2D bulk), charge
and spin density wave, incompressible state reminescent of
fractional quantum Hall effect in 2D (for reviews see
Refs.~\onlinecite{ReimannRMP,Rontani01}).

It is important to understand where the dominant correlation effects
arise from. In QDs the ``external'' confinement potential originates
either from band mismatch or from the self-consistent field (SCF)
due to doping charges. In both cases, the total field modulation
which confines a few free carriers is smooth and can often be
approximated by a parabolic potential in two dimensions.
Therefore, the single-particle (SP) energy gaps of the free carrier
states are uniform in a broad energy range. This is in contrast with
natural atoms, where the electron confinement is provided by the
singular nuclear potential. Furthermore, while the kinetic energy
term scales as $r_s^{-2}$, $r_s$ being the dimensionless parameter
measuring the average distance between electrons, the Coulomb energy
scales as $r_s^{-1}$. Contrary to natural atoms, in
semiconductor QDs the Coulomb-to-kinetic-energy ratio can be rather
large (even larger than one order of magnitude), the smaller the
carrier density the larger the ratio. This causes Coulomb
correlation to severely mix many different SP configurations, or
Slater determinants (SDs). Therefore, we expect that any successful
robust computational strategy in QDs, if flexible enough to address
different correlation regimes, must not rely on truncation of the
Fock space of SDs obtained by filling a given SP basis with $N$
electrons.

The theoretical understanding of QD electronic states in a vaste
class of devices is based on the envelope function and effective
mass model.\cite{Bastard,Cardona} Here, changes in the Bloch states,
the eigenfunctions of the bulk semiconductor, brought about by
``external'' potentials other than the perfect crystal potential,
are taken into account by a slowly varying (envelope) function which
multiplies the fast oscillating periodic part of the Bloch states.
This decoupling of fast and slow Fourier components of the wave
function is valid provided the modulation of the external potential
is slow on the scale of the lattice constant. Then, the theory
allows for calculating such envelope functions from an ``effective''
Schr\"odinger equation where only the external potentials appear,
while the unperturbed crystalline potential enters as a renormalized
electron mass, i.e.~the ``effective'' value $m^*$ replaces the free
electron mass. Therefore, SP states can be calculated in a
straightforward way once compositional and geometrical parameters
are known. This approach was proved to be remarkably accurate by
spectroscopy experiments for weakly confined QDs; \cite{Rontani04,ReimannRMP}
for strongly confined systems, such as certain classes of
self-assembled QDs, atomistic methods might be
necessary.\cite{Zunger99}

Several different theoretical approaches have been employed in the
solution of the few-electron, fully interacting effective-mass
Hamiltonian of the QD system \cite{ReimannRMP,Rontani01}
\begin{equation}
H = \sum_{i}^{N}H_{0}(i)+\frac{1}{2}\sum_{i\neq j}\frac{e^{2}}
{\kappa|\boldsymbol{r_{i}}-\boldsymbol{r_{j}}|},
\label{eq:HI}
\end{equation}
with
\begin{equation}
H_{0}(i)\,=\,\frac{1}{2m^{*}}\Big[\boldsymbol{p}-\frac{e}{c}
\boldsymbol{A}(\boldsymbol{r})\Big]^{2}\,+\,V(\boldsymbol{r}).
\label{eq:HSP}
\end{equation}
Here, $N$ is the number of free conduction band electrons (or free
valence band holes) localized in the device ``active'' area, $e$,
$m^*$, and $\kappa$ are respectively the carrier charge, effective
mass, and static relative dielectric constant of the host
semiconductor, $\boldsymbol{r}$ is the position of the electron,
$\boldsymbol{p}$ is its canonically conjugated momentum,
$\boldsymbol{A}(\boldsymbol{r})$ is the vector potential associated
with an external magnetic field. The effective potential
$V(\boldsymbol{r})$ describes the carrier confinement due to the
electrostatic interaction with the environment. We ignore the
(usually small) Zeeman term coupling spin with magnetic field; this
small perturbation can always be added \emph{a posteriori}.

The eigenvalue problem associated to the Hamiltonian (\ref{eq:HI})
has been tackled mainly via Hartree-Fock (HF) method, density
functional theory, configuration interaction (CI), and quantum Monte
Carlo (QMC) (for reviews see
Refs.~\onlinecite{Jacak,Tapah,ReimannRMP,Rontani01}). Each method
has its own merits and drawbacks; mean-field methods, for example,
are simple and may treat a large number of carriers, a common
situation in many experimental situations. On the other hand, CI and
QMC methods allow for the treatment of Coulomb correlation with
arbitrary numerical precision, and, therefore, represent the natural
choice for studies interested in strongly correlated (low density or
high magnetic field) regimes which are nowadays attainable in very
high quality samples. In addition, CI is a straightforward approach
with respect to QMC, and gives access to both ground and lowest
excited states at the same time and with comparable accuracy, which
allows for the calculation of various response functions to be
directly compared with experiments. The main limitations within the
CI approach, when applied to the strongly correlated regimes attainable
in artificial atoms, are that truncation of the Fock space
--- one of the standard procedures in traditional CI methods of
quantum chemistry (the so-called {\em CI with singles, doubles},
etc.\cite{Bauschlicher90,Raghavachari96,Shavitt98,Sherril99,Jensen99,libronecorni}) 
--- generally gives large errors or even
qualitatively wrong results, and that the Fock space dimension
grows exponentially with $N$.

Here we present a newly developed CI code \cite{manzoni} with performances
comparable to QMC calculations for a broad range of electron
densities, covering the transition from liquid to solid electron
phases. The implementation is independent of the SP orbital basis
and, therefore, of the details of the device. We do not assume any
truncation of the Fock space of SDs generated from the chosen SP
basis [{\em full-CI} 
\cite{Bauschlicher90,Raghavachari96,Shavitt98,Sherril99,Jensen99,libronecorni} 
(FCI)]. 
The large eigenvalue problem
is reduced to its minimal size by exploiting the symmetry of the
effective-mass interacting Hamiltonian (\ref{eq:HI}), including
square total spin $(S^2$). \cite{atenziun} The resulting Hamiltonian,
which represents the Coulomb interaction in the many-electron basis,
is diagonalized via a parallel version of the Lanczos algorithm. 

The motivation for developing a new code, instead of using CI
codes already available in the computational quantum chemistry
literature, was the difficulty of handling such codes, optimized for
traditional {\em ab initio} applications, to treat effective or
model problems such as a few electrons in artificial atoms.
Therefore, our code has been developed entirely from scratch and specifically 
designed for quantum dot applications. One peculiar feature is that
it builds the whole Hamiltonian in the many-electron basis and it
stores it in computer memory (see Sec.~\ref{s:code}).
This procedure presently constitutes a limitation with respect to
state-of-the-art CI algorithms 
\cite{Bauschlicher90,Raghavachari96,Shavitt98,Sherril99,Jensen99,libronecorni,Roos72,Siegbahn84,Knowles84,Bauschlicher87,Olsen88,Knowles89,Harrison91,Caballol92,Povill93,Sakai93,Daudey93,Szalay95,Hanrath97,Stephan98,Amor98,Stampfuss99,Stampfuss00,Ansaloni00,Lischka01,Ivanic03,Stampfuss05} 
of Quantum Chemistry based on the
{\em direct} technique, which avoids the storage of the 
Hamiltonian (cf.~Sec.~\ref{s:discussion}). The implementation of the
direct CI algorithm is left to future work.

In order to demonstrate the
effectiveness of our code, we present a benchmark calculation in the
strongly correlated regime {\em par excellence}, namely the
crossover region between Fermi liquid and Wigner crystallization,
and we compare results with available QMC data. 
Since the FCI code gives access to both wave functions and energies of the
first excited states, we analize the signatures of crystallization
in the excitation spectrum. We also discuss the
performance of the algorithm in a parallel environment.

The paper is organized as follows: First the general code structure
is illustrated (Sec.~\ref{s:code}), then possible SP bases and
related convergence problems are considered (Sec.~\ref{s:SP}).
Section \ref{s:performance}
analyzes the code performance in a parallel environment. A benchmark
physical application in demanding correlation regimes is considered
in Sec.~\ref{s:physics}, where results are compared with QMC and CI
data taken from literature. Section \ref{s:yrast} discusses
the peculiar features of the excitation spectrum in the
strongly correlated regime.
Finally, some general considerations are
addressed (Sec.~\ref{s:discussion}), followed by the concluding
section \ref{s:conclusion}. 
Appendix \ref{a:CSF} reviews the construction of the many-electron
Hilbert space and related issues, while
Appendix \ref{a:Coulombmatrix} provides
the explicit expression of the 2D Coulomb matrix elements for the
so-called Fock-Darwin SP orbitals.

\section{Code structure}\label{s:code}

The interacting Hamiltonian (\ref{eq:HI}) can be written in second
quantized form \cite{Landau} on the basis of a complete and
orthonormal set of SP orbitals
$\left\{\phi_a\!(\boldsymbol{r})\right\}$:
\begin{equation}
\mathcal{H}=\sum_{a b \sigma}\varepsilon_{ab}\,c^{\dagger}_{a\sigma}
c_{b \sigma} + \frac{1}{2}
\sum_{a b c d}\sum_{\sigma \sigma^{'}}
V_{a b c d}\,
c^{\dagger}_{a \sigma} c^{\dagger}_{b \sigma^{'}}
c_{c \sigma^{'}}c_{d \sigma}.
\label{eq:HII}
\end{equation}
Here $c^{\dagger}_{a\sigma}$ ($c_{a\sigma}$) creates (destroys) an
electron in the spin-orbital
$\phi_a\!(\boldsymbol{r})\chi_{\sigma}\!\left(s\right)$,
$\chi_{\sigma}\!\left(s\right)$ is the spinor eigenvector of the
spin $z$-component with eigenvalue $\sigma$ ($\sigma=\pm1/2$), the
label $a$ stands for a certain set of orbital quantum numbers,
$\varepsilon_{ab}$ and $V_{a b c d}$ are the one- and two-body
matrix elements, respectively, of the Hamiltonian (\ref{eq:HI}). The
latter are defined as follows:
\begin{subequations}\label{eq:paramdef}
\begin{eqnarray}
\varepsilon_{ab} & = & \int\!\!\text{d}\,\boldsymbol{r}\,
\phi^{*}_{a}\!(\boldsymbol{r})\,H_{0}(\boldsymbol{r})
\,\phi_{b}(\boldsymbol{r}), \\
V_{abcd} & = &
\int\!\!\text{d}\,\boldsymbol{r}\!\!\int\!\!
\text{d}\,\boldsymbol{r^{'}}\,
\phi^{*}_{a}\!(\boldsymbol{r})\,
\phi^{*}_{b}\!(\boldsymbol{r^{'}}) \nonumber\\
&&\times\quad\frac{e^{2}}
{\kappa|\boldsymbol{r}-\boldsymbol{r^{'}}|}\,
\phi_{c}(\boldsymbol{r^{'}})\, \phi_{d}(\boldsymbol{r}).
\label{eq:paramdefcoul}
\end{eqnarray}
\end{subequations}

The adopted FCI strategy to find the ground- and excited-state
energies and wave functions of $N$ interacting electrons, described
by the Hamiltonian (\ref{eq:HI}) or (\ref{eq:HII}), proceeds in
three separate steps:
\begin{itemize}
\item[{\tt step 1}] A specific set of $N_{\text{SP}}$ single-particle
orbitals,
$\left\{\phi_a\!(\boldsymbol{r})\right\}$, is chosen.
$N_{\text{SP}}$ should be large enough to guarantee as much
completeness of the SP space as possible. The parameters
$\varepsilon_{ab}$ and $V_{abcd}$ for a given device [Eqs.
(\ref{eq:paramdef})] are computed numerically only once, and then
stored on disk. In case $\phi_a\!(\boldsymbol{r})$ is an
eigenfunction of the SP Hamiltonian (\ref{eq:HSP}) the matrix
$\varepsilon_{ab}$ acquires a diagonal form,
$\varepsilon_{ab}=\varepsilon_a\delta_{ab}$. In a few (but
important) cases, the SP orbitals are eigenfunctions of a
particularly simple SP Hamiltonian $H_0$, and analytic expressions
for $\varepsilon_{ab}$ and $V_{abcd}$ can be derived; one example is
discussed in Sec.~\ref{s:FD}.
\item[{\tt step 2}] The many-electron Hilbert
(Fock) space is built up. The $N$-electron wave function,
$\left|\Psi_N\right>$, is expanded as a linear combination of
configurational state functions
\cite{Pauncz79,McWeeny92,Jensen99,libronecorni} (CSFs),
$\left|\psi_i\right>$, namely primitive combinations of SDs which
are simultaneously eigenvectors of the spatial symmetry group of
Hamiltonian (\ref{eq:HI}), the square total spin $S^2$, and the
projection of the spin along the $z$-axis $S_z$:
\begin{equation}
\left|\Psi_N\right> = \sum_i a_i \left|\psi_i\right>.
\label{eq:CIexpansion}
\end{equation}
On such a basis set, the matrix representation of the Hamiltonian
(\ref{eq:HII}) acquires a block diagonal form, where each block is
identified by a different irreducible representation of the spatial
and spin symmetry group (the latter is labeled by $S^2$, $S_z$). The
algorithm for building the whole set of CSFs starting from the SP
basis set is described in detail in App.~A; 
here we are
only concerned with the fact that CSFs are linear combination of
SDs, $\left|\Phi_j\right>$, obtained by filling in, in all possible
ways consistent with symmetry requirements, the $2N_{\text{SP}}$
spin-orbitals with $N$ electrons:
\begin{equation}
\left|\psi_i\right> = \sum_j b_{ij} \left|\Phi_j\right>,
\end{equation}
the $b_{ij}$'s being the Clebsch-Gordan coefficients. These do not
depend on the device, and can be calculated once (see
App.~A) and stored on disk.
\item[{\tt step 3}] Hamiltonian matrix elements in the CSF representation
are calculated. Since several CSFs $\left|\psi_i\right>$ may consist
of different linear combinations of the same SDs
$\left|\Phi_j\right>$, it is convenient to compute and store once
and for all the non-null matrix elements of Hamiltonian
(\ref{eq:HII}) between all possible SDs,
$\left<\Phi_j\right|\mathcal{H}\left|\Phi_{j'}\right>$;
subsequently, the matrix elements between CSFs,
$\left<\psi_i\right|\mathcal{H}\left|\psi_{i'}\right>$, can be
simply evaluated using the coefficients $b_{ij}$ previously stored:
\begin{equation}
\left<\psi_i\right|\mathcal{H}\left|\psi_{i'}\right>
= \sum_{jj'} b^*_{ij}\, b_{i'j'}
\left<\Phi_j\right|\mathcal{H}\left|\Phi_{j'}\right>.
\label{eq:matrixH}
\end{equation}
Then, the secular equation
\begin{equation}
\left(\mathcal{H}-E_N\mathcal{I}\right)\left|\Psi_N\right> = 0,
\end{equation}
where $\mathcal{I}$ is the identity operator, is solved by means of
the Lanczos package ARPACK, \cite{ARPACK} designed to handle the
eigenvalue problem of large size sparse matrices. Eventually,
eigenenergies $E_N$ and eigenvectors $a_i$ are found and saved,
ready for post-processing (wave function analysis, computation of
charge density, pair correlation functions, response functions,
etc).
\end{itemize}

\begin{figure}
\begin{picture}(300,180)
\put(10,0){\epsfig{file=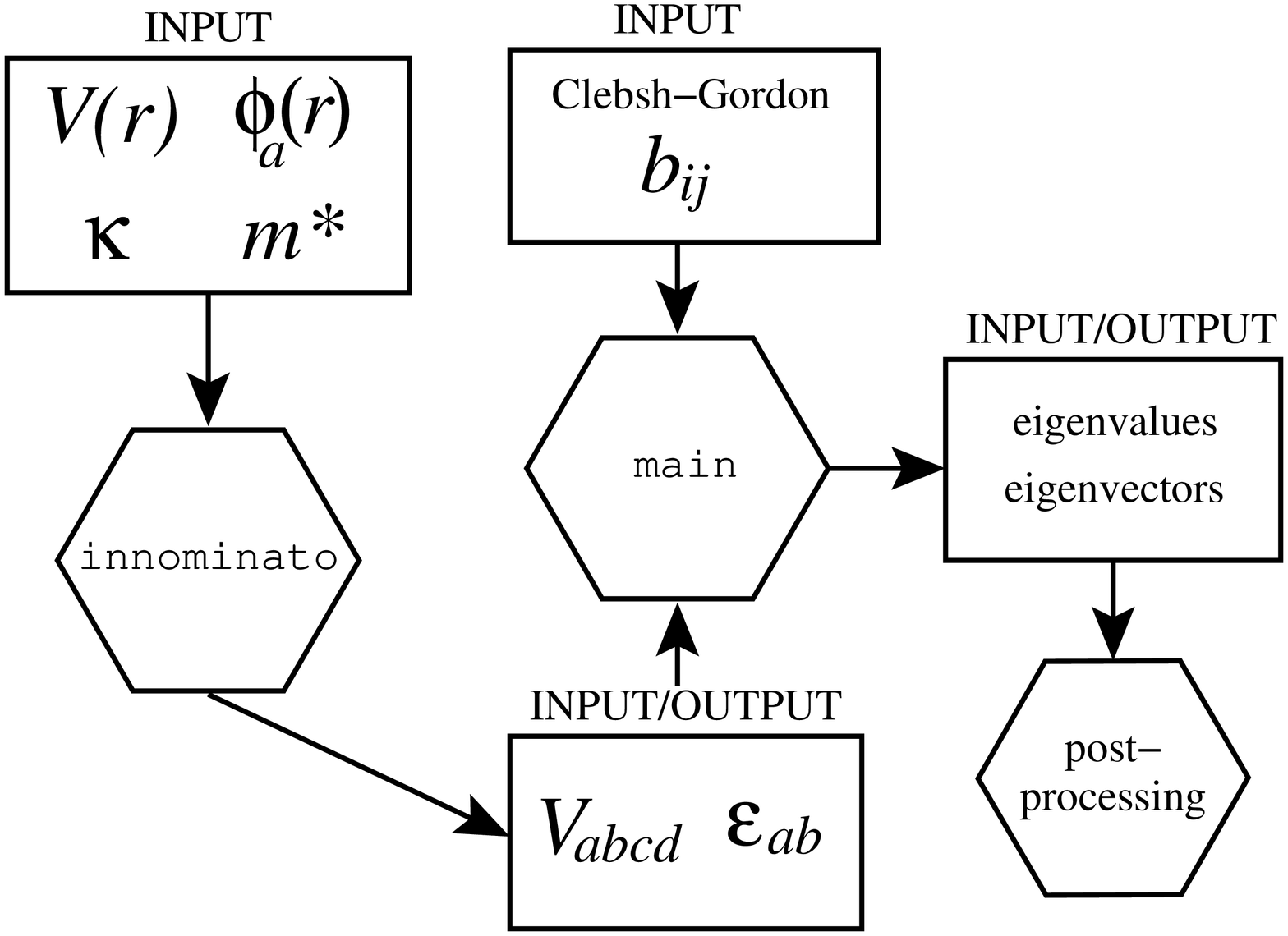,width=3.0in,,angle=0}}
\end{picture}
\caption{
Pictorial synopsis of the {\tt DONRODRIGO} suite of
FCI codes.
}
\label{fig:schema}
\end{figure}

{\tt step 1-3} are implemented in the suite of programs {\tt
DONRODRIGO}, \cite{manzoni} named after the characters of Alessandro
Manzoni's literary masterpiece \emph{I promessi sposi}. Each step
corresponds to a different high-level routine, providing maximum
code flexibility. Specifically, {\tt step 1} is an independent
program on its own, the {\tt INNOMINATO} code: The idea is that, to
describe different QD devices, it is sufficient to modify only the
form of the confinement potential, $V(\boldsymbol{r})$, and field
components appearing in the SP Hamiltonian (\ref{eq:HSP}). Indeed,
the confinement potential and the SP basis set only affect
parameters $\varepsilon_{ab}$ and $V_{abcd}$. Such parameters are
passed, as input data, to {\tt step 2-3} (cf. Fig.~\ref{fig:schema})
which constitute the core of the main program, and do not depend on
the form of the SP Hamiltonian and basis set. {\tt step 2-3} are by
far the most intensive computational parts and are parallelized via
the MPI interface: the Hamiltonian matrix elements
$\left<\Phi_j\right|\mathcal{H}\left|\Phi_{j'}\right>$ are
distributed among all processors. Only non-null elements are
allocated in local memory, and the parallel version of ARPACK
routine is used for diagonalization. Matrix element computation is
optimized by implementing a binary representation for CSFs: each CSF
corresponds to a couple of 64-bit integers, where each bit represent
the occupation of a SP orbital (App.~A2), 
and highly
efficient bit-per-bit logical operations are used to determine the
sign of the matrix element, depending on the anti-commuting behavior
of fermionic operators. An example is given in
Sec.~A3 of the Appendix.

Figure \ref{fig:schema} is the graphical synopsis of the interface
between codes belonging to the suite {\tt DONRODRIGO}, including
post-processing routines, which actually consitute a rich collection
of codes capable of calculating experimentally accessible quantities
and response functions. Presently, we have already implemented wave
function analysis, computation of charge density and pair
correlation functions,
\cite{Rontani01,Rontani02,RontaniNato,Bellucci03prl} dynamical form
factor, \cite{Rontani02} purely electronic Raman excitation
spectrum, \cite{Rontani02,Garcia05} single-electron excitation
transport spectrum, \cite{Rontani04,Bellucci03prl} spectral density
weight (one-particle Green's function),
\cite{Rontani04,Rontani05b,Rontani05c} relaxation time of excited
states via acoustic phonon emission. \cite{Bertoni04,Bertoni05}

\section{Single particle basis}\label{s:SP}

So far, in applications, among all possible SP basis sets
$\left\{\phi_a\!(\boldsymbol{r})\right\}$ we chose the
eigenfunctions of the SP Hamiltonian $H_0(\boldsymbol{r})$. In this
section we review the two most common cases.

\subsection{Fock-Darwin states}\label{s:FD}

Two common classes of QD devices are the so-called ``planar'' and
``vertical'' QDs. Planar QDs are obtained starting from a quantum
well semiconductor heterostructure, where conduction electrons are
free to move in the plane parallel to the interfaces, therefore
realizing a two dimensional electron gas. The second step is to
deposit electrodes on top of the structure, which laterally deplete
the electron gas forming a puddle of carriers, i.e., the dot
\cite{Ashoori96}. By appropriate engineering the top gates, more
than one QD can also be formed, coupled by quantum mechanical
tunneling. Vertical QDs are included in cylindrical mesa pillars
built by laterally etching a quantum well or a heterojunction.
\cite{Tarucha96,Rontani04} In this case, devices with two or more
tunnel coupled QDs can be obtained starting from coupled quantum
wells or heterojunctions.

\begin{figure}
\begin{picture}(300,180)
\put(35,160){\epsfig{file=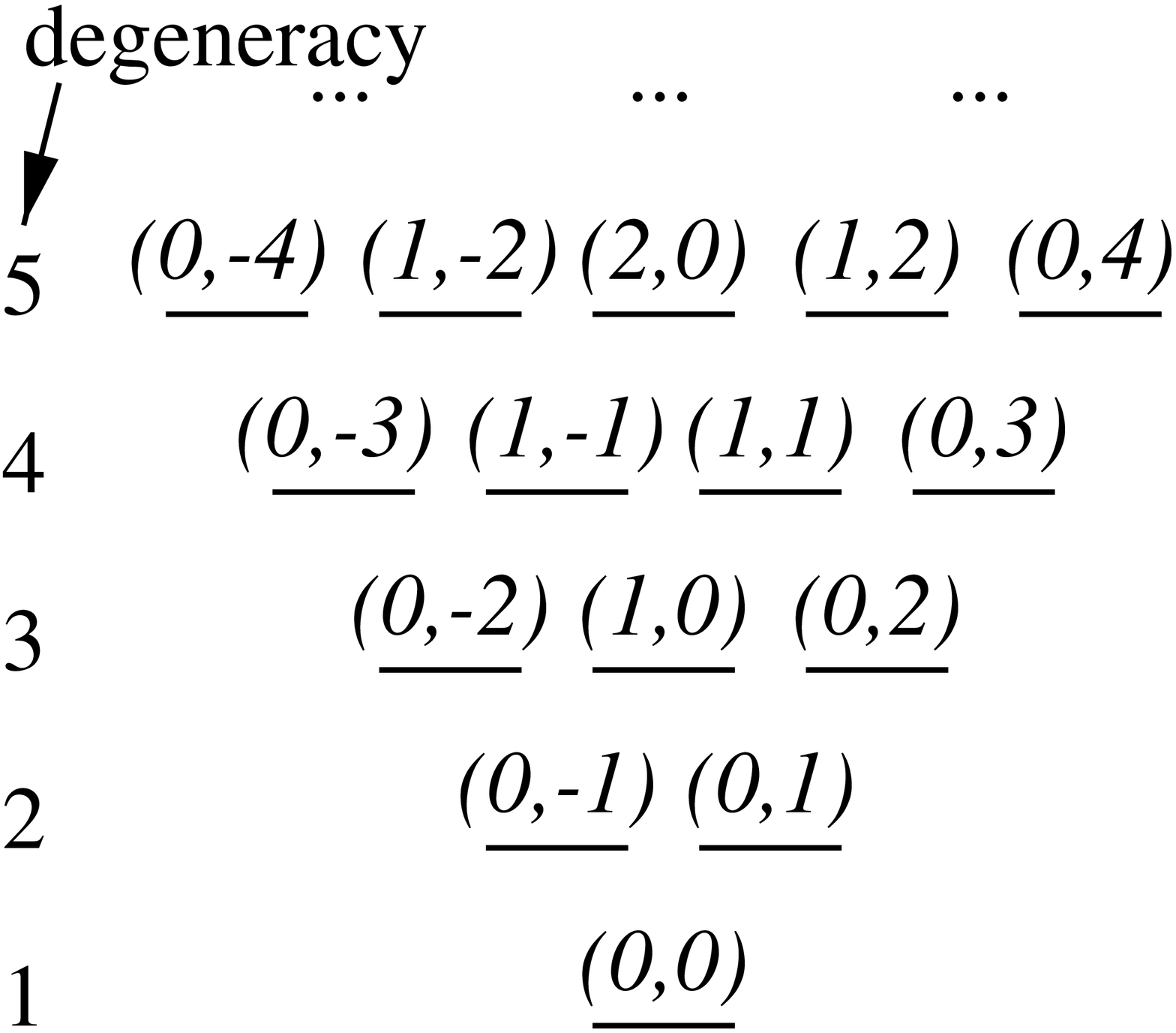,width=2.2in,,angle=-90}}
\end{picture}
\caption{ Shell structure of the SP Fock-Darwin energy spectrum, and
associated degeneracies. Each shell corresponds to the energy
\protect{$\hbar\omega_0 (N_{\text{shell}}+1)$}, where
\protect{$N_{\text{shell}}=2n+\left|m\right|$} is fixed, and
\protect{$(n,m)$} are the radial and azimuthal quantum numbers,
respectively. The degeneracy of each shell (not taking into account
the spin) is $N_{\text{shell}}+1$.} \label{f:FDshell}
\end{figure}

In the above devices, the in-plane confinement is much weaker than
the confinement in the underlying quantum well or heterojunction. It
is commonly accepted \cite{ReimannRMP} that the confinement
potential for the above devices is accurately described by
\begin{equation}
V(\boldsymbol{r}) =  \frac{1}{2}m^*\omega_0^2\left(x^2+y^2\right)
+ V(z),
\label{eq:VSP}
\end{equation}
where $\omega_0$ is the natural frequency of a 2D harmonic trap and
$V(z)$ describes the profile of a single or multiple quantum well
along the growth axis $z$. The eigenfunctions of the Hamiltonian
(\ref{eq:HSP}), with the potential given by (\ref{eq:VSP}) and
possibly a homogeneous static magnetic field applied parallel to
$z$, are of the type
\begin{equation}
\phi_{nmi}(\boldsymbol{r}) = \varphi_{nm}(x,y)\chi_i(z),
\end{equation}
namely the motion is decoupled between the $x-y$ plane and the $z$
axis. Here, the $\varphi_{nm}$'s, known as Fock-Darwin orbitals
\cite{Jacak}  (see below), are eigenfunctions of the in-plane part
of the SP Hamiltonian with eigenvalues
\begin{equation}
\varepsilon_{nm} = \hbar\Omega\left(2n+\left|m\right|+1\right)
-\hbar\omega_c/2,
\label{eq:eigenFD}
\end{equation}
$n$ and $m$ being respectively the radial and azimuthal quantum
numbers ($n=0,1,2,\ldots$ $m=0,\pm1,\pm2,\ldots$),
$\Omega=(\omega_0^2+\omega_c^2/4)^{1/2}$, $\omega_c=eB/m^*c$, $B$
being the magnetic field, and the $\chi_i(z)$'s ($i=1,2,3,\ldots$)
are eigenfunctions of the confinement potential along $z$. At zero
field the Fock-Darwin energies (\ref{eq:eigenFD}) display a
characteristic shell structure, the orbital degeneracy increasing
linearly with the shell number (Fig.~\ref{f:FDshell}).

The explicit expression of Fock Darwin orbitals is
\begin{eqnarray}\label{e:F-D}
\varphi_{nm}(\rho,\vartheta)&=&
{\ell}^{-\left(\left|m\right|+1\right)/4}
\sqrt{\frac{n!}{\pi\left( n+\left|m\right|\right)!}}
\,\rho^{\left|m\right|}\, {\rm e}^{-\left(\rho/\ell\right)^2/2}
\nonumber\\
&&\times
\quad L_n^{\left|m\right|}
\!\!\left[ \left(\rho/\ell\right)^2\right]
{\rm e}^{-{\rm i}m\vartheta},
\label{eq:FDexplicit}
\end{eqnarray}
where $(\rho,\vartheta)$ are polar coordinates and
$L^{\left|m\right|}_n(\xi)$
is a Generalized Laguerre Polynomial.
The typical lateral extension is given by the characteristic
dot radius $\ell = (\hbar/m^*\omega_0)^{1/2}$, $\ell$ being the
mean square root of $\rho$ on the Fock-Darwin ground state $\varphi_{00}$.

In usual applications only the first one or two confined states
along $z$ are important, since the localization in the growth
direction is much stronger than in the plane,
\cite{Rontani01,Rontani02,RontaniNato,Rontani04} and it is often
possible to assume a mirror symmetry with respect to the plane
crossing the origin, \cite{Rontani04} $V(z)=V(-z)$. In the above
simple model the spatial symmetry group of the device, $D_{\infty
h}$, is abelian and therefore SDs (CSFs) transforming according to
its irreducible representations may be straightforwardly built.
\cite{Slater63} The presence of a vertical magnetic field makes the
system chiral and reduces the symmetry to $C_{\infty h}$.

The most obvious advantage for choosing as SP basis orbitals the
eigenfunctions of the SP Hamiltonian itself is that they represent a
natural and simple starting point with regards to the physics of the
problem, allowing for both fast convergence in SP space and easy
symmetrization of many-electron wave functions. Besides, in the two
dimensional case Coulomb matrix elements are known analytically
(cf.~App.~B), while in the three dimensional
case [Eq.~(\ref{eq:VSP})] one can limit the effort of a six
dimensional integral evalutation to only a three dimensional
numerical integration, two variables being along the $z$-axis and
one along the radial relative coordinate, respectively, after
transformation to cylindrical coordinates in the in-plane
center-of-mass frame. Therefore, the evalutation of the Coulomb
integrals (\ref{eq:paramdefcoul}) is easily feasible --- the serial
version of the {\tt INNOMINATO} routine contributes a negligible
part to the total computation time
--- and makes
the usage of Fock-Darwin orbitals in the QD context competitive with
the implementation, e.g., of a gaussian basis.
\cite{Brasken00,Corni03}

\subsection{Numerical single particle states}\label{s:numerical}

Exploiting spatial symmetry and nearly parabolic confinemente is not
always possible. For example, self-assembled QDs grown by the
Stranski-Krastanov mechanism in strained materials, like in several
III-V materials \cite{Bennett97}, grow in highly non symmetric
shapes. Another case where symmetry is lost is by application of a
magnetic field with arbitrary direction.
\cite{Bellucci03prl,Bellucci03physe,Bellucci04} In both cases, the
no analytic solutions are available. In these unfavorable cases we
solve in a completely numerical way the SP Schr\"odinger equation
and obtain Coulomb matrix elements via Fourier transformation of
(\ref{eq:paramdefcoul}) in the reciprocal space (for more details
see Ref.~\onlinecite{Bellucci04}). Note that the Coulomb integrals
in (\ref{eq:paramdefcoul}) become complex as far as the field is
tilted with respect to the $z$ axis. In this general case the
algorithm is computationally demanding already at the SP level,
requiring extensive code parallelization.

\subsection{A test case --- Convergence issues}\label{s:test}

In order to assess quantitatively both completeness and convergence
rates for the Fock-Darwin SP basis, we set up a test case, which we
will systematically refer to in the remainder of the paper. We
therefore consider $N$ electrons in a 2D harmonic trap
[Eq.~(\ref{eq:VSP}) with $V(z)=0$] as the density is progressively
reduced, namely the lateral dimension $\ell$ increases as $N$ is
kept fixed. This test case is a significant benchmark since: (i)
path integral QMC calculations are available,
\cite{Egger99,Egger99E,Reusch03} providing ``exact'' data concerning
energies and wave functions of the ground state; (ii) as far as
$\ell$ increases, the dot goes progressively into a strong
correlation regime, which is more and more demanding with respect to
the size of the many-electron space needed; (iii) the model is
simple and the parameters of Eqs.~(\ref{eq:paramdef}) may be
computed analytically.

\begin{figure}
\begin{picture}(300,180)
\put(10,0){\epsfig{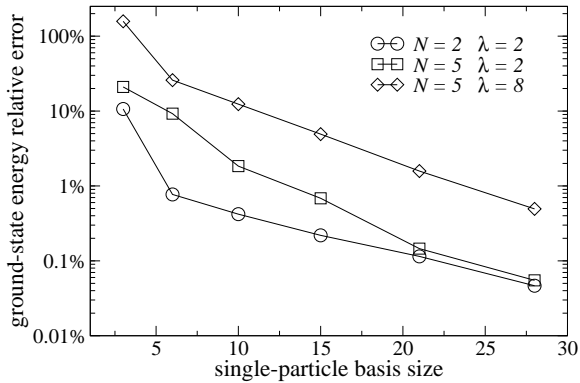}}
\end{picture}
\caption{ Relative error of the ground state energy for the test
case defined in Sec.~\ref{s:test}, as a function of the size of the
SP basis, for different values of the number of electrons, $N$, and
the dimensionless interaction strength, $\lambda$,
for a two dimensional harmonic trap. The reference values
are taken from Refs.~\onlinecite{Egger99E,Reusch03}. }
\label{fig:conv}
\end{figure}
We measure the strength of correlation by means of the dimensionless
parameter \cite{Egger99} $\lambda = \ell/a^*_{\text{B}}$ [$a^*_{\text{B}}
=\hbar^2\kappa/(m^*e^2)$ is the effective Bohr radius of the dot],
which is the QD analog to the density parameter $r_s$ in extended systems.
As a rough indication, consider that for $\lambda \approx 2$ or lower
the electronic ground state is liquid, while well above $\lambda = 8$
electrons form a ``crystallized'' phase, reminescent of the
Wigner crystal in the bulk. In the latter phase electrons are
localized in space and arrange themselves
in a geometrically ordered configuration
such that electrostatic repulsion is minimized. \cite{Egger99}

In Fig.~\ref{fig:conv} we study the completeness of the SP basis,
plotting the relative error of the ground state energy, as a
function of the basis size, in different correlation regimes. The
``exact'' reference values are taken to be the QMC data,
\cite{Egger99E,Reusch03} and the calculated values correspond to
adding successive Fock-Darwin orbital shells of increasing energy to
the SP basis (for a total amount of 3, 6, 10, 15, 21, 28 orbitals,
respectively). We see that, in the liquid phase ($\lambda=2$), a
basis made of 15 orbitals is sufficient to guarantee a precision of
one part over one hundred. The error, however, depends sensitively
on $N$: the smaller the electron number, the higher the precision.
The reason is that, since the confinement potential is soft,
electrons tend to move towards the outer QD region as $N$ increases,
and therefore higher energy orbital shells are needed to build wave
functions with larger lateral extension. A basis made of 21
orbitals, however, is enough to obtain a similar precision for both
two- and five-electron ground states (around 0.1\%). Figure
\ref{fig:conv} also shows that, as we move towards the high
correlation regime, i.e., as $\lambda$ increases, the error rapidly
increases if the SP basis size is kept fixed. For example, when
$\lambda=8$ the error for the $N=5$ ground state using 21 orbitals
is about 2\%, an order-of-magnitude larger than for the liquid
regime at $\lambda=2$. Therefore, $\lambda$ turns out to be a
crucial parameter for the convergence of the SP basis.

\section{Code parallelization and performances}\label{s:performance}

Since the main program of {\tt DONRODRIGO} (here referred to as {\tt
DONRODRIGO} itself) is  the most computationally intensive part of
the suite (cf.~Sec.~\ref{s:code}), it has been parallelized to take
advantage of parallel architectures. In this section the
parallelization strategies and the results of parallel benchmark are
discussed.

As anticipated in Sec.~\ref{s:code}, the code has been parallelized
using the message passing paradigm based on MPI interface. This
choice has been guided both by the possibility of having a portable
code, that could run from small departmental clusters to large
supercomputers, and, most of all, by the possibility of using the
PARPACK library, namely the parallel version of the ARPACK library
\cite{ARPACK} used in the scalar code to solve the eigenproblem,
which in turn uses MPI. The package is designed to compute a few
eigenvalues and eigenvectors of a general square matrix, and it is
most appropriate for sparse matrices (see the ARPACK user guide
\cite{ARPACK}). The algorithm reduces to a
Lanczos process (the Implicitly Restarted Lanczos Method) in the
present case of a symmetric matrix, and only requires the action of
the matrix on a vector. For a comparative analysis
of the algorithm and its performances 
with regards to other methods see Ref.~\onlinecite{refARPACK}.

Specifically, {\tt DONRODRIGO} has been parallelized as follows: In
the first part of the code the matrix elements of the Hamiltonian
$\left<\psi_i\right|\mathcal{H}\left|\psi_{i'}\right>$ to be
computed are distributed to the available processes, which then
compute all the elements in parallel. Once the distributed
Hamiltonian matrix has been built, the eigenvalues and eigenvectors
are computed using the PARPACK library. Note that in the computation
of the Hamiltonian no communication is needed between processors.
The PARPACK library manages the communications between processors by
its own, and the calling code {\tt DONRODRIGO} has only to handle
the distribution of data as required by the PARPACK itself. Together
with the data distribution, PARPACK requires that the user writes a
few parallel support subroutines. Specifically, the computation of
the product between the Hamiltonian matrix and a vector is needed.
Within this scheme both data and computations are distributed among
all processes, and then the parallel version of {\tt DONRODRIGO}
scales as the number of processors, in terms of both wallckock time
and local memory allocation.

To evaluate the performance of the code a benchmark has been
run on two parallel machines, with very different architectures:
a Linux cluster (CLX, equipped with IBM x335, PIV 3.06 GHz, dual processor
nodes and Myricom interconnection) and an IBM AIX parallel supercomputer
(SP4, equipped with p690 nodes, containing 32 Power4 1.3 GHz processors
each, and IBM ``Federation'' high performance switch interconnections).

\begin{table}
\begin{tabular}{ccc}
  \hline\hline
    & \raisebox{-1.5ex}{~subspace~} & \raisebox{-1.5ex}{non-null} \\
  ~\raisebox{1.5ex}{$S$}~ & size & ~$\left<\psi_i\right|
   \mathcal{H}\left|\psi_{i'}\right>$~\\\hline
  0 & 8018 & 3338976 \\
  1 & 11461 & 7090087 \\
  2 & 3493 & 709953 \\
  \hline
\end{tabular}
\caption{Subspace dimensions and non-null matrix elements for the
$M=0$ sector of the $N=4$ problem with $N_{\text{SP}}=36$. }
\label{tab-small-test}
\end{table}
The test case chosen for the benchmark is the problem of $N=4$ and
$N=5$ electrons in a 2D harmonic trap (see Sec.~\ref{s:test}), with
$N_{\text{SP}}=36$. Requirements of such test in terms of memory and
computations are far from typical needs of our most demanding
applications (see, e.g., Refs.~\onlinecite{Rontani05b,Rontani05c}
and Sec.~\ref{s:physics}). A first small test calculation has been
chosen where the effect of communications over the computations is
maximized. It is the case of the $M=0$ subspace for $N=4$, where the
initial SD space with linear size $N_{\text{SD}}=22972$,
corresponding to $S_z=0$, is rearranged giving three CSF subspaces
for $S=0,1,2$. The sizes and numbers of non-null matrix elements
$\left<\psi_i\right|\mathcal{H}\left|\psi_{i'}\right>$ are given in
Table~\ref{tab-small-test}. The job is small enough to run on a
single processor.
\begin{figure}
\begin{picture}(300,170)
\put(10,0){\epsfig{file=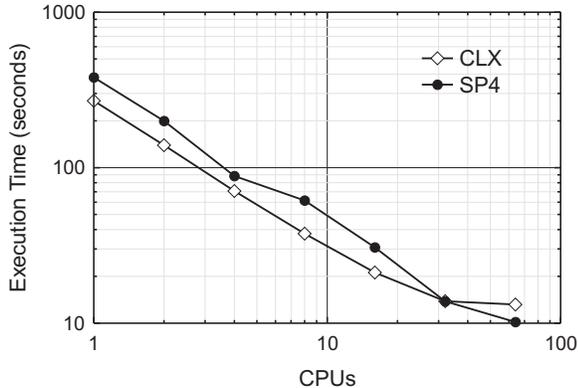,width=3.0in,,angle=0}}
\end{picture}
\caption{ Overall execution time of {\tt DONRODRIGO} on CLX and SP4,
as a function of the number of processors, for a test case with
$N=4$, $N_{\text{SP}}=36$, $M=0$, $N_{\text{SD}}=22972$. The size
and number of non-null matrix elements
$\left<\psi_i\right|\mathcal{H}\left|\psi_{i'}\right>$ of each
subspace is given in Table~\ref{tab-small-test}.}
\label{fig:performance}
\end{figure}
Even if communications between processes dominate,
nevertheless the execution time (Fig.~\ref{fig:performance}) displays a good
scalability up to 64 processors on both machines, giving the user
a large flexibility in the choice of the number of processors.

\begin{figure}
\begin{picture}(300,195)
\put(10,0){\epsfig{file=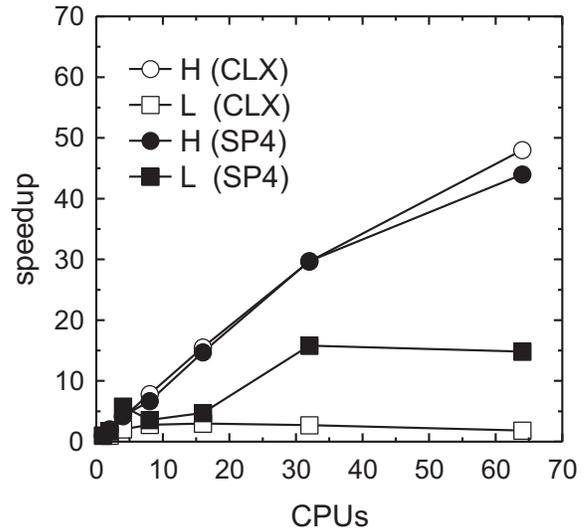,width=3.0in,,angle=0}}
\end{picture}
\caption{ Speedup vs number of CPUs (ratio of the execution time
when running on one processor to the time when using $x$ CPUs) on
CLX and SP4, for the computation of the Hamiltonian elements (H),
including I/O, and for the computation of eigenvalues and
eigenvectors using PARPACK libraries (L), respectively. Same test
case as in Fig.~\ref{fig:performance}. } \label{fig:speedup}
\end{figure}

The speedup (see Fig.~\ref{fig:speedup}) of the computation of the
Hamiltonian elements (H) scales almost linearly as expected,
since there are no communications but those used to collect the data.
For 64 processors the speedup shown in Fig.~\ref{fig:speedup} is
sublinear because the data distribution across processors
turns out to be unbalanced, since the size of the job becomes
unrealistically small.
The computation of the eigenvalues and eigenvectors (L) for this test
case is much less expensive than the computation
of the Hamiltonian (roughly 1:10), and its scalability (especially on CLX)
is very small, since the execution time is already very small on
one processor.

\begin{table}
\begin{tabular}{ccccc}
  \hline\hline
  $S$ & \multicolumn{2}{c}{subspace size} & \multicolumn{2}{c}{non-null
  $\left<\psi_i\right|\mathcal{H}\left|\psi_{i'}\right>$}\\\hline
      & $M=9$ & $M=0$ & $M=9$ & $M=0$ \\
  1/2 & 63077 & 123418 & 72197095 & 1.889862$\cdot 10^8$ \\
  3/2 & 46062 & 91896 & 46136336 & 1.118364$\cdot 10^8$ \\
  5/2 & 9896 & 20370& 2507738 & 6212517 \\
  \hline
\end{tabular}
\caption{Subspace dimensions and non-null matrix elements for the
$M=0$ and $M=9$ sectors of the $N=5$ Hilbert space with
$N_{\text{SP}}=36$. }
\label{tab-large-test}
\end{table}
In order to evaluate the parallel performance as the computational
load of jobs increases, we consider two additional test jobs for
$N=5$, corresponding, in order of increasing computational load, to
$M=9$ and $M=0$, respectively. In both cases {\tt DONRODRIGO}
rearranges the SD space into three CSF subspaces corresponding to
$S=1/2, 3/2, 5/2$, respectively. The size of each CSF subspace  and
the number of non-null Hamiltonian matrix elements is given in Table
\ref{tab-large-test}.
\begin{figure}
\begin{picture}(300,155)
\put(10,0){\epsfig{file=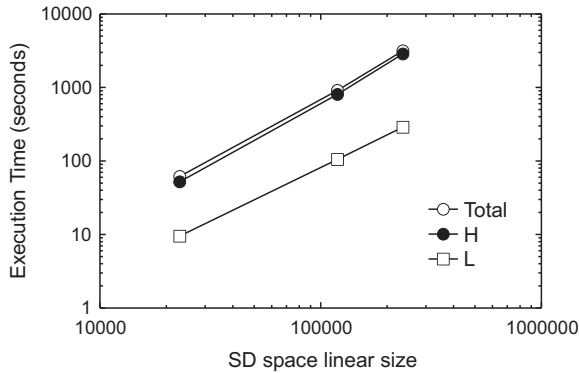,width=3.0in,,angle=0}}
\end{picture}
\caption{ Execution time on 8 SP4 CPUs vs SD space linear size for
selected jobs (see text and Table~\ref{tab-large-test}). The three
lines refer to the overall execution time (Total), the time spent to
compute Hamiltonian matrix elements (H), including I/O, and that to
compute eigenvalues and eigenvectors using PARPACK libraries (L). }
\label{fig:jobsize}
\end{figure}
In Fig.~\ref{fig:jobsize} the execution time on eight SP4 CPUs
for the three jobs mentioned above is displayed as a function of the job size.
The execution time as a function of the linear size of the
SD space scales roughly as a power law, with exponent close to two. Note
that both the computation of the Hamiltonian matrix (H) and the solution of the
eigenproblem (L) have similar behavior. This is possible because the
Lanczos method is applied, which scales with a lower exponent than a direct
diagonalization method. The ratio between the eigenproblem part (L) and the
Hamiltonian part (H) slowly decreases as the size of the system increases.
This implies that the parallelism becomes progressively more effective as the
system size increases, since part H displays a better
speedup than part L when the number of
processors increases (cf.~Fig.~\ref{fig:speedup}).

\section{Two-dimensional quantum dot in very
strong correlation regimes}\label{s:physics}

In this section we demonstrate the power and reliability of {\tt
DONRODRIGO} in treating the few-electron problem in QDs for our test
case (cf.~Sec.~\ref{s:test}).
There have been many CI calculations for few-electron QDs in
different geometries and conditions.
\cite{BryantPRL,MaksymPRL,Merkt91,Pfannkuche93,Hawrylak93,MacDonald93,Yang93,Palacios94,Pfannkuche95,Wojs96,Maksym96,Bruce97,Kouwenhoven97,Eto97,Ezaki97,Goldmann99,Creffield99,ReimannPRB,Brasken00,Manninen01,Mikhailov02,Mikhailov02bis,Yang02,Guclu02,Guclu03,Cha03,Tavernier03,Corni03,Yannouleas03,Chung04,Anisimovas04,Szafran04a,Szafran04b,Korkusinski04,Manninen05,Garcia05,Rontani05a,Rontani05b,Rontani05c,Palacios95,Oh96,Imamura96,Imamura98,Imamura99,Tokura99,Tokura00,Moreno00,Rontani01,Hu01,Harju02,Rontani02,RontaniNato,Bellucci03physe,Bellucci03prl,Rontani04,Jacob04,Wunsch05,Ota05}
Here we consider $N$ electrons in a 2D harmonic trap, as the density
is progressively diluted, namely going from a Fermi liquid behavior
(small $\lambda$) to a Wigner molecule regime (large $\lambda$).
Such a case is particularly interesting since it is the object of a
{\em querelle} between different theoretical methods (HF,
\cite{Yannouleas99} QMC, \cite{Egger99} CI \cite{ReimannPRB}) about
what value of $\lambda$ corresponds to the onset of
``crystallization'' and what the spin polarization of the
six-electron ground state is (see the introduction of
Ref.~\onlinecite{ReimannPRB} for a discussion of the physics behind
this problem).

A complete analysis of the physics emerging from our results is beyond the
aim of this paper. In this section we just focus on the convergence and
accuracy of ground state energies, for different $N$ and Hilbert
space sectors, in a wide range of $\lambda$. A possible source of
confusion in the above querelle is the fact that different authors
use different conventions to identify density values, like $\lambda$
\cite{Egger99,Reusch03,Mikhailov02bis,Mikhailov02} or $r_s$
\cite{Yannouleas99,ReimannPRB} or other. \cite{Pederiva00}
Therefore, it is not always clear which energy values, taken from
different papers, should be compared. Here we consider path integral
QMC data \cite{Egger99,Egger99E,Reusch03} for $2\le N \le 8$ and $2 \le
\lambda \le 10$ and CI data obtained from a cutoff procedure applied
to the SD space \cite{Mikhailov02bis,Mikhailov02} for $N=3,4$ and $2
\le \lambda \le 20$. Note that the extreme values of density and $N$
considered here cannot be reached simultaneously, since they are well
beyond the limits of FCI calculations performed until now for
QDs. For example, the lower density limit for the $N=6$ CI
calculation of Ref.~\onlinecite{ReimannPRB} was $\lambda \approx
3.5$, while Mikhailov \cite{Mikhailov02bis,Mikhailov02} considered a
large density range but only up to $N=4$.

\begingroup
\squeezetable
\begin{table}
\begin{tabular}{cccc|ccc|cc}
    & $\lambda$ &$M$&$S$& \multicolumn{3}{c}{\colorbox{grigio}{SP set
dimension}}\vline & Ref.~\onlinecite{Egger99,Egger99E}
& Ref.~\onlinecite{Reusch03}
  \\\hline
  \multirow{11}{*}{\begin{sideways} \fbox{\textbf{\large $N=2$ }}
  \end{sideways}} &  &
 &  & \colorbox{grigio}{21} & \colorbox{grigio}{28} &
  \colorbox{grigio}{36} &  &  \\
    & \textbf{2} & 0 & 0 & 3.7338 & 3.7312 & 3.7295 &   &        \\[-1ex]
    &            & 1 & 2 & 4.1437 & 4.1431 & 4.1427 &   &        \\\cline{2-9}
    & \textbf{4} & 0 & 0 & &  & 4.8502 &   & 4.893(7) \\[-1ex]
    &            & 1 & 2 & &  & 5.1203 &   & 5.118(8) \\\cline{2-9}
    & \textbf{6} & 0 & 0 & &  & 5.7850 &   &          \\[-1ex]
    &            & 1 & 2 & &  & 5.9910 &   &          \\\cline{2-9}
    & \textbf{8} & 0 & 0 & 6.6185 & 6.6185 & 6.6185 &   &        \\[-1ex]
    &            & 1 & 2 & 6.7873 & 6.7873 & 6.7873 &   &        \\\cline{2-9}
    & \textbf{10}& 0 & 0 & &  & 7.3840 &   &          \\[-1ex]
    &            & 1 & 2 & &  & 7.5286 &   &          \\\hline
  \multirow{11}{*}{\begin{sideways} \fbox{\textbf{\large $N=3$ }}
  \end{sideways}}
 &  &  &  &  & \colorbox{grigio}{21} &
\colorbox{grigio}{36}
  &  &  \\
   & \textbf{2} & 1 & 1 & & 8.1755 & 8.1671 & 8.16(3) &           \\[-1ex]
   &            & 0 & 3 & & 8.3244 & 8.3224 & 8.37(1) &           \\\cline{2-9}
   & \textbf{4} & 1 & 1 & & 11.046 & 11.043 & 11.05(1)& 11.055(8) \\[-1ex]
   &            & 0 & 3 & & 11.055 & 11.053 & 11.05(2)& 11.050(10)\\\cline{2-9}
   & \textbf{6} & 1 & 1 & & 13.468 & 13.467 &         &           \\[-1ex]
   &            & 0 & 3 & & 13.439 & 13.438 & 13.43(1)&           \\\cline{2-9}
   & \textbf{8} & 1 & 1 & & 15.641 & 15.634 &         &           \\[-1ex]
   &            & 0 & 3 & & 15.597 & 15.595 & 15.59(1)&           \\\cline{2-9}
   & \textbf{10}& 1 & 1 & & 17.652 & 17.630 &         &           \\[-1ex]
   &            & 0 & 3 & & 17.600 & 17.588 & 17.60(1)&           \\\hline
  \multirow{20}{*}{\begin{sideways} \fbox{\textbf{\large $N=4$ }}
 \end{sideways}}
  &  &  &  &  &
   \colorbox{grigio}{21}& \colorbox{grigio}{36} &   &
 \\
    & \textbf{2} & 0 & 2 & & & 13.626 & 13.78(6) &            \\[-1ex]
    &            & 2 & 0 & & & 13.771 &          &            \\[-1ex]
    &            & 0 & 0 & & & 13.848 &          &            \\[-1ex]
    &            & 2 & 4 & & & 14.256 & 14.30(5) &            \\\cline{2-9}
    & \textbf{4} & 0 & 2 & & & 19.035 & 19.15(4) & 19.104(6)  \\[-1ex]
    &            & 0 & 0 & & & 19.146 &          &            \\[-1ex]
    &            & 2 & 0 & & & 19.170 &          &            \\[-1ex]
    &            & 2 & 4 & & & 19.359 & 19.42(1) & 19.34(1)   \\\cline{2-9}
    & \textbf{6} & 0 & 2 & & 23.641 & 23.598  & 23.62(2)  &   \\[-1ex]
    &            & 0 & 0 & & 23.691 & 23.650  &           &   \\[-1ex]
    &            & 2 & 4 & & 23.870 & 23.805  & 23.790(12)&   \\\cline{2-9}
    &  &  &  & & \colorbox{grigio}{32} & \colorbox{grigio}{36}  &   &
 \\
    & \textbf{8} & 0 & 2 & & 27.677 & 27.675  & 27.72(1)   &   \\[-1ex]
    &            & 0 & 0 & & 27.702 & 27.699  &            &   \\[-1ex]
    &            & 2 & 4 & &        & 27.828  & 27.823(11) &   \\\cline{2-9}
    & \textbf{10}& 0 & 2 & &        & 31.429  & 31.48(2)   &   \\[-1ex]
    &            & 0 & 0 & &        & 31.441  &            &   \\[-1ex]
    &            & 2 & 4 & &        & 31.553  & 31.538(12) &   \\\hline
\end{tabular}
\caption{Comparison between FCI ground-states energies (units of
$\hbar\omega_0$)  obtained via {\tt DONRODRIGO} and QMC results
taken from the literature for $2\le N \le 4$ electrons in a 2D
harmonic trap. Few-electron states belong to different Hilbert space
sectors classified according to their total orbital angular momentum
$\hbar M$ ($z$-component) and total square spin $\hbar^2 S(S+1)$
(units of 1/2). The parameter $\lambda$ is the ratio between the
characteristic dot radius $\ell$ and the effective Bohr radius (see
text). In the middle column, we report results corresponding to
diferent SP basis set, as indicated in the cells with gray
background.}
\label{tab1}
\end{table}
\endgroup

\begingroup
\squeezetable
\begin{table}
\begin{tabular}{cccc|ccc|cc}
   & $\lambda$ &$M$&$S$& \multicolumn{3}{c}{\colorbox{grigio}
{SP set dimension}}\vline & Ref.~\onlinecite{Egger99,Egger99E} &
  Ref.~\onlinecite{Reusch03} \\\hline
  \multirow{19}{*}{\begin{sideways} \fbox{\textbf{\large $N=5$ }}
  \end{sideways}}
  &  &  &  &
  \colorbox{grigio}{21} & \colorbox{grigio}{28} & \colorbox{grigio}{36} &   &
 \\
    & \textbf{2} & 1 & 1 & 20.36 & 20.34 & 20.33 & 20.30(8) &          \\[-1ex]
    &            & 2 & 3 & 20.64 & 20.62 & 20.61 & 20.71(8) &          \\[-1ex]
    &            & 1 & 3 & 20.92 & 20.90 & 20.90 &          &          \\[-1ex]
    &            & 0 & 5 & 21.15 & 21.13 & 21.13 & 21.29(6) &     \\\cline{2-9}
    & \textbf{4} & 1 & 1 & 29.01 & 28.96 & 28.94 & 29.09(6) & 29.01(2) \\[-1ex]
    &            & 2 & 3 & 29.21 & 29.11 & 29.10 & 29.15(6) & 29.12(2) \\[-1ex]
    &            & 0 & 5 & 29.44 & 29.31 & 29.30 & 29.22(7) & 29.33(2)
                                                                  \\\cline{2-9}
    & \textbf{6} & 1 & 1 &       &       & 36.22 & 36.26(4) &          \\[-1ex]
    &            & 2 & 3 &       &       & 36.34 & 36.35(4) &          \\[-1ex]
    &            & 0 & 5 &       &       & 36.42 & 36.44(3) &     \\\cline{2-9}
    &  &  &  &  \colorbox{grigio}{28} & \colorbox{grigio}{32} &
    \colorbox{grigio}{36} &   & \\
    & \textbf{8} & 1 & 1 & 42.98 & 42.80 & 42.77 & 42.77(4) &   \\[-1ex]
    &            & 2 & 3 & 43.04 & 42.91 & 42.86 & 42.82(2) &   \\[-1ex]
    &            & 0 & 5 & 43.01 & 42.93 & 42.88 & 42.86(4) &   \\\cline{2-9}
    & \textbf{10}& 1 & 1 &       &       & 48.84 & 48.76(2) &   \\[-1ex]
    &            & 1 & 3 &       &       & 48.91 & 48.78(3) &   \\[-1ex]
    &            & 2 & 3 &       &       & 48.93 &          &   \\[-1ex]
    &            & 0 & 5 &       &       & 48.91 & 48.79(2) &   \\\hline
  \multirow{21}{*}{\begin{sideways} \fbox{\textbf{\large $N=6$ }}
  \end{sideways}}
  &  &  &  & \colorbox{grigio}{21} & \colorbox{grigio}{28}
  & \colorbox{grigio}{36} &   & \\
    & \textbf{2} & 0 & 0 &  28.03 &       & 27.98  &  &          \\[-1ex]
    &            & 1 & 2 &  28.36 &       & 28.30  &  &          \\[-1ex]
    &            & 0 & 4 &  28.54 &       & 28.48  &  &          \\[-1ex]
    &            & 0 & 6 &  28.98 &       & 28.93  &  &          \\\cline{2-9}
    & \textbf{4} & 0 & 0 &  40.74 & 40.54 & 40.45  &  & 40.53(1) \\[-1ex]
    &            & 1 & 2 &  40.96 &       &        &  & 40.62(2) \\[-1ex]
    &            & 0 & 4 &  41.06 & 40.71 & 40.66  &  & 40.69(2) \\[-1ex]
    &            & 0 & 6 &  41.29 & 40.88 & 40.85  &  & 40.83(4) \\\cline{2-9}
    & \textbf{6} & 0 & 0 &        & 51.35 & 51.02  &          &  \\[-1ex]
    &            & 0 & 4 &        & 51.38 & 51.15  &          &  \\[-1ex]
    &            & 0 & 6 &        & 51.46 & 51.25  &          &  \\\cline{2-9}
    &  &  &  & \colorbox{grigio}{26} & \colorbox{grigio}{28} &
   \colorbox{grigio}{36} &   & \\
    & \textbf{8} & 0 & 0 &  61.44 & 61.35 & 60.64  &          &  \\[-1ex]
    &            & 1 & 2 &  61.53 & 61.37 & 60.71  & 60.37(2) &  \\[-1ex]
    &            & 0 & 4 &  61.43 & 61.32 & 60.73  &          &  \\[-1ex]
    &            & 0 & 6 &  61.47 & 61.38 & 60.80  & 60.42(2) &  \\\cline{2-9}
    & \textbf{10} & 0 & 0 & & & 69.74 &  &   \\[-1ex]
    &             & 0 & 4 & & & 69.81 &  &   \\[-1ex]
    &             & 0 & 6 & & & 69.86 &  &   \\\hline
  \multirow{6}{*}{\begin{sideways} \fbox{\textbf{\large $N=7$ }}
  \end{sideways}}
  &  & &  &  \colorbox{grigio}{15} & \colorbox{grigio}{21} & &  & \\
    & \textbf{4} & 3 & 7 & 57.02 & 55.20 & &  &           \\[-1ex]
    &            & 0 & 5 & 56.53 & 54.93 & &  & 53.93(5)  \\[-1ex]
    &            & 1 & 3 & 56.55 & 54.78 & &  & 53.80(2)  \\[-1ex]
    &            & 0 & 1 & 56.60 & 54.69 & &  &           \\[-1ex]
    &            & 2 & 1 &       & 54.68 & &  & 53.71(2)  \\\hline
  \multirow{14}{*}{\begin{sideways} \fbox{\textbf{\large $N=8$ }}
  \end{sideways}}
    &  &  &  & \colorbox{grigio}{15} & \colorbox{grigio}{21} &   &   & \\
    & \textbf{2} & 0 & 2 &        & 47.14 &  & 46.5(2) &          \\[-1ex]
    &            & 0 & 0 &        & 47.26 &  &         &          \\[-1ex]
    &            & 0 & 4 &        & 47.58 &  & 46.9(3) &          \\[-1ex]
    &            & 0 & 6 &        & 48.19 &  & 47.4(3) &          \\[-1ex]
    &            & 0 & 8 &        & 48.48 &  & 48.3(2) &          \\\cline{2-9}
    & \textbf{4} & 0 & 2 &  73.54 & 70.48 &     &         & 68.44(1)  \\[-1ex]
    &            & 0 & 0 &  73.63 & 70.57 &     &         & 68.52(2)  \\[-1ex]
    &            & 0 & 4 &  73.72 & 70.65 &     & 68.3(2) &           \\[-1ex]
    &            & 0 & 6 &  74.17 & 70.73 &     & 68.5(2) &           \\[-1ex]
    &            & 0 & 8 &  74.15 & 70.83 &     & 69.2(1) &           \\[-1ex]
    &            & 1 & 0 &  73.44 &       &     &         &           \\[-1ex]
    &            & 1 & 2 &  73.44 &       &     &         & 68.44(1) \\\hline
\end{tabular}
\caption{Same comparison as in Table \ref{tab1}
for $5\le N\le 8$ electrons.}
\label{tab2}
\end{table}
\endgroup
In Tables \ref{tab1} and \ref{tab2} we compare our FCI results for
ground state energies, for $2\le\lambda\le 10$ and $2\le N\le 8$ and
different values of $S$ and $M$, with QMC data taken from
Refs.~\onlinecite{Egger99,Egger99E,Reusch03}. The energies are given
in units of $\hbar\omega_0$. Progressively larger sets of SP
orbitals are considered. Note that, contrary to available QMC
energies, which only refer to different values of $S$, here in
addition we are able to assign values of $M$. QMC data for a given
value of $S$ are compared with FCI data with $M$ corresponding to the
lowest energy for the same value of $S$. For $2\le N\le 5$ we find
an almost perfect agreement between FCI and QMC data in the full
range of $\lambda$; in several cases FCI results are slightly lower
in energy than corresponding QMC data. Since the FCI calculation
takes into account all possible SDs that can be
made from a given SP orbital set, the accuracy of the calculation
solely depends on the size of the SP basis. In this regard, we note
that a SP basis with $N_{\text{SP}}=21$ is large enough to achieve
full convergence of results, at least for $N\le 4$. For $N=5$, a
good compromise is $N_{\text{SP}}=28$ (Table \ref{tab2}). For $N\ge
6$, the larger $\lambda$, the larger the required $N_{\text{SP}}$
to achieve convergence. For $N=6$ we were able to obtain well
converged results in the full $\lambda$ range only using
$N_{\text{SP}}=36$. The agreement is excellent up to $\lambda
\approx 8$ (the slight discrepancy between FCI and QMC data is five
parts per thousand for $\lambda=8$). For $N\ge 7$ we were only able
to use $N_{\text{SP}}=21$, and of course the FCI-QMC discrepancy gets
worse. However, even in this case, the agreement between FCI and QMC
data is at least qualitatively correct. In fact, FCI and QMC agree on
the value of $S$ for the absolute ground state. Indeed, a wrong
prediction for $S$ would be a typical indication of bad convergence.

\begin{figure}
\begin{picture}(300,180)
\put(10,0){\epsfig{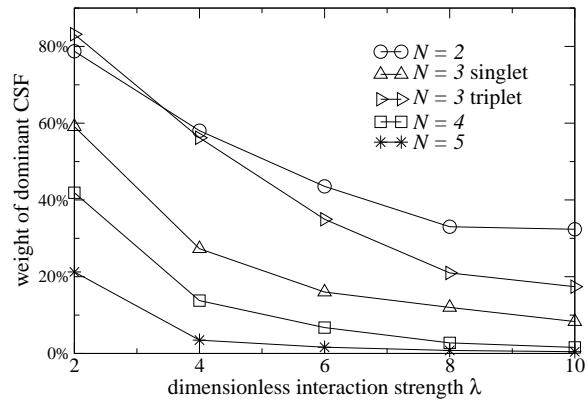}}
\end{picture}
\caption{ Weight of the CSF with the largest probability in the
linear expansion of the interacting ground state, in a two
dimensional harmonic trap, as a function of the dimensionless
interaction strength $\lambda$, defined in the main text, for
different values of the number of electrons, $N$. Note that, for
$N=3$, since there is a crossing in energy between singlet and
triplet lowest-energy states in the range $4<\lambda<6$,
both states are depicted. }
\label{fig:singleCSF}
\end{figure}

The above discussion shows that: (i) FCI gives results with accuracy
comparable to that of QMC (and in addition provides the wave
function of both ground and excited states) (ii) FCI calculation
becomes more demanding as $\lambda$ (and $N$, of course) increases.

To gain a deeper insight into point (ii), we analyzed the ground
state wave function for different values of $\lambda$ and $N$,
focusing on the CSF with the largest weight in the FCI linear
expansion (\ref{eq:CIexpansion}) of the interacting wave function.
Figure \ref{fig:singleCSF} shows that the weight of the dominant CSF
dramatically decreases as either $\lambda$ or $N$ increase. For
example, at $\lambda=2$, the dominant CSF weight changes from
$\approx 80\%$ for $N=2$ to $\approx 20\%$ for $N=5$. The weight
loss is even more dramatic as $\lambda$ increases: e.g., for $N\ge
4$ the decrease is larger than one order of magnitude as $\lambda$
goes from 2 to 10. This trend clearly illustrates the evolution
between two limiting cases: as $\lambda \to 0$, the dot turns into a
non-interacting system, and just one CSF is the exact eigenstate of
the ideal Fermi gas, with weight equal to 100\%. What CSF is the
ground state is dictated by the {\em Aufbau} principle, according to
which the lowest-energy SP orbitals are filled in with $N$
electrons. As $\lambda \to \infty$, the system evolves into the
limit of complete crystallization: electrons are perfectly localized
in space, therefore the SP basis, and consequently the CSF basis,
turn out to be extremely ineffective in representing the Wigner
molecule, unless one uses a special SP basis of localized gaussians,
or similar kinds of orbitals.

The loss of weight shown in Fig.~\ref{fig:singleCSF} depends of
course on the chosen SP basis. Therefore we expect that, by means of
a proper unitary transformation of SP orbitals, like that provided
by a self-consistent HF calculation, weights of dominant CSFs will
be in general larger than those shown in Fig.~\ref{fig:singleCSF}.
However, except for a shift of weight values, the above discussion
will not be qualitatively affected. Note that past experience in HF
calculations showed that convergence of the HF self-consistent cycle
is not necessarily achieved at low densities. \cite{Rontani99}

\begin{figure}
\begin{picture}(300,180)
\put(10,0){\epsfig{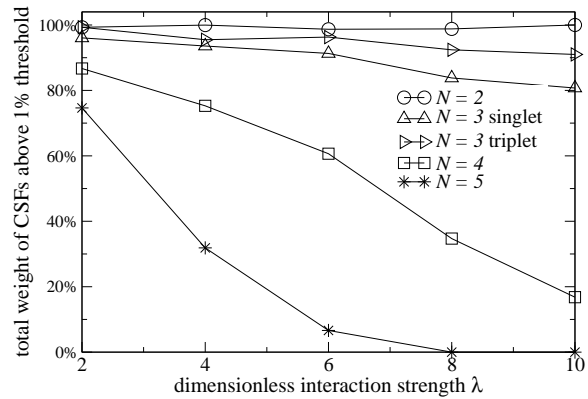}}
\end{picture}
\caption{ Total weight of CSFs whose probability in the linear
expansion of the ground state is larger than 1\% as a function of
the dimensionless interaction strength $\lambda$ for different
values of $N$. The system is the same as in
Fig.~\ref{fig:singleCSF}.} \label{fig:h-f}
\end{figure}

Figure \ref{fig:singleCSF} also shows that the triplet ground state,
for $N=3$, is relatively less affected by interaction than the
singlet ground state (see also Fig.~\ref{fig:h-f}), namely less CSFs
are needed for the triplet than for the singlet in the FCI linear
expansion (\ref{eq:CIexpansion}) of the wave function. The reason is
that in the triplet exchange interaction is sufficient to keep
electrons far apart as Coulomb repulsion between electrons
increases; the same effect can be obtained for the singlet only in
virtue of a deeper correlation hole around each electron. Such
behavior is obtained by mixing more CSFs, namely the singlet
increases its correlation with respect to the triplet.

Not only the dominant CSF loses weight as $\lambda \to \infty$, but
also more and more CSFs are needed to build up the exact wave
function. This is illustrated by Fig.~\ref{fig:h-f}, where we show
the sum of weights of ``significant'' CSFs as a function of
$\lambda$ and for different values of $N$. Here by ``significant
CSF'' we mean a configuration $\left|\psi_i\right>$ whose weight
$\left|a_i\right|^2$ is larger than 1\%
[cf.~Eq.~(\ref{eq:CIexpansion})]. Figure \ref{fig:h-f} clearly
demonstrates that, as $\lambda$ increases, the set of significant
CSFs is progressively emptied. From such result we infer that, in
regimes of strong correlation, only the \emph{full} CI approach is a
reliable algorithm. Therefore, cutoff procedures like those of
Ref.~\onlinecite{ReimannPRB}, where the full space of SDs is
filtered and truncated by considering the kinetic energy per single
SD, are potentially dangerous in strong correlation regimes where
they might not be able to warrant convergence.\cite{notaReimann}

\begingroup
\squeezetable
\begin{table}
\begin{tabular}{cccc|c|c}
   & $\lambda$ &$M$&$S$&
{\tt DONRODRIGO} & Refs.~\onlinecite{Mikhailov02bis,Mikhailov02}
   \\\hline
  \multirow{14}{*}{\begin{sideways} \fbox{\textbf{\large $N=3$ }}
  \end{sideways}}
    & \textbf{2}  & 1 & 1 &  8.1633 & 8.1651  \\
    &             & 0 & 3 &  8.3217 & 8.3221  \\\cline{2-6}
    & \textbf{4}  & 1 & 1 & 11.0423 & 11.0422 \\
    &             & 0 & 3 & 11.0527 & 11.0527 \\\cline{2-6}
    & \textbf{6}  & 1 & 1 & 13.4666 & 13.4658 \\
    &             & 0 & 3 & 13.4380 & 13.4373 \\\cline{2-6}
    & \textbf{8}  & 1 & 1 & 15.6344 & 15.6334 \\
    &             & 0 & 3 & 15.5948 & 15.5938 \\\cline{2-6}
    & \textbf{10} & 1 & 1 & 17.6293 & 17.6279 \\
    &             & 0 & 3 & 17.5877 & 17.5863 \\\cline{2-6}
    & \textbf{15} & 1 & 1 & 22.1127 &         \\
    &             & 0 & 3 & 22.0754 &         \\\cline{2-6}
    & \textbf{20} & 1 & 1 & 26.1184 &         \\
    &             & 0 & 3 & 26.0863 &         \\\hline
  \multirow{14}{*}{\begin{sideways} \fbox{\textbf{\large $N=4$ }}
  \end{sideways}}
    & \textbf{2}  & 0 & 2 & 13.6195 & 13.6180 \\
    &             & 2 & 4 & 14.2544 & 14.2535 \\\cline{2-6}
    & \textbf{4}  & 0 & 2 & 19.0335 & 19.0323 \\
    &             & 2 & 4 & 19.3578 & 19.3565 \\\cline{2-6}
    & \textbf{6}  & 0 & 2 & 23.5975 & 23.5958 \\
    &             & 2 & 4 & 23.8041 & 23.8025 \\\cline{2-6}
    & \textbf{8}  & 0 & 2 & 27.6715 & 27.6696 \\
    &             & 2 & 4 & 27.8222 & 27.8203 \\\cline{2-6}
    & \textbf{10} & 0 & 2 & 31.4148 & 31.4120 \\
    &             & 2 & 4 & 31.5352 & 31.5323 \\\cline{2-6}
    & \textbf{15} & 0 & 2 & 39.8197 & 39.8163 \\
    &             & 2 & 4 & 39.9054 & 39.8970 \\\cline{2-6}
    & \textbf{20} & 0 & 2 & 47.3443 & 47.4002 \\
    &             & 2 & 4 & 47.4153 & 47.4013 \\\hline
\end{tabular}
\caption{Comparison between different CI calculations for $N=3,4$.
FCI ground-state energies (units of $\hbar\omega_0$) from {\tt
DONRODRIGO} were obtained using a SP basis set corresponding to the
first ten lowest-energy Fock-Darwin shells (55 orbitals) and no
truncation of the many-electron space. Results taken from
Refs.~\onlinecite{Mikhailov02bis,Mikhailov02} were obtained
considering a larger number of SP orbitals but with a cutoff
procedure on the kinetic energy per single SD. $S$ is given in units
of 1/2.}
\label{tab3}
\end{table}
\endgroup
This is illustrated in Table \ref{tab3}, where we compare CI results
obtained via two different algorithms. The first column in Table
\ref{tab3} shows ground state energies, for $2\le \lambda \le 20$
and $N=3,4$, obtained from FCI {\tt DONRODRIGO} runs. Here
$N_{\text{SP}}=55$ was considered, and no cutoff procedure was
implemented. The second column collects data taken from
Refs.~\onlinecite{Mikhailov02bis,Mikhailov02}. In those papers
$N_{\text{SP}}> 55$ but only certain SDs are retained, namely those
whose kinetic energy per single SD is smaller than a certain cutoff.
For example, in Ref.~\onlinecite{Mikhailov02} the largest size of
the SD space with $N=4$, $S=1$, $M=0$ ($S=2$, $M=2$) considered was
24348 (8721) while in our calculation was 67225 (15659). The
comparison between the two sets of data gives excellent agreement,
since results systematically have at least four significant digits
identical (energies in the second column tend to have a slightly
lower energy due to the larger $N_{\text{SP}}$ which could be used
for such a small number of electrons). However, there is one
important exception, namely at very low density ($\lambda=20$,
$N=4$, $S=1$, $M=0$), where our result is 47.34 against 47.40. This
can be explained by the loss of angular correlation once one
truncates the SD space as it is done in
Ref.~\onlinecite{Mikhailov02}.

\section{Normal modes of the Wigner molecule}\label{s:yrast}

In order to illustrate the wealth of information 
provided by the FCI method with respect
to QMC techniques, in this section we give an example of
excitation spectrum calculation. In fact, the access to excited states
is a difficult task for QMC calculations, which routinely
focus on ground-state properties.\cite{ReimannRMP} Nevertheless,
the knowledge of excited-state energies and wave functions 
is of the utmost interest from both the theoretical and experimental
points of view. Indeed, several useful dynamical response functions
can be computed from excited states.
Moreover, many experimental
techniques allow nowadays to access QD excitation spectrum, such as
single-electron non-linear tunneling experiments,\cite{Rontani04,Ota05}
far-infrared spctroscopy,\cite{Jacak} inelastic light 
scattering.\cite{Garcia05} In addition, the theoretical
analysis of the energy spectrum gives a precious insight into
QD physics.

\begin{figure}
\begin{picture}(300,180)
\put(10,0){\epsfig{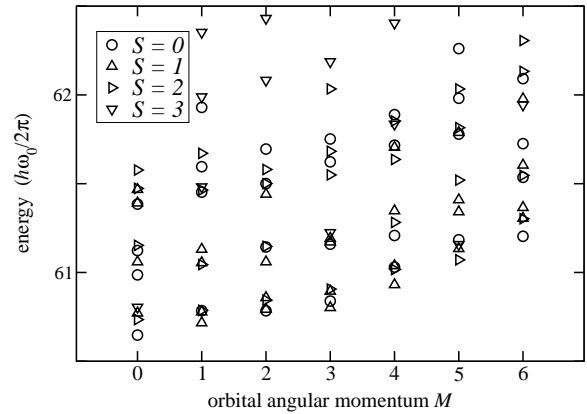}}
\end{picture}
\caption{Excitation energies of a six-electron quantum dot
in the Wigner regime ($\lambda=8$) for different values of
the total orbital angular momentum, $M$, and total spin, $S$. 
Energies are in units of $\hbar\omega_0$.}\label{fig:yrast}
\end{figure}

Figure \ref{fig:yrast} displays our FCI results for the low-energy
region of the excitation spectrum 
of a six-electron quantum dot in the Wigner regime ($\lambda=8$),  
for different values of the total orbital angular momentum
and spin multiplicities.
The plot of the lowest energies as a function of the
angular momentum is known
as {\em yrast line} in nuclear physics,\cite{ReimannRMP} and
provides several hints on the crystallized electron phase.

We see from Fig.~\ref{fig:yrast} that the ground state is the non-degenerate
spin singlet state with $M=0$, which is found to be the 
lowest-energy state in the whole range $0\le \lambda \le 10$.
The absence of any level crossing as the density is progressively
diluted (i.e.~$\lambda$ increases) implies that the crystallization
process evolves in a continuos manner, consistently with the finite-size
character of the system. Nevertheless, several features of 
Fig.~\ref{fig:yrast} demonstrate the formation of a Wigner molecule
in the dot. 

First, the three possible spin multiplets 
other than $S=0$, namely $S=1,2,3$,
lie very close in energy to the ground state. 
In general, at the lowest energies for any given $M$,
several spin multiplets in Fig.~\ref{fig:yrast}
appear as almost degenerate.

The overall behavior is well explained by invoking
electron crystallization. In the Wigner limit, the Hamiltonian
of the system turns into a classical quantity, since the 
kinetic energy term in it may be
neglected with respect to the Coulomb term. Therefore, 
only commuting operators (the electron positions) appear
in the Hamiltonian. In this regime the spin,
which has no classical counterpart, becomes 
irrelevant:\cite{Rontani05a} spin-dependent
energies show a tendency to degeneracy. This can also be understood 
in the following way: if electrons sit at some lattice sites with
unsubstatial overlap of their localized wave functions, then the
total energy must not depend on the relative orientation of
neighboring spins.

Note also that the proximity of the four possible spin multiplets 
at the lowest energies occurs for both $M=0$ and
for $M=5$. Such a period of five units on the $M$ axis identifies
a magic number,\cite{MaksymPRL,Koskinen01} whose origin is brought about
by the internal spatial symmetry of 
the interacting wave function.\cite{Ruan95} In fact, when electrons form
a stable Wigner molecule, they arrange themselves into a
{\em five-fold} symmetry configuration, where charges are localized 
at the corners of a regular pentagon plus one electron at the 
center.\cite{Bolton93,Bedanov94,Rontani02}

A second distinctive signature of crystallization is the appearance
of {\em rotational bands},\cite{Koskinen01} 
which we identify in Fig.~\ref{fig:yrast} 
as those bunches of levels, separated by energy gaps of about 
$0.2\hbar\omega_0$, that increase monotonically as $M$ increases. 
We are able to distinguish in Fig.~\ref{fig:yrast}
at least three bands, where each band is composed of different
spin multiplets. Such bands are called ``rotational'' since they
can be identified with the quantized levels $E_{\text{rot}}(M)$ 
of a rigid two-dimensional top, given by the formula
\begin{displaymath}
E_{\text{rot}}(M)=\frac{\hbar^2}{2I}M^2,
\end{displaymath}
where $I$ is the moment of inertia of the top.\cite{Koskinen01} These
excitations may be thought of as the ``normal modes'' of
the Wigner molecule rotating as a whole in the $xy$ plane
around the vertical symmetry axis parallel to $z$.

\section{Discussion}\label{s:discussion}

In the previous sections we demonstrated that {\tt DONRODRIGO} is a
flexible and well performing code, suitable to treat
strongly-correlated few-electron problems in quantum dots to the
same degree of accuracy as QMC calculations. With respect to the
latter, the FCI method also provides full access to excited states.

A major achievement of {\tt DONRODRIGO} is the
independence of core routines from both Hamiltonian and SP basis.
Differently from other CI codes, the SP basis is not optimized by a
SCF calculation, before the actual FCI calculation. Of course the
FCI results are independent from the SP basis, but the question
arises whether a proper SP basis could improve the efficiency of the
FCI calculation, at the time cost of a preliminary SCF calculation.
However, the usage of non self-consistent SP orbitals does not seem
to be a serious drawback, since {\tt DONRODRIGO} was designed to
treat a wide range of different correlation regimes, especially
those for which SCF calculations are not expected to converge. An
example of the latter circumstance is the case of very large dots
($\lambda$ large). Of course in such regime a large SP basis and a
full CI diagonalization are both necessary requirements (see
discussion in Secs.~\ref{s:SP} and \ref{s:physics}). Another major
advantage is the code simple and transparent structure, which
allowed for easy parallelization and straightforward coding of
post-processing routines.

Although very efficient also in highly correlated regimes and with
very good scaling properties in parallel architectures, our code
still lacks some smartnesses of pre-existing quantum chemistry CI
codes. One major drawback is the requirement of large amounts of
memory, since, for each Hilbert space sector, all matrix elements
$\left<\Phi_j\right|\mathcal{H}\left|\Phi_{j'}\right>$ are stored
(cf.~Sec.~\ref{s:code}). An alternative strategy is given by
so-called {\em direct} methods, 
\cite{Bauschlicher90,Raghavachari96,Shavitt98,Sherril99,Jensen99,libronecorni,Roos72,Siegbahn84,Knowles84,Bauschlicher87,Olsen88,Knowles89,Harrison91,Caballol92,Povill93,Sakai93,Daudey93,Szalay95,Hanrath97,Stephan98,Amor98,Stampfuss99,Stampfuss00,Ansaloni00,Lischka01,Ivanic03,Stampfuss05}
where required
matrix elements are computed {\em on the fly} during each step of
recursive diagonalization of the many-electron Hamiltonian. The
latter approach  constitutes an interesting direction for future
work, together with possible interfaces with other CI codes.

\section{Conclusion}\label{s:conclusion}

We presented a new high performance full CI code specifically
designed for quantum dot applications. With respect to previous CI
codes, {\tt DONRODRIGO} is powerful enough to treat demanding
correlation regimes with the same degree of accuracy as QMC method.
In addition, the code provides full information about ground and
excited states, its only limitation being the number of electrons.
Its flexible structure allows for inclusion of different device
geometries and external fields, plus easy post-processing code
development.

{\tt DONRODRIGO} is available, upon request
and on the basis of scientific collaboration,
at {\tt http://www.s3.infm.it/donrodrigo}.

\section*{ACKNOWLEDGMENTS}

Andrea Bertoni and Filippo Troiani contributed in a relevant way to
the implementation of the SP basis in an arbitrary magnetic field.
We acknowledge valuable discussions with Stefano Corni. This paper
is supported by MIUR-FIRB RBAU01ZEML, MIUR-COFIN 2003020984, INFM
Supercomputing Project 2004 and 2005 (Iniziativa Trasversale INFM
per il Calcolo Parallelo), Italian Ministry of Foreign Affairs
(Ministero degli Affari Esteri, Direzione Generale per la Promozione
e la Cooperazione Culturale).

\appendix

\section{Configurational state functions}\label{a:CSF}

In this Appendix we illustrate how  CSFs are built. The method is
straightforward, namely one writes the $\hat{S}^2$ operator in a
suitable matrix form and diagonalizes it numerically, storing the
eigenvectors (the Clebsch-Gordan coefficients $b_{ij}$) once and for
all. This approach is mentioned for example in
Ref.~\onlinecite{Pauncz79}, together with other possible strategies.
However, it turns out that the approach is non standard for actual
CI code implementations \cite{libronecorni}, and we summarize here
the algorithm for completeness. By means of simple examples we show
how it is implemented in {\tt DONRODRIGO}, following Slater.
\cite{Slater63} In this section we distinguish operators from
eigenvalues by means of the ``hat'' symbol.

\subsection{Building the configurational state function space}

Let us consider a SP basis made of $N_{\text{SP}}$ orbitals. The
corresponding SDs are built occupying such orbitals with $N$
electrons in all possible ways, taking the two possible spin
orientations into account. $N_{\text{SD}}={2N_{\text{SP}} \choose
N}$ is the total number of SDs so obtained; each SD is trivially
associated with an eigenvalue of $\hat{S}_{z}$. Note that many SDs
may be associated to the same value of $S_z$. The SD set
$\left|\Phi_i\right> \equiv \left|i\right>$, $i=1,2,\ldots,
N_{\text{SD}}$ constitutes a basis for both operators $\hat{S}_{z}$
and $\hat{\mathcal{H}}$. Therefore, the energy eigenvalues can be
obtained by diagonalization of
\begin{equation}\label{HSz}
\left< j,S_{zj}\right|\hat{\mathcal{H}}\left|i,S_{zi}\right> \qquad
i,j=1,\ldots,N_{\text{SD}}.
\end{equation}
If $S_{zi}\neq S_{zj}$ the matrix element (\ref{HSz}) is zero, i.e.,
$\hat{\mathcal{H}}$ is block diagonal in the different $S_z$
sectors. Since both operators $\hat{S}_{z}$ and $\hat{S}^{2}$
commute with the Hamiltonian, as well as with each other, it is also
possible to build a Fock space whose basis vectors are
simultaneously eigenstates of operators $\hat{S}_{z}$ and
$\hat{S}^{2}$. In this new basis, $\hat{\mathcal{H}}$ has a block
diagonal representation, labelled by $(S_z,S)$, with dimensions of
the blocks smaller than in the previous case. Below we show how one
can find simultaneous eigenstates of both $\hat{S}_{z}$ and
$\hat{S}^{2}$.

The search of the eigenstates of $\hat{S}^{2}$ starting from those
of $\hat{S}_{z}$ can be traced to that of a unitary transformation
\begin{equation}
\left|i\right> \rightarrow \widetilde{\left|i\right>},
\end{equation}
where $\left|i\right>$ are eigenstates of $\hat{S}_{z}$ and
$\widetilde{\left|i\right>}$ of $\hat{S}_{z}$ and $\hat{S}^{2}$
simultaneously. It is possible to represent the vectors of the
second basis on those of the first through a suitable linear
combination with (Clebsch-Gordan) coefficients $b_{ij}$,
\begin{equation}
\widetilde{\left|i\right>}=\sum_{j=1}^{N_{\text{SD}}}
b_{ij}\left|j\right>, \label{eq:CG}
\end{equation}
under the unitariness condition
 \begin{equation}
\sum_{j=1}^{N_{\text{SD}}}b_{ij}^*b_{lj}=\delta_{il}.
\end{equation}
The problem of finding the Clebsch-Gordan coefficients can be
greatly reduced by considering only the spins of singly occupied
orbitals, which, in general, are much less than the total number of
electrons. This is based on the following property: \emph{the SDs
with only empty or doubly occupied orbitals are eigenstates of
$\hat{S}^{2}$ with $S=0$.} To see this, consider the identity
($\hbar=1$)
\begin{equation}
\hat{S}^{2}\,=\,(\hat{S}_{x}+i\hat{S}_{y})(\hat{S}_{x}-i\hat{S}_{y})
+\hat{S}^{2}_{z}-\hat{S}_{z}, \label{squadro}
\end{equation}
which does not depend on $N$. Furthermore, let
$\left|\uparrow\right>\equiv \hat{c}^{\dagger}_{\uparrow}
\!\left|0\right>$ and $\left|\downarrow\right>\equiv
\hat{c}^{\dagger}_{\downarrow}\!\left|0\right>$ denote the spinors
related to a generic orbital. Then the following relations hold:
\begin{eqnarray}
& (\hat{S}_{x}-i\hat{S}_{y})\left|\uparrow\right>
=\left|\downarrow\right>, \nonumber \\
& (\hat{S}_{x}+i\hat{S}_{y})\left|\downarrow\right>
=\left|\uparrow\right>, \nonumber \\
& (\hat{S}_{x}-i\hat{S}_{y})\left|\downarrow\right>=0, \nonumber \\
& (\hat{S}_{x}+i\hat{S}_{y})\left|\uparrow\right>=0, \nonumber \\
& \hat{S}_{z}\left|\uparrow\right>=1/2\left|\uparrow\right>, \nonumber \\
& \hat{S}_{z}\left|\downarrow\right>=-1/2\left|\downarrow\right>.
\nonumber
\end{eqnarray}
The operator $(\hat{S}_{x}-i\hat{S}_{y})$ is a  step-down operator:
when acting on a given SD, the result is a sum of SDs each having
one of the original up spins turned down. For example,
\begin{equation}
  (\hat{S}_{x}-i\hat{S}_{y}) \left|\uparrow\uparrow\downarrow\right> =
\left|\downarrow\uparrow\downarrow\right> +
\left|\uparrow\downarrow\downarrow\right> ,
\end{equation}
where $ \left|\uparrow\uparrow\downarrow\right>\equiv
\hat{c}^{\dagger}_{a\uparrow}\hat{c}^{\dagger}_{b\uparrow}
\hat{c}^{\dagger}_{c\downarrow}\!\left|0\right>$. Analogously, the
operator $(\hat{S}_{x}+i\hat{S}_{y})$ is a step-up operator, turning
$\downarrow$ spins into $\uparrow$ spins. The above operators in
second quantized form, for just one orbital, are:
 \begin{eqnarray}
&
(\hat{S}_{x}+i\hat{S}_{y})\equiv\hat{S}^{+}=\hat{c}^{\dagger}_{\uparrow}
\hat{c}_{\downarrow}, \nonumber \\
&
(\hat{S}_{x}-i\hat{S}_{y})\equiv\hat{S}^{-}=\hat{c}^{\dagger}_{\downarrow}
\hat{c}_{\uparrow}, \nonumber \\
&
\hat{S}_{z}=1/2\left[\hat{c}^{\dagger}_{\uparrow}\hat{c}_{\uparrow}-
\hat{c}^{\dagger}_{\downarrow}\hat{c}_{\downarrow}\right]. \nonumber
\end{eqnarray}
We now consider two electrons on the same orbital. The corresponding
SD,
$\hat{c}^{\dagger}_{\uparrow}\hat{c}^{\dagger}_{\downarrow}\left|0\right>$,
is an eigenstate of $\hat{S}_{z}$ with $S_{z}=0$. To see that it is
also an eigenstate of $\hat{S}^{2}$ with $S=0$ it is sufficient to
apply (\ref{squadro}) using the definition of $\hat{S}^{+}$ and
$\hat{S}^{-}$. One verifies immediately that
\begin{equation}
\hat{S}^{+}\hat{S}^{-}\hat{c}^{\dagger}_{\uparrow}
\hat{c}^{\dagger}_{\downarrow}\left|0\right>=
\hat{c}^{\dagger}_{\uparrow}\hat{c}_{\downarrow}\hat{c}^{\dagger}_{\downarrow}
\hat{c}_{\uparrow}\hat{c}^{\dagger}_{\uparrow}
\hat{c}^{\dagger}_{\downarrow}\left|0\right>=0.
\end{equation}
In the general case, $N$ electrons are distributed over
$N_{\text{SP}}$ orbitals, so that $N\le 2N_{\text{SP}}$. The
operator $\hat{\mathbf{S}}$ is
\begin{equation}
\hat{\mathbf{S}}=\hat{\mathbf{S}}(1)+\hat{\mathbf{S}}(2)+\ldots
+\hat{\mathbf{S}}(N_{\text{SP}}),
\end{equation}
where $\hat{\mathbf{S}}(a) \equiv
[\hat{S}_{x}(a),\hat{S}_{y}(a),\hat{S}_{z}(a)]$ acts on the $a$-th
orbital. Therefore, we can write
\begin{equation}
\hat{S}^{2}\,=\,\sum_{a,b=1}^{N_{\text{SP}}}
\hat{c}^{\dagger}_{b\uparrow}\hat{c}_{b\downarrow}
\hat{c}^{\dagger}_{a\downarrow}\hat{c}_{a\uparrow}+
\hat{S}^{2}_{z}-\hat{S}_{z}. \label{eq:explicit}
\end{equation}
Again, we ask whether a generic SD, which is eigenstate of
$\hat{S}_{z}$ with some orbitals $a$ doubly occupied or otherwise
empty, is also eigenstate of $\hat{S}^{2}$. The answer is
affirmative, since, in the term $\hat{S}^{+}\hat{S}^{-}$ explicited
in Eq.~(\ref{eq:explicit}), when first $\hat{S}^{-}$ acts on the SD,
it destroys a spin $\uparrow$ in the full $a$-th orbital and creates
a spin $\downarrow$ on the same orbital; the total contribution of
$\hat{S^{+}}\hat{S^{-}}$, however, is zero, since the $a$-th orbital
is already occupied by a spin $\downarrow$ electron. Therefore, it
is proved that the SD is eigenstate of $\hat{S}^{2}$ with $S=0$.

\subsection{Examples}\label{a:examples}

Let us consider the case of SDs having some singly occupied
orbitals. For definiteness, let us focus on the example $N=4$ and
$N_{\text{SP}}=4$ ($a=0,\ldots,3$), with two electrons in orbital
$0$, zero in $2$, and one in each orbital $1$ and $3$, with opposite
spin:
\begin{equation}
\left|i_{1}\right>=\hat{c}^{\dagger}_{0\uparrow}\hat{c}^{\dagger}_{0\downarrow}
\hat{c}^{\dagger}_{1\uparrow}\hat{c}^{\dagger}_{3\downarrow}\left|0\right>.
\end{equation}
The operator $\hat{S}_{z}$, explicitly, is
\begin{equation}
\hat{S}_{z}=\,\frac{1}{2} \sum_{a=0}^3
\left(\hat{c}^{\dagger}_{a\uparrow}\hat{c}_{a\uparrow}
-\hat{c}^{\dagger}_{a\downarrow}\hat{c}_{a\downarrow}\right).
\end{equation}
Evidently, $\hat{S}_{z}\left|i_{1}\right>=0\left|i_{1}\right>$. The
goal is to build linear combinations of SDs, including
$\left|i_{1}\right>$, that are eigenstates of $\hat{S}^{2}$. To this
aim one needs to identify all the SDs which couple to
$\left|i_{1}\right>$ through the operator $\hat{S}^{2}$ and
represent $\hat{S}^{2}$ in this set. In the present case, it is
sufficient to consider only SDs with  $S_{z}=0$. It is possible to
establish \emph{a priori} what these SDs are since they differ from
$\left|i_{1}\right>$ only for the spin orientation of the electrons
on the same singly occupied orbitals ($1$ and $3$ in the present
example), while the doubly occupied or empty orbitals ($0$ and $2$,
respectively) remain unchanged. This is evident from
Eq.~(\ref{eq:explicit}), using similar arguments as above; for a
formal proof see, e.g., Ref.~\onlinecite{libronecorni}. This way,
starting from $\left|i_{1}\right>$, we build the \emph{equivalence
class} $\cal{S}$ (of which $|i_{1}\rangle$ is the representative) of
all SDs having the same doubly occupied and empty orbitals, and
differing only in the spin configurations of same singly occupied
orbitals, with the same value of $S_z$. In our example, the only
other SD belonging to $\cal{S}$ is
\begin{equation}
\left|i_{2}\right>\,=\,\hat{c}^{\dagger}_{0\uparrow}
\hat{c}^{\dagger}_{0\downarrow}
 \hat{c}^{\dagger}_{1\downarrow}
\hat{c}^{\dagger}_{3\uparrow}\left|0\right>. \label{i2}
\end{equation}

The idea is that we only need to find the Clebsch-Gordan
coefficients of the SDs belonging to the same equivalence class
$\cal{S}$. In this way we reduce the problem to the vectorial sum of
$n$ angular momenta of value $1/2$, where $n$ is the number of
unpaired electrons, and usually $n \ll N$. We approach this problem
by first writing down $\hat{S}^2$ as a matrix, and then
diagonalizing it numerically: the eigenvectors are the requested
Clebsch-Gordan coefficients. In the case considered above, the two
elements belonging to $\cal{S}$ are
\begin{equation}
\begin{array}{c} \phantom{\mbox{\tiny 1}} \\ | \end{array} \!\!
\begin{array}{c} \mbox{\tiny 1} \\ \uparrow \end{array} \!\!
\begin{array}{c} \mbox{\tiny 3} \\ \downarrow \end{array} \!\!
\begin{array}{c} \phantom{\mbox{\tiny 1}} \\ \rangle \,, \end{array}
\qquad
\begin{array}{c} \phantom{\mbox{\tiny 1}} \\ | \end{array} \!\!
\begin{array}{c} \mbox{\tiny 1} \\ \downarrow \end{array} \!\!
\begin{array}{c} \mbox{\tiny 3} \\ \uparrow \end{array} \!\!
\begin{array}{c} \phantom{\mbox{\tiny 1}} \\ \rangle \,, \end{array}
\end{equation}
where according to the above prescription we only refer to the spins
of singly occupied orbitals, 1 and 3, as indicated. On this basis,
the matrix form of $\hat{S}^2$, using Eq.~(\ref{eq:explicit}), is
\begin{equation}
\begin{pmatrix}
 1 & 1 \cr
 1 & 1 \cr
\end{pmatrix},
\label{eq:matrix}
\end{equation}
whose eigenvalues correspond to the singlet ($S=0$) and triplet
($S=1$) states. The corresponding eigenvectors provide us the
desired Clebsch-Gordan coefficients $b_{ij}$, $i,j=1,2$
[cf.~Eq.~(\ref{eq:CG})]. Specifically, the eigenstates of
$\hat{S}^2$ obtained from the SDs belonging to the equivalence class
$\cal{S}$ are:
\begin{equation}
\widetilde{\left|{i}_{1}\right>}=\frac{1}{\sqrt{2}}
\left|i_{1}\right>+\frac{1}{\sqrt{2}}\left|i_{2}\right> \qquad
(S=1),
\end{equation}
\begin{equation}
\widetilde{\left|{i}_{2}\right>}=\frac{1}{\sqrt{2}}\left|i_{1}\right>
-\frac{1}{\sqrt{2}}\left|i_{2}\right> \qquad (S=0).
\end{equation}
Note that, for the triplet state $S=1$, there are other CSFs
degenerate in energy with $S_z=\pm 1$, namely
$\left|\uparrow\uparrow\right>$ and
$\left|\downarrow\downarrow\right>$ (incidentally, the CSFs with
$S=\pm n$ are always single SDs). Such CSFs are redundant in
applications and are ignored by {\tt DONRODRIGO}. In general, the
subspace diagonalization with $S_z=0$ (or $S_z=1/2$ if $n$ is odd)
provides all spin eigenvalues allowed by symmetry. Such sectors are
the only ones considered by the code.

As another example, let us consider the case with $N_{\text{SP}}=15$
(the SP orbital index runs from 0 to 14) and $n=3$ singly occupied
orbitals ($N$ can be, of course, much larger). Let us now focus on a
given SD, e.g.,
\begin{displaymath}
\hat{c}^{\dagger}_{0\uparrow} \hat{c}^{\dagger}_{0\downarrow}
\hat{c}^{\dagger}_{3\uparrow} \hat{c}^{\dagger}_{5\uparrow}
\hat{c}^{\dagger}_{7\uparrow}
 \hat{c}^{\dagger}_{7\downarrow}
\hat{c}^{\dagger}_{14\downarrow}\left|0\right>
\end{displaymath}
($N=7$). The above SD corresponds to a state with two doubly
occupied orbitals (0 and 7), two singly occupied orbitals with spin
up (3 and the 5), one singly occupied orbital with spin down (14).
In {\tt DONRODRIGO} a SD is uniquely identified by a couple of
8-byte integers in binary representation. The first (second) integer
labels the spin-up (-down) orbitals: each of the 64 bits represents
a SP orbital; if a bit is set to 1 then the orbital is occupied by
one electron. Therefore, the above SD is coded as
\begin{equation} \label{eq:SDexample}
\begin{array}{c} \mbox{\tiny 0} \\ 1 \\ 1 \end{array} \!\!
\begin{array}{c} \mbox{\tiny 1} \\ 0 \\ 0 \end{array} \!\!
\begin{array}{c} \mbox{\tiny 2} \\ 0 \\ 0 \end{array} \!\!
\begin{array}{c} \mbox{\tiny 3} \\ 1 \\ 0 \end{array} \!\!
\begin{array}{c} \mbox{\tiny 4} \\ 0 \\ 0 \end{array} \!\!
\begin{array}{c} \mbox{\tiny 5} \\ 1 \\ 0 \end{array} \!\!
\begin{array}{c} \mbox{\tiny 6} \\ 0 \\ 0 \end{array} \!\!
\begin{array}{c} \mbox{\tiny 7} \\ 1 \\ 1 \end{array} \!\!
\begin{array}{c} \mbox{\tiny 8} \\ 0 \\ 0 \end{array} \!\!
\begin{array}{c} \mbox{\tiny 9} \\ 0 \\ 0 \end{array} \!\!
\begin{array}{c} \mbox{\tiny 10} \\ 0 \\ 0 \end{array} \!\!
\begin{array}{c} \mbox{\tiny 11} \\ 0 \\ 0 \end{array} \!\!
\begin{array}{c} \mbox{\tiny 12} \\ 0 \\ 0 \end{array} \!\!
\begin{array}{c} \mbox{\tiny 13} \\ 0 \\ 0 \end{array} \!\!
\begin{array}{c} \mbox{\tiny 14} \\ 0 \\ 1 \end{array} \,.
\end{equation}
The corresponding equivalence class $\cal{S}$ uniquely associated to
(\ref{eq:SDexample}) is coded as
\begin{equation}\label{eq:CSFexample}
\begin{array}{c} \mbox{\tiny 0} \\ 0 \\ 1 \end{array} \!\!
\begin{array}{c} \mbox{\tiny 1} \\ 0 \\ 0 \end{array} \!\!
\begin{array}{c} \mbox{\tiny 2} \\ 0 \\ 0 \end{array} \!\!
\begin{array}{c} \mbox{\tiny 3} \\ 1 \\ 0 \end{array} \!\!
\begin{array}{c} \mbox{\tiny 4} \\ 0 \\ 0 \end{array} \!\!
\begin{array}{c} \mbox{\tiny 5} \\ 1 \\ 0 \end{array} \!\!
\begin{array}{c} \mbox{\tiny 6} \\ 0 \\ 0 \end{array} \!\!
\begin{array}{c} \mbox{\tiny 7} \\ 0 \\ 1 \end{array} \!\!
\begin{array}{c} \mbox{\tiny 8} \\ 0 \\ 0 \end{array} \!\!
\begin{array}{c} \mbox{\tiny 9} \\ 0 \\ 0 \end{array} \!\!
\begin{array}{c} \mbox{\tiny 10} \\ 0 \\ 0 \end{array} \!\!
\begin{array}{c} \mbox{\tiny 11} \\ 0 \\ 0 \end{array} \!\!
\begin{array}{c} \mbox{\tiny 12} \\ 0 \\ 0 \end{array} \!\!
\begin{array}{c} \mbox{\tiny 13} \\ 0 \\ 0 \end{array} \!\!
\begin{array}{c} \mbox{\tiny 14} \\ 1 \\ 0 \end{array} \,,
\end{equation}
where the first integer labels the singly occupied orbitals: if a
bit is set to 1 then the orbital is occupied by one electron. The
second integer represents doubly occupied orbitals, using the same
convention.

Independently from the number of doubly occupied orbitals, it will
be sufficient to solve the equivalent Clebsch-Gordan problem for the
following three determinants,
\begin{equation}
\begin{array}{c} \phantom{\mbox{\tiny 1}} \\ | \end{array} \!\!
\begin{array}{c} \mbox{\tiny 3} \\ \uparrow \end{array} \!\!
\begin{array}{c} \mbox{\tiny 5} \\ \uparrow \end{array} \!\!
\begin{array}{c} \mbox{\tiny 14} \\ \downarrow \end{array} \!\!
\begin{array}{c} \phantom{\mbox{\tiny 1}} \\ \rangle \end{array}
\begin{array}{c} \phantom{\mbox{\tiny 1}} \\ | \end{array} \!\!
\begin{array}{c} \mbox{\tiny 3} \\ \downarrow \end{array} \!\!
\begin{array}{c} \mbox{\tiny 5} \\ \uparrow \end{array} \!\!
\begin{array}{c} \mbox{\tiny 14} \\ \uparrow \end{array} \!\!
\begin{array}{c} \phantom{\mbox{\tiny 1}} \\ \rangle \end{array}
\begin{array}{c} \phantom{\mbox{\tiny 1}} \\ | \end{array} \!\!
\begin{array}{c} \mbox{\tiny 3} \\ \uparrow \end{array} \!\!
\begin{array}{c} \mbox{\tiny 5} \\ \downarrow \end{array} \!\!
\begin{array}{c} \mbox{\tiny 14} \\ \uparrow \end{array} \!\!
\begin{array}{c} \phantom{\mbox{\tiny 1}} \\ \rangle, \end{array}
\end{equation}
where the first ket, $\left| \uparrow\uparrow\downarrow \right>$,
represents (\ref{eq:SDexample}). Note that for all kets $S_z=1/2$.
Therefore, we need to diagonalize $\hat{S}^{2}$ --- a $3\times 3$
matrix --- as we did with (\ref{eq:matrix}). One obtains 3
eigenvalues, $1/2,1/2,3/2$, and the associated Clebsch-Gordan
eigenvectors. Eventually, the equivalence class
(\ref{eq:CSFexample}) is associated with an additional index
labelling the pertinent Clebsch-Gordan eigenspace (in this case,
either the quadruplet or one of the two doublets). These two
entities
--- the equivalence class (\ref{eq:CSFexample}) and the
latter multiplet index --- identify the CSFs.

More generally, a $N$-electron SD, $\left|i\right>$, with $n'$
electrons in doubly occupied orbitals and $n$ in singly occupied
orbitals ($n+n'=N$), is uniquely associated with an equivalence
class ${\cal{S}}_i$. The class ${\cal{S}}_i$ is the same for all and
only those SDs which differ from $\left|i\right>$ for the spin
orientation of the $n$ electrons, keeping $S_z$ constant. {\tt
DONRODRIGO} scans the whole SD space $\{\left|i\right>\}_i$
exhaustively ($i=1,\ldots,N_{\text{SD}}$), rearranging it as a set
of different equivalence classes ${\cal{S}}_i$. Within each class
${\cal{S}}_i$, the problem is now equivalent to the vectorial sum of
$n$ spins, the maximum possible value of $n$ ($S$) being $N$
($N/2$). For large values of $n$, where the solution of matrices
like (\ref{eq:matrix}) is not possible by simple analytical methods,
$\hat{S}^2$ is diagonalized by a standard numerical routine. In such
a way Clebsch-Gordan coefficients up to $n=15$ (corresponding to a
$6435\times 6435$ matrix) were obtained and stored. By means of
associating each class ${\cal{S}}_i$ with all possible multiplet
indices, all CSFs are built.

As an example taken from an actual application,
\cite{Rontani05a,Rontani05b,Rontani05c} we consider the reduction of
the many-electron Hilbert space in our test case (Sec.~IIIC)
of a 2D harmonic trap with $N=6$. Truncating the SP basis to the
first 8 shells of Fock-Darwin orbitals ($N_{\text{SP}}=36$), we
first reduce the SD space in blocks corresponding to different
values of the total orbital angular momentum, $M$ (see Table
\ref{tab0}). Focusing on $M=0$, the dimension of the SD space
corresponding to $S_z=0$ is $N_{\text{SD}}=2459910$. From such
starting point {\tt DONRODRIGO} extracts 190380 equivalence classes,
allowing to build a CSF space divided in singlet ($S=0$, dimension
661300), triplet ($S=1$, dimension 1131738), quintuplet ($S=2$,
dimension 568896), and septuplet ($S=3$ dimension 97976) sectors,
respectively. The sum of sector dimensions is equal to
$N_{\text{SD}}$. Table \ref{tab0} is a larger list of CSF subspaces
partitioning the initial SD space for different values of $M$.
\begin{table}
\begin{tabular}{ccccccc}
 \hline\hline
 $M$ & $S=0$  &   $S=1$ &  $S=2$ & $S=3$ & $N_{\text{SD}}$ & ${\cal{S}}_i$
 \\\hline
  0  & 661300 & 1131738 & 568896 & 97976 & 2459910 & 190380 \\
  1  & 656476 & 1123952 & 564697 & 97221 & 2442346 & 189000 \\
  2  & 643242 & 1100391 & 552661 & 95043 & 2391337 & 185624 \\
  3  & 621112 & 1062496 & 532877 & 91493 & 2307978 & 179728 \\
  4  & 591897 & 1011245 & 506545 & 86741 & 2196428 & 172093 \\
  5  & 555754 & 949079  & 474285 & 80960 & 2060078 & 162429 \\
  6  & 514945 & 877805  & 437690 & 74407 & 1904847 & 151662
     \\\hline
\end{tabular}
\caption{Partitioning of the SD space into CSF subspaces for
selected values of the total orbital angular momentum $M$. We
consider $N=6$ and a SP basis made of Fock-Darwin orbitals
corresponding to the first 8 shells ($N_{\text{SP}}=36$). $S$ is the
total spin, $N_{\text{SD}}$ is the dimension of the SD space, and
${\cal{S}}_i$ is the number of equivalence classes from which CSFs
are built.} \label{tab0}
\end{table}

\subsection{Computing matrix elements}\label{s:bit-per-bit}

Since any state may be written as a linear combination of CSFs
$\left|\psi_{i}\right>$, any matrix element of the generic operator
$\hat{O}$ is a linear combination of the matrix elements
$\left<\psi_i\right|\hat{O}\left|\psi_{i'}\right>$. Such elements in
turn are computed by expanding the CSFs on the SD basis
$\left|\Phi_{j}\right>$ [Eq.~(7)]:
\begin{equation}
\left<\psi_i\right|\hat{O}\left|\psi_{i'}\right> = \sum_{jj'}
b^*_{ij}\, b_{i'j'}
\left<\Phi_j\right|\hat{O}\left|\Phi_{j'}\right>. \label{eq:matrixO}
\end{equation}

The usage of the binary representation of SDs, as in
Eq.~(\ref{eq:SDexample}), allows for extremely efficient bit-per-bit
operations when evaluating matrix elements. Consider, e.g., the
matrix element
\begin{equation}
\left<\Phi_f\right| c^{\dagger}_{a \uparrow} c^{\dagger}_{b
\downarrow} c_{c \downarrow}c_{d \uparrow}  \left|\Phi_i\right>,
\label{eq:matrixexemplum}
\end{equation}
which might represent a term in the Coulomb interaction. The ket
\begin{equation}
c^{\dagger}_{a \uparrow} c^{\dagger}_{b \downarrow} c_{c
\downarrow}c_{d \uparrow}  \left|\Phi_i\right> \label{eq:partial}
\end{equation}
will differ from $\left|\Phi_i\right>$ at most in the occupancy of
one orbital with spin up and one with spin down. Translated into the
binary representation, this implies that in either the spin-up or
-down integers identifying the ket (\ref{eq:partial}), a couple of
bits at most will be swapped with respect to $\left|\Phi_i\right>$,
their values being changed from $(0 1)$ to $(1 0)$ or vice versa (or
no change at all). The matrix element (\ref{eq:matrixexemplum}) will
differ from zero only if the ket (\ref{eq:partial}) is equal to
$\left|\Phi_f\right>$, but for the sign: this is efficiently checked
by performing a bit-per-bit {\tt xor} operation between the
same-spin integers representing $\left|\Phi_i\right>$ and
$\left|\Phi_f\right>$, respectively. Only if the number of resulting
bits set to {\tt true} (after the {\tt xor} operation per each spin)
is equal to 0 or 2, the matrix element (\ref{eq:matrixexemplum}) may
be non null. Eventually, the sign of (\ref{eq:matrixexemplum}) is
determined by counting how many bits set to 1 occur between those
swapped. With those and similar tricks all types of matrix elements
are easily evaluated.

The above procedure is especially advantageous for post-processing,
\cite{Rontani01,Rontani02,RontaniNato,Bellucci03prl,Garcia05,Rontani04,Bertoni04,Bertoni05,Rontani05a,Rontani05b,Rontani05c}
since the evalutation of various matrix elements is particularly
transparent and simple once physical operators are expressed in
their second quantized form.

\section{Coulomb matrix elements of Fock-Darwin states for the 2D
harmonic trap}\label{a:Coulombmatrix}

We report, for the sake of completeness, the explicit expressions
of the Coulomb
matrix elements $V_{abcd}$ referred to Fock-Darwin orbitals (Sec.~IIIA),
which have been used in the calculations of few-electron states in a
2D harmonic trap presented in the this paper (Sec.~V).
For a full derivation see Refs.~\onlinecite{Anisimovas98,RontaniPhD}.

Since each Fock-Darwin orbital [Eq.~(12)] is identified
by the radial and azimuthal quantum numbers $n$ and $m$,
respectively, the generic
Coulomb matrix element is identified by eight indices:
$V_{n_1m_1,n_2m_2,n_3m_3,n_4m_4}$. Among such indices, only seven
are independent, in virtue of total orbital angular momentum
conservation. The explicit expression is:
\begin{eqnarray}
&&V_{n_1m_1,n_2m_2,n_3m_3,n_4m_4}=\delta_{m_1+m_2,m_3+m_4}
\,\frac{e^2}{\kappa\ell}\nonumber\\ &&\quad\times\quad
\left[\prod_{i=1}^4\frac{n_i!}{\left(n_i+\left|m_i\right|\right)!}\right]^{1/2}
\!\!\sum_{\left( 4\right) j=0}^{n}
\frac{\left(-1\right)^{j_1+j_2+j_3+j_4}}
{j_1!j_2!j_3!j_4!}
\nonumber\\&&\quad\quad\times\quad
\prod_{l=1}^4\!{n_l+\left|m_l\right| \choose n_l-j_l}
\;2^{-G/2-1/2}\!
\sum_{\left( 4\right)\ell=0}^{\gamma}
\!\delta_{\ell_1+\ell_2,\ell_3+\ell_4}\nonumber\\
&&\quad\quad\quad\times\quad\prod_{t=1}^4\!{\gamma_t \choose \ell_t}
\left( -1\right)^{\gamma_2+\gamma_3-\ell_2-\ell_3}
\,\Gamma\!\left(\Lambda/2+1\right) \nonumber\\
&&\quad\quad\quad\quad\times\quad
\Gamma\!\left(\left[G-\Lambda+1\right]/2\right).
\label{eq:Venorme}
\end{eqnarray}
Here, $\Gamma(\xi)$ denotes the Gamma function and we use the
following conventions:
\begin{itemize}
\item[(i)] The shorthand
\begin{displaymath}
\sum_{\left( 4\right) j=0}^{n}\equiv \sum_{j_1=0}^{n_1}
\sum_{j_2=0}^{n_2}\sum_{j_3=0}^{n_3}\sum_{j_4=0}^{n_4}
\end{displaymath}
has been used.
\item[(ii)] The $\gamma$'s are defined as:
\begin{eqnarray*}
\gamma_1&=& j_1+j_4+\left(\left|m_1\right|+m_1\right)/2+
\left(\left|m_4\right|-m_4\right)/2,\nonumber\\
\gamma_2&=& j_2+j_3+\left(\left|m_2\right|+m_2\right)/2+
\left(\left|m_3\right|-m_3\right)/2,\nonumber\\
\gamma_3&=& j_2+j_3+\left(\left|m_2\right|-m_2\right)/2+
\left(\left|m_3\right|+m_3\right)/2,\nonumber\\
\gamma_4&=& j_1+j_4+\left(\left|m_1\right|-m_1\right)/2+
\left(\left|m_4\right|+m_4\right)/2.
\end{eqnarray*}
\item[(iii)] $G$ and $\Lambda$ are defined as:
\begin{eqnarray*}
G&=&\gamma_1+\gamma_2+\gamma_3+\gamma_4,\nonumber\\
\Lambda &=&\ell_1+\ell_2+\ell_3+\ell_4.
\end{eqnarray*}
\end{itemize}
%


\begin{thebibliography}{133}
\expandafter\ifx\csname natexlab\endcsname\relax\def\natexlab#1{#1}\fi
\expandafter\ifx\csname bibnamefont\endcsname\relax
  \def\bibnamefont#1{#1}\fi
\expandafter\ifx\csname bibfnamefont\endcsname\relax
  \def\bibfnamefont#1{#1}\fi
\expandafter\ifx\csname citenamefont\endcsname\relax
  \def\citenamefont#1{#1}\fi
\expandafter\ifx\csname url\endcsname\relax
  \def\url#1{\texttt{#1}}\fi
\expandafter\ifx\csname urlprefix\endcsname\relax\def\urlprefix{URL }\fi
\providecommand{\bibinfo}[2]{#2}
\providecommand{\eprint}[2][]{\url{#2}}

\bibitem[{\citenamefont{Jacak et~al.}(1998)\citenamefont{Jacak, Hawrylak, and
  W\'ojs}}]{Jacak}
\bibinfo{author}{\bibfnamefont{L.}~\bibnamefont{Jacak}},
  \bibinfo{author}{\bibfnamefont{P.}~\bibnamefont{Hawrylak}}, \bibnamefont{and}
  \bibinfo{author}{\bibfnamefont{A.}~\bibnamefont{W\'ojs}},
  \emph{\bibinfo{title}{Quantum dots}} (\bibinfo{publisher}{Springer},
  \bibinfo{address}{Berlin}, \bibinfo{year}{1998}).

\bibitem[{\citenamefont{Bimberg et~al.}(1998)\citenamefont{Bimberg, Grundmann,
  and Ledentsov}}]{Bimberg}
\bibinfo{author}{\bibfnamefont{D.}~\bibnamefont{Bimberg}},
  \bibinfo{author}{\bibfnamefont{M.}~\bibnamefont{Grundmann}},
  \bibnamefont{and} \bibinfo{author}{\bibfnamefont{N.~N.}
  \bibnamefont{Ledentsov}}, \emph{\bibinfo{title}{Quantum dot
  heterostructures}} (\bibinfo{publisher}{Wiley},
  \bibinfo{address}{Chichester}, \bibinfo{year}{1998}).

\bibitem[{\citenamefont{Woggon}(1997)}]{Woggon}
\bibinfo{author}{\bibfnamefont{U.}~\bibnamefont{Woggon}},
  \emph{\bibinfo{title}{Optical properties of semiconductor quantum dots}}
  (\bibinfo{publisher}{Springer}, \bibinfo{address}{Berlin},
  \bibinfo{year}{1997}).

\bibitem[{\citenamefont{Chakraborty}(1999)}]{Tapah}
\bibinfo{author}{\bibfnamefont{T.}~\bibnamefont{Chakraborty}},
  \emph{\bibinfo{title}{Quantum dots - A survey of the properties of artificial
  atoms}} (\bibinfo{publisher}{North-Holland}, \bibinfo{address}{Amsterdam},
  \bibinfo{year}{1999}).

\bibitem[{\citenamefont{Maksym and Chakraborty}(1990)}]{MaksymPRL}
\bibinfo{author}{\bibfnamefont{P.~A.} \bibnamefont{Maksym}} \bibnamefont{and}
  \bibinfo{author}{\bibfnamefont{T.}~\bibnamefont{Chakraborty}},
  \bibinfo{journal}{Phys.\ Rev.\ Lett.} \textbf{\bibinfo{volume}{64}},
  \bibinfo{pages}{108} (\bibinfo{year}{1990}).

\bibitem[{\citenamefont{Kastner}(1993)}]{Kastner}
\bibinfo{author}{\bibfnamefont{M.~A.} \bibnamefont{Kastner}},
  \bibinfo{journal}{Phys.~Today} \textbf{\bibinfo{volume}{46}},
  \bibinfo{pages}{24} (\bibinfo{year}{1993}).

\bibitem[{\citenamefont{Ashoori}(1996)}]{Ashoori96}
\bibinfo{author}{\bibfnamefont{R.}~\bibnamefont{Ashoori}},
  \bibinfo{journal}{Nature (London)} \textbf{\bibinfo{volume}{379}},
  \bibinfo{pages}{413} (\bibinfo{year}{1996}).

\bibitem[{\citenamefont{Tarucha et~al.}(1996)\citenamefont{Tarucha, Austing,
  Honda, van~der Hage, and Kouwenhoven}}]{Tarucha96}
\bibinfo{author}{\bibfnamefont{S.}~\bibnamefont{Tarucha}},
  \bibinfo{author}{\bibfnamefont{D.~G.} \bibnamefont{Austing}},
  \bibinfo{author}{\bibfnamefont{T.}~\bibnamefont{Honda}},
  \bibinfo{author}{\bibfnamefont{R.~J.} \bibnamefont{van~der Hage}},
  \bibnamefont{and} \bibinfo{author}{\bibfnamefont{L.~P.}
  \bibnamefont{Kouwenhoven}}, \bibinfo{journal}{Phys.\ Rev.\ Lett.}
  \textbf{\bibinfo{volume}{77}}, \bibinfo{pages}{3613} (\bibinfo{year}{1996}).

\bibitem[{\citenamefont{Rontani et~al.}(2003)\citenamefont{Rontani, Goldoni,
  and Molinari}}]{RontaniNato}
\bibinfo{author}{\bibfnamefont{M.}~\bibnamefont{Rontani}},
  \bibinfo{author}{\bibfnamefont{G.}~\bibnamefont{Goldoni}}, \bibnamefont{and}
  \bibinfo{author}{\bibfnamefont{E.}~\bibnamefont{Molinari}}, in
  \emph{\bibinfo{booktitle}{New directions in mesoscopic physics (towards
  nanoscience)}}, edited by
  \bibinfo{editor}{\bibfnamefont{R.}~\bibnamefont{Fazio}},
  \bibinfo{editor}{\bibfnamefont{V.~F.} \bibnamefont{Gantmakher}},
  \bibnamefont{and} \bibinfo{editor}{\bibfnamefont{Y.}~\bibnamefont{Imry}}
  (\bibinfo{publisher}{Kluwer}, \bibinfo{address}{Dordrecht},
  \bibinfo{year}{2003}), vol. \bibinfo{volume}{125} of
  \emph{\bibinfo{series}{NATO Science Series II: Physics and Chemistry}}, p.
  \bibinfo{pages}{361}.

\bibitem[{\citenamefont{Rontani et~al.}(1998)\citenamefont{Rontani, Rossi,
  Manghi, and Molinari}}]{Rontani98}
\bibinfo{author}{\bibfnamefont{M.}~\bibnamefont{Rontani}},
  \bibinfo{author}{\bibfnamefont{F.}~\bibnamefont{Rossi}},
  \bibinfo{author}{\bibfnamefont{F.}~\bibnamefont{Manghi}}, \bibnamefont{and}
  \bibinfo{author}{\bibfnamefont{E.}~\bibnamefont{Molinari}},
  \bibinfo{journal}{Appl.~Phys.~Lett.} \textbf{\bibinfo{volume}{72}},
  \bibinfo{pages}{957} (\bibinfo{year}{1998}).

\bibitem[{\citenamefont{Bryant}(1987)}]{BryantPRL}
\bibinfo{author}{\bibfnamefont{G.~W.} \bibnamefont{Bryant}},
  \bibinfo{journal}{Phys.\ Rev.\ Lett.} \textbf{\bibinfo{volume}{59}},
  \bibinfo{pages}{1140} (\bibinfo{year}{1987}).

\bibitem[{\citenamefont{Garc\'{\i}a et~al.}(2005)\citenamefont{Garc\'{\i}a,
  Pellegrini, Pinczuk, Rontani, Goldoni, Molinari, Dennis, Pfeiffer, and
  West}}]{Garcia05}
\bibinfo{author}{\bibfnamefont{C.~P.} \bibnamefont{Garc\'{\i}a}},
  \bibinfo{author}{\bibfnamefont{V.}~\bibnamefont{Pellegrini}},
  \bibinfo{author}{\bibfnamefont{A.}~\bibnamefont{Pinczuk}},
  \bibinfo{author}{\bibfnamefont{M.}~\bibnamefont{Rontani}},
  \bibinfo{author}{\bibfnamefont{G.}~\bibnamefont{Goldoni}},
  \bibinfo{author}{\bibfnamefont{E.}~\bibnamefont{Molinari}},
  \bibinfo{author}{\bibfnamefont{B.~S.} \bibnamefont{Dennis}},
  \bibinfo{author}{\bibfnamefont{L.~N.} \bibnamefont{Pfeiffer}},
  \bibnamefont{and} \bibinfo{author}{\bibfnamefont{K.~W.} \bibnamefont{West}},
  \bibinfo{journal}{Phys.\ Rev.\ Lett.} \textbf{\bibinfo{volume}{95}},
  \bibinfo{pages}{266806} (\bibinfo{year}{2005}).

\bibitem[{\citenamefont{Rontani et~al.}(2004)\citenamefont{Rontani, Amaha,
  Muraki, Manghi, Molinari, Tarucha, and Austing}}]{Rontani04}
\bibinfo{author}{\bibfnamefont{M.}~\bibnamefont{Rontani}},
  \bibinfo{author}{\bibfnamefont{S.}~\bibnamefont{Amaha}},
  \bibinfo{author}{\bibfnamefont{K.}~\bibnamefont{Muraki}},
  \bibinfo{author}{\bibfnamefont{F.}~\bibnamefont{Manghi}},
  \bibinfo{author}{\bibfnamefont{E.}~\bibnamefont{Molinari}},
  \bibinfo{author}{\bibfnamefont{S.}~\bibnamefont{Tarucha}}, \bibnamefont{and}
  \bibinfo{author}{\bibfnamefont{D.~G.} \bibnamefont{Austing}},
  \bibinfo{journal}{Phys.\ Rev.\ B} \textbf{\bibinfo{volume}{69}},
  \bibinfo{pages}{85327} (\bibinfo{year}{2004}).

\bibitem[{\citenamefont{Rontani et~al.}(2001)\citenamefont{Rontani, Troiani,
  Hohenester, and Molinari}}]{Rontani01}
\bibinfo{author}{\bibfnamefont{M.}~\bibnamefont{Rontani}},
  \bibinfo{author}{\bibfnamefont{F.}~\bibnamefont{Troiani}},
  \bibinfo{author}{\bibfnamefont{U.}~\bibnamefont{Hohenester}},
  \bibnamefont{and} \bibinfo{author}{\bibfnamefont{E.}~\bibnamefont{Molinari}},
  \bibinfo{journal}{Solid State Commun.} \textbf{\bibinfo{volume}{119}},
  \bibinfo{pages}{309} (\bibinfo{year}{2001}), \bibinfo{note}{special Issue on
  Spin Effects in Mesoscopic Systems}.

\bibitem[{\citenamefont{Rontani et~al.}(2002)\citenamefont{Rontani, Goldoni,
  Manghi, and Molinari}}]{Rontani02}
\bibinfo{author}{\bibfnamefont{M.}~\bibnamefont{Rontani}},
  \bibinfo{author}{\bibfnamefont{G.}~\bibnamefont{Goldoni}},
  \bibinfo{author}{\bibfnamefont{F.}~\bibnamefont{Manghi}}, \bibnamefont{and}
  \bibinfo{author}{\bibfnamefont{E.}~\bibnamefont{Molinari}},
  \bibinfo{journal}{Europhys.\ Lett.} \textbf{\bibinfo{volume}{58}},
  \bibinfo{pages}{555} (\bibinfo{year}{2002}).

\bibitem[{\citenamefont{Reimann and Manninen}(2002)}]{ReimannRMP}
\bibinfo{author}{\bibfnamefont{S.~M.} \bibnamefont{Reimann}} \bibnamefont{and}
  \bibinfo{author}{\bibfnamefont{M.}~\bibnamefont{Manninen}},
  \bibinfo{journal}{Rev.~Mod.~Phys.} \textbf{\bibinfo{volume}{74}},
  \bibinfo{pages}{1283} (\bibinfo{year}{2002}).

\bibitem[{\citenamefont{Goldoni et~al.}(2005)\citenamefont{Goldoni, Troiani,
  Rontani, Bellucci, Molinari, and Hohenester}}]{GoldoniNato}
\bibinfo{author}{\bibfnamefont{G.}~\bibnamefont{Goldoni}},
  \bibinfo{author}{\bibfnamefont{F.}~\bibnamefont{Troiani}},
  \bibinfo{author}{\bibfnamefont{M.}~\bibnamefont{Rontani}},
  \bibinfo{author}{\bibfnamefont{D.}~\bibnamefont{Bellucci}},
  \bibinfo{author}{\bibfnamefont{E.}~\bibnamefont{Molinari}}, \bibnamefont{and}
  \bibinfo{author}{\bibfnamefont{U.}~\bibnamefont{Hohenester}}, in
  \emph{\bibinfo{booktitle}{Quantum Dots: Fundamentals, Applications, and
  Frontiers}}, edited by \bibinfo{editor}{\bibfnamefont{B.~A.}
  \bibnamefont{Joyce}},
  \bibinfo{editor}{\bibfnamefont{P.}~\bibnamefont{Kelires}},
  \bibinfo{editor}{\bibfnamefont{A.}~\bibnamefont{Naumovets}},
  \bibnamefont{and} \bibinfo{editor}{\bibfnamefont{D.~D.}
  \bibnamefont{Vvedensky}} (\bibinfo{publisher}{Springer},
  \bibinfo{year}{2005}), vol. \bibinfo{volume}{190} of
  \emph{\bibinfo{series}{NATO Science Series II: Mathematics, Physics and
  Chemistry}}, p. \bibinfo{pages}{269}.

\bibitem[{\citenamefont{Grabert and Devoret}(1992)}]{Grabert92}
\bibinfo{author}{\bibfnamefont{H.}~\bibnamefont{Grabert}} \bibnamefont{and}
  \bibinfo{author}{\bibfnamefont{M.~H.} \bibnamefont{Devoret}},
  \emph{\bibinfo{title}{Single charge tunneling: Coulomb blockade phenomena in
  nanostructures}} (\bibinfo{publisher}{Plenum}, \bibinfo{address}{New York},
  \bibinfo{year}{1992}), vol. \bibinfo{volume}{294} of
  \emph{\bibinfo{series}{NATO ASI series B: physics}}.

\bibitem[{\citenamefont{Arakawa and Sakaki}(1982)}]{Arakawa82}
\bibinfo{author}{\bibfnamefont{Y.}~\bibnamefont{Arakawa}} \bibnamefont{and}
  \bibinfo{author}{\bibfnamefont{H.}~\bibnamefont{Sakaki}},
  \bibinfo{journal}{Appl.~Phys.~Lett.} \textbf{\bibinfo{volume}{40}},
  \bibinfo{pages}{939} (\bibinfo{year}{1982}).

\bibitem[{\citenamefont{Kirstaedter et~al.}(1994)\citenamefont{Kirstaedter,
  Ledentsov, Grundmann, Bimberg, Ustinov, Ruvimov, Maximov, Kopev, Alferov,
  Richter et~al.}}]{Kirstaedter94}
\bibinfo{author}{\bibfnamefont{N.}~\bibnamefont{Kirstaedter}},
  \bibinfo{author}{\bibfnamefont{N.~N.} \bibnamefont{Ledentsov}},
  \bibinfo{author}{\bibfnamefont{M.}~\bibnamefont{Grundmann}},
  \bibinfo{author}{\bibfnamefont{D.}~\bibnamefont{Bimberg}},
  \bibinfo{author}{\bibfnamefont{V.~M.} \bibnamefont{Ustinov}},
  \bibinfo{author}{\bibfnamefont{S.~S.} \bibnamefont{Ruvimov}},
  \bibinfo{author}{\bibfnamefont{M.~V.} \bibnamefont{Maximov}},
  \bibinfo{author}{\bibfnamefont{P.~S.} \bibnamefont{Kopev}},
  \bibinfo{author}{\bibfnamefont{Z.~I.} \bibnamefont{Alferov}},
  \bibinfo{author}{\bibfnamefont{U.}~\bibnamefont{Richter}},
  \bibnamefont{et~al.}, \bibinfo{journal}{Electron.~Lett.}
  \textbf{\bibinfo{volume}{30}}, \bibinfo{pages}{1416} (\bibinfo{year}{1994}).

\bibitem[{\citenamefont{Edwards et~al.}(1995)\citenamefont{Edwards, Niu,
  Georgakis, and de~Lozanne}}]{Edwards95}
\bibinfo{author}{\bibfnamefont{H.~L.} \bibnamefont{Edwards}},
  \bibinfo{author}{\bibfnamefont{Q.}~\bibnamefont{Niu}},
  \bibinfo{author}{\bibfnamefont{G.~A.} \bibnamefont{Georgakis}},
  \bibnamefont{and} \bibinfo{author}{\bibfnamefont{A.~L.}
  \bibnamefont{de~Lozanne}}, \bibinfo{journal}{Phys.\ Rev.\ B}
  \textbf{\bibinfo{volume}{52}}, \bibinfo{pages}{5714} (\bibinfo{year}{1995}).

\bibitem[{\citenamefont{Michalet et~al.}(2005)\citenamefont{Michalet, Pinaud,
  Bentolila, Tsay, Doose, Li, Sundaresan, Wu, Gambhir, and Weiss}}]{Michalet05}
\bibinfo{author}{\bibfnamefont{X.}~\bibnamefont{Michalet}},
  \bibinfo{author}{\bibfnamefont{F.~F.} \bibnamefont{Pinaud}},
  \bibinfo{author}{\bibfnamefont{L.~A.} \bibnamefont{Bentolila}},
  \bibinfo{author}{\bibfnamefont{J.~M.} \bibnamefont{Tsay}},
  \bibinfo{author}{\bibfnamefont{S.}~\bibnamefont{Doose}},
  \bibinfo{author}{\bibfnamefont{J.~J.} \bibnamefont{Li}},
  \bibinfo{author}{\bibfnamefont{G.}~\bibnamefont{Sundaresan}},
  \bibinfo{author}{\bibfnamefont{A.~M.} \bibnamefont{Wu}},
  \bibinfo{author}{\bibfnamefont{S.~S.} \bibnamefont{Gambhir}},
  \bibnamefont{and} \bibinfo{author}{\bibfnamefont{S.}~\bibnamefont{Weiss}},
  \bibinfo{journal}{Science} \textbf{\bibinfo{volume}{307}},
  \bibinfo{pages}{538} (\bibinfo{year}{2005}).

\bibitem[{\citenamefont{Bailey et~al.}(2004)\citenamefont{Bailey, Smith, and
  Nie}}]{Bailey04}
\bibinfo{author}{\bibfnamefont{R.~E.} \bibnamefont{Bailey}},
  \bibinfo{author}{\bibfnamefont{A.~M.} \bibnamefont{Smith}}, \bibnamefont{and}
  \bibinfo{author}{\bibfnamefont{S.}~\bibnamefont{Nie}},
  \bibinfo{journal}{Physica E} \textbf{\bibinfo{volume}{25}},
  \bibinfo{pages}{1} (\bibinfo{year}{2004}).

\bibitem[{\citenamefont{Loss and DiVincenzo}(1998)}]{Loss98}
\bibinfo{author}{\bibfnamefont{D.}~\bibnamefont{Loss}} \bibnamefont{and}
  \bibinfo{author}{\bibfnamefont{D.~P.} \bibnamefont{DiVincenzo}},
  \bibinfo{journal}{Phys.\ Rev.\ A} \textbf{\bibinfo{volume}{57}},
  \bibinfo{pages}{120} (\bibinfo{year}{1998}).

\bibitem[{\citenamefont{Troiani et~al.}(2000)\citenamefont{Troiani, Hohenester,
  and Molinari}}]{FilippoQIP}
\bibinfo{author}{\bibfnamefont{F.}~\bibnamefont{Troiani}},
  \bibinfo{author}{\bibfnamefont{U.}~\bibnamefont{Hohenester}},
  \bibnamefont{and} \bibinfo{author}{\bibfnamefont{E.}~\bibnamefont{Molinari}},
  \bibinfo{journal}{Phys.\ Rev.\ B} \textbf{\bibinfo{volume}{62}},
  \bibinfo{pages}{RC2263} (\bibinfo{year}{2000}).

\bibitem[{\citenamefont{Biolatti et~al.}(2000)\citenamefont{Biolatti, Iotti,
  Zanardi, and Rossi}}]{Biolatti00}
\bibinfo{author}{\bibfnamefont{E.}~\bibnamefont{Biolatti}},
  \bibinfo{author}{\bibfnamefont{R.~C.} \bibnamefont{Iotti}},
  \bibinfo{author}{\bibfnamefont{P.}~\bibnamefont{Zanardi}}, \bibnamefont{and}
  \bibinfo{author}{\bibfnamefont{F.}~\bibnamefont{Rossi}},
  \bibinfo{journal}{Phys.\ Rev.\ Lett.} \textbf{\bibinfo{volume}{85}},
  \bibinfo{pages}{5647} (\bibinfo{year}{2000}).

\bibitem[{\citenamefont{Bastard}(1998)}]{Bastard}
\bibinfo{author}{\bibfnamefont{G.}~\bibnamefont{Bastard}},
  \emph{\bibinfo{title}{Wave mechanics applied to semiconductor
  heterostructures}} (\bibinfo{publisher}{Les Editions de Physique},
  \bibinfo{address}{Les Ulis, France}, \bibinfo{year}{1998}).

\bibitem[{\citenamefont{Yu and Cardona}(1996)}]{Cardona}
\bibinfo{author}{\bibfnamefont{P.~Y.} \bibnamefont{Yu}} \bibnamefont{and}
  \bibinfo{author}{\bibfnamefont{M.}~\bibnamefont{Cardona}},
  \emph{\bibinfo{title}{Fundamentals of semiconductors}}
  (\bibinfo{publisher}{Springer}, \bibinfo{address}{Berlin},
  \bibinfo{year}{1996}).

\bibitem[{\citenamefont{Wang and Zunger}(1999)}]{Zunger99}
\bibinfo{author}{\bibfnamefont{L.-W.} \bibnamefont{Wang}} \bibnamefont{and}
  \bibinfo{author}{\bibfnamefont{A.}~\bibnamefont{Zunger}},
  \bibinfo{journal}{Phys. Rev. B} \textbf{\bibinfo{volume}{59}},
  \bibinfo{pages}{15806} (\bibinfo{year}{1999}).

\bibitem[{\citenamefont{Bauschlicher et~al.}(1990)\citenamefont{Bauschlicher,
  Langhoff, and Taylor}}]{Bauschlicher90}
\bibinfo{author}{\bibfnamefont{C.~W.} \bibnamefont{Bauschlicher}},
  \bibinfo{author}{\bibfnamefont{S.~R.} \bibnamefont{Langhoff}},
  \bibnamefont{and} \bibinfo{author}{\bibfnamefont{P.~R.}
  \bibnamefont{Taylor}}, \bibinfo{journal}{Adv.~Chem.~Phys.}
  \textbf{\bibinfo{volume}{77}}, \bibinfo{pages}{103} (\bibinfo{year}{1990}).

\bibitem[{\citenamefont{Raghavachari and Anderson}(1996)}]{Raghavachari96}
\bibinfo{author}{\bibfnamefont{K.}~\bibnamefont{Raghavachari}}
  \bibnamefont{and} \bibinfo{author}{\bibfnamefont{J.~B.}
  \bibnamefont{Anderson}}, \bibinfo{journal}{J.~Phys.~Chem.}
  \textbf{\bibinfo{volume}{100}}, \bibinfo{pages}{12960}
  (\bibinfo{year}{1996}).

\bibitem[{\citenamefont{Shavitt}(1998)}]{Shavitt98}
\bibinfo{author}{\bibfnamefont{I.}~\bibnamefont{Shavitt}},
  \bibinfo{journal}{Mol.~Phys.} \textbf{\bibinfo{volume}{94}},
  \bibinfo{pages}{3} (\bibinfo{year}{1998}).

\bibitem[{\citenamefont{Sherril and Schaefer}(1999)}]{Sherril99}
\bibinfo{author}{\bibfnamefont{C.~D.} \bibnamefont{Sherril}} \bibnamefont{and}
  \bibinfo{author}{\bibfnamefont{H.~F.} \bibnamefont{Schaefer}},
  \bibinfo{journal}{Advances in Quantum Chemistry}
  \textbf{\bibinfo{volume}{34}}, \bibinfo{pages}{143} (\bibinfo{year}{1999}).

\bibitem[{\citenamefont{Jensen}(1999)}]{Jensen99}
\bibinfo{author}{\bibfnamefont{F.}~\bibnamefont{Jensen}},
  \emph{\bibinfo{title}{Introduction to computational chemistry}}
  (\bibinfo{publisher}{Wiley}, \bibinfo{address}{Chichester},
  \bibinfo{year}{1999}).

\bibitem[{\citenamefont{Helgaker et~al.}(2000)\citenamefont{Helgaker,
  J{\o}rgensen, and Olsen}}]{libronecorni}
\bibinfo{author}{\bibfnamefont{T.}~\bibnamefont{Helgaker}},
  \bibinfo{author}{\bibfnamefont{P.}~\bibnamefont{J{\o}rgensen}},
  \bibnamefont{and} \bibinfo{author}{\bibfnamefont{J.}~\bibnamefont{Olsen}},
  \emph{\bibinfo{title}{Molecular electronic-structure theory}}
  (\bibinfo{publisher}{Wiley}, \bibinfo{address}{Chichester},
  \bibinfo{year}{2000}).

\bibitem[{man()}]{manzoni}
\bibinfo{note}{{\tt DONRODRIGO} is available, upon request and on the basis of
  scientific cooperation, at {\tt http://www.s3.infm.it/donrodrigo}.}

\bibitem[{ate()}]{atenziun}
\bibinfo{note}{The only CI calculation in QDs exploting $S^2$ symmetry we are
  aware of is Ref.~\onlinecite{Wensauer04}.}

\bibitem[{\citenamefont{Roos}(1972)}]{Roos72}
\bibinfo{author}{\bibfnamefont{B.}~\bibnamefont{Roos}},
  \bibinfo{journal}{Chem.~Phys.~Lett.} \textbf{\bibinfo{volume}{15}},
  \bibinfo{pages}{153} (\bibinfo{year}{1972}).

\bibitem[{\citenamefont{Siegbahn}(1984)}]{Siegbahn84}
\bibinfo{author}{\bibfnamefont{P.~E.~M.} \bibnamefont{Siegbahn}},
  \bibinfo{journal}{Chem.~Phys.~Lett.} \textbf{\bibinfo{volume}{109}},
  \bibinfo{pages}{417} (\bibinfo{year}{1984}).

\bibitem[{\citenamefont{Knowles and Handy}(1984)}]{Knowles84}
\bibinfo{author}{\bibfnamefont{P.~J.} \bibnamefont{Knowles}} \bibnamefont{and}
  \bibinfo{author}{\bibfnamefont{N.~C.} \bibnamefont{Handy}},
  \bibinfo{journal}{Chem.~Phys.~Lett.} \textbf{\bibinfo{volume}{111}},
  \bibinfo{pages}{315} (\bibinfo{year}{1984}).

\bibitem[{\citenamefont{Bauschlicher}(1987)}]{Bauschlicher87}
\bibinfo{author}{\bibfnamefont{C.~W.} \bibnamefont{Bauschlicher}},
  \bibinfo{journal}{Theo.~Chim.~Acta} \textbf{\bibinfo{volume}{71}},
  \bibinfo{pages}{263} (\bibinfo{year}{1987}).

\bibitem[{\citenamefont{Olsen et~al.}(1988)\citenamefont{Olsen, Roos,
  J{\o}rgensen, and Jensen}}]{Olsen88}
\bibinfo{author}{\bibfnamefont{J.}~\bibnamefont{Olsen}},
  \bibinfo{author}{\bibfnamefont{B.~O.} \bibnamefont{Roos}},
  \bibinfo{author}{\bibfnamefont{P.}~\bibnamefont{J{\o}rgensen}},
  \bibnamefont{and} \bibinfo{author}{\bibfnamefont{H.~J.~A.}
  \bibnamefont{Jensen}}, \bibinfo{journal}{J.~Chem.~Phys.}
  \textbf{\bibinfo{volume}{89}}, \bibinfo{pages}{2185} (\bibinfo{year}{1988}).

\bibitem[{\citenamefont{Knowles and Handy}(1989)}]{Knowles89}
\bibinfo{author}{\bibfnamefont{P.~J.} \bibnamefont{Knowles}} \bibnamefont{and}
  \bibinfo{author}{\bibfnamefont{N.~C.} \bibnamefont{Handy}},
  \bibinfo{journal}{J.~Chem.~Phys.} \textbf{\bibinfo{volume}{91}},
  \bibinfo{pages}{2396} (\bibinfo{year}{1989}).

\bibitem[{\citenamefont{Harrison}(1991)}]{Harrison91}
\bibinfo{author}{\bibfnamefont{R.~J.} \bibnamefont{Harrison}},
  \bibinfo{journal}{J.~Chem.~Phys.} \textbf{\bibinfo{volume}{94}},
  \bibinfo{pages}{5021} (\bibinfo{year}{1991}).

\bibitem[{\citenamefont{Caballol and Malrieu}(1992)}]{Caballol92}
\bibinfo{author}{\bibfnamefont{R.}~\bibnamefont{Caballol}} \bibnamefont{and}
  \bibinfo{author}{\bibfnamefont{J.~P.} \bibnamefont{Malrieu}},
  \bibinfo{journal}{Chem.~Phys.~Lett.} \textbf{\bibinfo{volume}{188}},
  \bibinfo{pages}{543} (\bibinfo{year}{1992}).

\bibitem[{\citenamefont{Povill et~al.}(1993)\citenamefont{Povill, Caballol,
  Rubio, and Malrieu}}]{Povill93}
\bibinfo{author}{\bibfnamefont{A.}~\bibnamefont{Povill}},
  \bibinfo{author}{\bibfnamefont{R.}~\bibnamefont{Caballol}},
  \bibinfo{author}{\bibfnamefont{J.}~\bibnamefont{Rubio}}, \bibnamefont{and}
  \bibinfo{author}{\bibfnamefont{J.~P.} \bibnamefont{Malrieu}},
  \bibinfo{journal}{Chem.~Phys.~Lett.} \textbf{\bibinfo{volume}{209}},
  \bibinfo{pages}{126} (\bibinfo{year}{1993}).

\bibitem[{\citenamefont{Sakai and Tanaka}(1993)}]{Sakai93}
\bibinfo{author}{\bibfnamefont{T.}~\bibnamefont{Sakai}} \bibnamefont{and}
  \bibinfo{author}{\bibfnamefont{K.}~\bibnamefont{Tanaka}},
  \bibinfo{journal}{Theo.~Chim.~Acta} \textbf{\bibinfo{volume}{85}},
  \bibinfo{pages}{451} (\bibinfo{year}{1993}).

\bibitem[{\citenamefont{Daudey et~al.}(1993)\citenamefont{Daudey, Heully, and
  Malrieu}}]{Daudey93}
\bibinfo{author}{\bibfnamefont{J.~P.} \bibnamefont{Daudey}},
  \bibinfo{author}{\bibfnamefont{J.~L.} \bibnamefont{Heully}},
  \bibnamefont{and} \bibinfo{author}{\bibfnamefont{J.~P.}
  \bibnamefont{Malrieu}}, \bibinfo{journal}{J.~Chem.~Phys.}
  \textbf{\bibinfo{volume}{99}}, \bibinfo{pages}{1240} (\bibinfo{year}{1993}).

\bibitem[{\citenamefont{Szalay and Bartlett}(1995)}]{Szalay95}
\bibinfo{author}{\bibfnamefont{P.~G.} \bibnamefont{Szalay}} \bibnamefont{and}
  \bibinfo{author}{\bibfnamefont{R.~J.} \bibnamefont{Bartlett}},
  \bibinfo{journal}{J.~Chem.~Phys.} \textbf{\bibinfo{volume}{103}},
  \bibinfo{pages}{3600} (\bibinfo{year}{1995}).

\bibitem[{\citenamefont{Hanrath and Engels}(1997)}]{Hanrath97}
\bibinfo{author}{\bibfnamefont{M.}~\bibnamefont{Hanrath}} \bibnamefont{and}
  \bibinfo{author}{\bibfnamefont{B.}~\bibnamefont{Engels}},
  \bibinfo{journal}{Chem.~Phys.} \textbf{\bibinfo{volume}{225}},
  \bibinfo{pages}{197} (\bibinfo{year}{1997}).

\bibitem[{\citenamefont{Stephan and Wenzel}(1998)}]{Stephan98}
\bibinfo{author}{\bibfnamefont{F.}~\bibnamefont{Stephan}} \bibnamefont{and}
  \bibinfo{author}{\bibfnamefont{W.}~\bibnamefont{Wenzel}},
  \bibinfo{journal}{J.~Chem.~Phys.} \textbf{\bibinfo{volume}{108}},
  \bibinfo{pages}{1015} (\bibinfo{year}{1998}).

\bibitem[{\citenamefont{Ben~Amor and Maynau}(1998)}]{Amor98}
\bibinfo{author}{\bibfnamefont{N.}~\bibnamefont{Ben~Amor}} \bibnamefont{and}
  \bibinfo{author}{\bibfnamefont{D.}~\bibnamefont{Maynau}},
  \bibinfo{journal}{Chem.~Phys.~Lett.} \textbf{\bibinfo{volume}{286}},
  \bibinfo{pages}{211} (\bibinfo{year}{1998}).

\bibitem[{\citenamefont{Stampfu{\ss} et~al.}(1999)\citenamefont{Stampfu{\ss},
  Wenzel, and Keiter}}]{Stampfuss99}
\bibinfo{author}{\bibfnamefont{P.}~\bibnamefont{Stampfu{\ss}}},
  \bibinfo{author}{\bibfnamefont{W.}~\bibnamefont{Wenzel}}, \bibnamefont{and}
  \bibinfo{author}{\bibfnamefont{H.}~\bibnamefont{Keiter}},
  \bibinfo{journal}{J.~Comp.~Chem.} \textbf{\bibinfo{volume}{20}},
  \bibinfo{pages}{1559} (\bibinfo{year}{1999}).

\bibitem[{\citenamefont{Stampfu{\ss} et~al.}(2000)\citenamefont{Stampfu{\ss},
  Hamacher, and Wenzel}}]{Stampfuss00}
\bibinfo{author}{\bibfnamefont{P.}~\bibnamefont{Stampfu{\ss}}},
  \bibinfo{author}{\bibfnamefont{K.}~\bibnamefont{Hamacher}}, \bibnamefont{and}
  \bibinfo{author}{\bibfnamefont{W.}~\bibnamefont{Wenzel}},
  \bibinfo{journal}{J.~Mol.~Struct.} \textbf{\bibinfo{volume}{506}},
  \bibinfo{pages}{99} (\bibinfo{year}{2000}).

\bibitem[{\citenamefont{Ansaloni et~al.}(2000)\citenamefont{Ansaloni,
  Bendazzoli, Evangelisti, and Rossi}}]{Ansaloni00}
\bibinfo{author}{\bibfnamefont{R.}~\bibnamefont{Ansaloni}},
  \bibinfo{author}{\bibfnamefont{G.~L.} \bibnamefont{Bendazzoli}},
  \bibinfo{author}{\bibfnamefont{S.}~\bibnamefont{Evangelisti}},
  \bibnamefont{and} \bibinfo{author}{\bibfnamefont{E.}~\bibnamefont{Rossi}},
  \bibinfo{journal}{Comp.~Phys.~Commun.} \textbf{\bibinfo{volume}{128}},
  \bibinfo{pages}{496} (\bibinfo{year}{2000}).

\bibitem[{\citenamefont{Lischka et~al.}(2001)\citenamefont{Lischka, Shepard,
  Pitzer, Shavitt, Dallos, Muller, Szalay, Seth, Kedziora, Yabushita
  et~al.}}]{Lischka01}
\bibinfo{author}{\bibfnamefont{H.}~\bibnamefont{Lischka}},
  \bibinfo{author}{\bibfnamefont{R.}~\bibnamefont{Shepard}},
  \bibinfo{author}{\bibfnamefont{R.~M.} \bibnamefont{Pitzer}},
  \bibinfo{author}{\bibfnamefont{I.}~\bibnamefont{Shavitt}},
  \bibinfo{author}{\bibfnamefont{M.}~\bibnamefont{Dallos}},
  \bibinfo{author}{\bibfnamefont{T.}~\bibnamefont{Muller}},
  \bibinfo{author}{\bibfnamefont{P.~G.} \bibnamefont{Szalay}},
  \bibinfo{author}{\bibfnamefont{M.}~\bibnamefont{Seth}},
  \bibinfo{author}{\bibfnamefont{G.~S.} \bibnamefont{Kedziora}},
  \bibinfo{author}{\bibfnamefont{S.}~\bibnamefont{Yabushita}},
  \bibnamefont{et~al.}, \bibinfo{journal}{Phys.~Chem.~Chem.~Phys.}
  \textbf{\bibinfo{volume}{3}}, \bibinfo{pages}{664} (\bibinfo{year}{2001}).

\bibitem[{\citenamefont{Ivanic}(2003)}]{Ivanic03}
\bibinfo{author}{\bibfnamefont{J.}~\bibnamefont{Ivanic}},
  \bibinfo{journal}{J.~Chem.~Phys.} \textbf{\bibinfo{volume}{119}},
  \bibinfo{pages}{9364} (\bibinfo{year}{2003}).

\bibitem[{\citenamefont{Stampfu{\ss} and Wenzel}(2005)}]{Stampfuss05}
\bibinfo{author}{\bibfnamefont{P.}~\bibnamefont{Stampfu{\ss}}}
  \bibnamefont{and} \bibinfo{author}{\bibfnamefont{W.}~\bibnamefont{Wenzel}},
  \bibinfo{journal}{J.~Chem.~Phys.} \textbf{\bibinfo{volume}{122}},
  \bibinfo{pages}{024110} (\bibinfo{year}{2005}).

\bibitem[{\citenamefont{Landau and Lifshitz}(1958)}]{Landau}
\bibinfo{author}{\bibfnamefont{L.~D.} \bibnamefont{Landau}} \bibnamefont{and}
  \bibinfo{author}{\bibfnamefont{E.~M.} \bibnamefont{Lifshitz}},
  \emph{\bibinfo{title}{Quantum mechanics -- Non relativistic theory}}
  (\bibinfo{publisher}{Pergamon Press}, \bibinfo{address}{Oxford},
  \bibinfo{year}{1958}).

\bibitem[{\citenamefont{Pauncz}(1979)}]{Pauncz79}
\bibinfo{author}{\bibfnamefont{R.}~\bibnamefont{Pauncz}},
  \emph{\bibinfo{title}{Spin eigenfunctions -- Construction and use}}
  (\bibinfo{publisher}{Plenum Press}, \bibinfo{address}{New York},
  \bibinfo{year}{1979}).

\bibitem[{\citenamefont{McWeeny}(1992)}]{McWeeny92}
\bibinfo{author}{\bibfnamefont{R.}~\bibnamefont{McWeeny}},
  \emph{\bibinfo{title}{Methods of molecular quantum mechanics}}
  (\bibinfo{publisher}{Academic Press}, \bibinfo{address}{London},
  \bibinfo{year}{1992}).

\bibitem[{\citenamefont{Lehoucq et~al.}(1997)\citenamefont{Lehoucq, Maschhoff,
  Sorensen, and Yang}}]{ARPACK}
\bibinfo{author}{\bibfnamefont{R.~B.} \bibnamefont{Lehoucq}},
  \bibinfo{author}{\bibfnamefont{K.}~\bibnamefont{Maschhoff}},
  \bibinfo{author}{\bibfnamefont{D.~C.} \bibnamefont{Sorensen}},
  \bibnamefont{and} \bibinfo{author}{\bibfnamefont{C.}~\bibnamefont{Yang}},
  \emph{\bibinfo{title}{ARPACK computer code}} (\bibinfo{year}{1997}),
  \bibinfo{note}{available at {\tt http://www.caam.rice.edu/software/ARPACK/}}.

\bibitem[{\citenamefont{Bellucci
  et~al.}(2004{\natexlab{a}})\citenamefont{Bellucci, Rontani, Troiani, Goldoni,
  and Molinari}}]{Bellucci03prl}
\bibinfo{author}{\bibfnamefont{D.}~\bibnamefont{Bellucci}},
  \bibinfo{author}{\bibfnamefont{M.}~\bibnamefont{Rontani}},
  \bibinfo{author}{\bibfnamefont{F.}~\bibnamefont{Troiani}},
  \bibinfo{author}{\bibfnamefont{G.}~\bibnamefont{Goldoni}}, \bibnamefont{and}
  \bibinfo{author}{\bibfnamefont{E.}~\bibnamefont{Molinari}},
  \bibinfo{journal}{Phys.\ Rev.\ B} \textbf{\bibinfo{volume}{69}},
  \bibinfo{pages}{RC201308} (\bibinfo{year}{2004}{\natexlab{a}}).

\bibitem[{\citenamefont{Rontani and Molinari}(2005)}]{Rontani05b}
\bibinfo{author}{\bibfnamefont{M.}~\bibnamefont{Rontani}} \bibnamefont{and}
  \bibinfo{author}{\bibfnamefont{E.}~\bibnamefont{Molinari}},
  \bibinfo{journal}{Phys.\ Rev.\ B} \textbf{\bibinfo{volume}{71}},
  \bibinfo{pages}{233106} (\bibinfo{year}{2005}).

\bibitem[{\citenamefont{Rontani and Molinari}(2006)}]{Rontani05c}
\bibinfo{author}{\bibfnamefont{M.}~\bibnamefont{Rontani}} \bibnamefont{and}
  \bibinfo{author}{\bibfnamefont{E.}~\bibnamefont{Molinari}},
  \bibinfo{journal}{Jp.~J.~Appl.~Phys.} \textbf{\bibinfo{volume}{45}}
  (\bibinfo{year}{2006}), \bibinfo{note}{available as cond-mat/0507688}.

\bibitem[{\citenamefont{Bertoni et~al.}(2004)\citenamefont{Bertoni, Rontani,
  Goldoni, Troiani, and Molinari}}]{Bertoni04}
\bibinfo{author}{\bibfnamefont{A.}~\bibnamefont{Bertoni}},
  \bibinfo{author}{\bibfnamefont{M.}~\bibnamefont{Rontani}},
  \bibinfo{author}{\bibfnamefont{G.}~\bibnamefont{Goldoni}},
  \bibinfo{author}{\bibfnamefont{F.}~\bibnamefont{Troiani}}, \bibnamefont{and}
  \bibinfo{author}{\bibfnamefont{E.}~\bibnamefont{Molinari}},
  \bibinfo{journal}{Appl.~Phys.~Lett.} \textbf{\bibinfo{volume}{85}},
  \bibinfo{pages}{4729} (\bibinfo{year}{2004}).

\bibitem[{\citenamefont{Bertoni et~al.}(2005)\citenamefont{Bertoni, Rontani,
  Goldoni, and Molinari}}]{Bertoni05}
\bibinfo{author}{\bibfnamefont{A.}~\bibnamefont{Bertoni}},
  \bibinfo{author}{\bibfnamefont{M.}~\bibnamefont{Rontani}},
  \bibinfo{author}{\bibfnamefont{G.}~\bibnamefont{Goldoni}}, \bibnamefont{and}
  \bibinfo{author}{\bibfnamefont{E.}~\bibnamefont{Molinari}},
  \bibinfo{journal}{Phys.~Rev.~Lett.} \textbf{\bibinfo{volume}{95}},
  \bibinfo{pages}{066806} (\bibinfo{year}{2005}).

\bibitem[{\citenamefont{Slater}(1963)}]{Slater63}
\bibinfo{author}{\bibfnamefont{J.~C.} \bibnamefont{Slater}},
  \emph{\bibinfo{title}{Quantum theory of molecules and solids, Vol.~1}}
  (\bibinfo{publisher}{McGraw-Hill}, \bibinfo{address}{New York},
  \bibinfo{year}{1963}).

\bibitem[{\citenamefont{Brask\'en et~al.}(2000)\citenamefont{Brask\'en,
  Lindberg, Sundholm, and Olsen}}]{Brasken00}
\bibinfo{author}{\bibfnamefont{M.}~\bibnamefont{Brask\'en}},
  \bibinfo{author}{\bibfnamefont{M.}~\bibnamefont{Lindberg}},
  \bibinfo{author}{\bibfnamefont{D.}~\bibnamefont{Sundholm}}, \bibnamefont{and}
  \bibinfo{author}{\bibfnamefont{J.}~\bibnamefont{Olsen}},
  \bibinfo{journal}{Phys.\ Rev.\ B} \textbf{\bibinfo{volume}{61}},
  \bibinfo{pages}{7652} (\bibinfo{year}{2000}).

\bibitem[{\citenamefont{Corni et~al.}(2003)\citenamefont{Corni, Brask\'en,
  Lindberg, Olsen, and Sundholm}}]{Corni03}
\bibinfo{author}{\bibfnamefont{S.}~\bibnamefont{Corni}},
  \bibinfo{author}{\bibfnamefont{M.}~\bibnamefont{Brask\'en}},
  \bibinfo{author}{\bibfnamefont{M.}~\bibnamefont{Lindberg}},
  \bibinfo{author}{\bibfnamefont{J.}~\bibnamefont{Olsen}}, \bibnamefont{and}
  \bibinfo{author}{\bibfnamefont{D.}~\bibnamefont{Sundholm}},
  \bibinfo{journal}{Phys.\ Rev.\ B} \textbf{\bibinfo{volume}{67}},
  \bibinfo{pages}{45313} (\bibinfo{year}{2003}).

\bibitem[{\citenamefont{Bennett et~al.}(1997)\citenamefont{Bennett, Shanabrook,
  Thibado, Whitman, and Magno}}]{Bennett97}
\bibinfo{author}{\bibfnamefont{B.~R.} \bibnamefont{Bennett}},
  \bibinfo{author}{\bibfnamefont{B.~V.} \bibnamefont{Shanabrook}},
  \bibinfo{author}{\bibfnamefont{P.~M.} \bibnamefont{Thibado}},
  \bibinfo{author}{\bibfnamefont{L.~J.} \bibnamefont{Whitman}},
  \bibnamefont{and} \bibinfo{author}{\bibfnamefont{R.}~\bibnamefont{Magno}},
  \bibinfo{journal}{J. Cryst. Growth} \textbf{\bibinfo{volume}{175/176}},
  \bibinfo{pages}{888} (\bibinfo{year}{1997}).

\bibitem[{\citenamefont{Bellucci
  et~al.}(2004{\natexlab{b}})\citenamefont{Bellucci, Rontani, Goldoni, Troiani,
  and Molinari}}]{Bellucci03physe}
\bibinfo{author}{\bibfnamefont{D.}~\bibnamefont{Bellucci}},
  \bibinfo{author}{\bibfnamefont{M.}~\bibnamefont{Rontani}},
  \bibinfo{author}{\bibfnamefont{G.}~\bibnamefont{Goldoni}},
  \bibinfo{author}{\bibfnamefont{F.}~\bibnamefont{Troiani}}, \bibnamefont{and}
  \bibinfo{author}{\bibfnamefont{E.}~\bibnamefont{Molinari}},
  \bibinfo{journal}{Physica E} \textbf{\bibinfo{volume}{22}},
  \bibinfo{pages}{482} (\bibinfo{year}{2004}{\natexlab{b}}).

\bibitem[{\citenamefont{Bellucci
  et~al.}(2004{\natexlab{c}})\citenamefont{Bellucci, Troiani, Goldoni, and
  Molinari}}]{Bellucci04}
\bibinfo{author}{\bibfnamefont{D.}~\bibnamefont{Bellucci}},
  \bibinfo{author}{\bibfnamefont{F.}~\bibnamefont{Troiani}},
  \bibinfo{author}{\bibfnamefont{G.}~\bibnamefont{Goldoni}}, \bibnamefont{and}
  \bibinfo{author}{\bibfnamefont{E.}~\bibnamefont{Molinari}},
  \bibinfo{journal}{Phys.\ Rev.\ B} \textbf{\bibinfo{volume}{70}},
  \bibinfo{pages}{205332} (\bibinfo{year}{2004}{\natexlab{c}}).

\bibitem[{\citenamefont{Egger et~al.}(1999{\natexlab{a}})\citenamefont{Egger,
  H{\"a}usler, Mak, and Grabert}}]{Egger99}
\bibinfo{author}{\bibfnamefont{R.}~\bibnamefont{Egger}},
  \bibinfo{author}{\bibfnamefont{W.}~\bibnamefont{H{\"a}usler}},
  \bibinfo{author}{\bibfnamefont{C.~H.} \bibnamefont{Mak}}, \bibnamefont{and}
  \bibinfo{author}{\bibfnamefont{H.}~\bibnamefont{Grabert}},
  \bibinfo{journal}{Phys.\ Rev.\ Lett.} \textbf{\bibinfo{volume}{82}},
  \bibinfo{pages}{3320} (\bibinfo{year}{1999}{\natexlab{a}}).

\bibitem[{\citenamefont{Egger et~al.}(1999{\natexlab{b}})\citenamefont{Egger,
  H{\"a}usler, Mak, and Grabert}}]{Egger99E}
\bibinfo{author}{\bibfnamefont{R.}~\bibnamefont{Egger}},
  \bibinfo{author}{\bibfnamefont{W.}~\bibnamefont{H{\"a}usler}},
  \bibinfo{author}{\bibfnamefont{C.~H.} \bibnamefont{Mak}}, \bibnamefont{and}
  \bibinfo{author}{\bibfnamefont{H.}~\bibnamefont{Grabert}},
  \bibinfo{journal}{Phys.\ Rev.\ Lett.} \textbf{\bibinfo{volume}{83}},
  \bibinfo{pages}{E462} (\bibinfo{year}{1999}{\natexlab{b}}).

\bibitem[{\citenamefont{Reusch and Egger}(2003)}]{Reusch03}
\bibinfo{author}{\bibfnamefont{B.}~\bibnamefont{Reusch}} \bibnamefont{and}
  \bibinfo{author}{\bibfnamefont{R.}~\bibnamefont{Egger}},
  \bibinfo{journal}{Europhys.\ Lett.} \textbf{\bibinfo{volume}{64}},
  \bibinfo{pages}{84} (\bibinfo{year}{2003}).

\bibitem[{\citenamefont{Lehoucq and Scott}(1996)}]{refARPACK}
\bibinfo{author}{\bibfnamefont{R.~B.} \bibnamefont{Lehoucq}} \bibnamefont{and}
  \bibinfo{author}{\bibfnamefont{J.~A.} \bibnamefont{Scott}}
  (\bibinfo{year}{1996}), \bibinfo{note}{preprint MCS-P547-1195, Argonne
  National Laboratory, available at {\tt
  ftp://info.mcs.anl.gov/pub/tech\_reports/reports/P547.ps.Z}}.

\bibitem[{\citenamefont{Merkt et~al.}(1991)\citenamefont{Merkt, Huser, and
  Wagner}}]{Merkt91}
\bibinfo{author}{\bibfnamefont{U.}~\bibnamefont{Merkt}},
  \bibinfo{author}{\bibfnamefont{J.}~\bibnamefont{Huser}}, \bibnamefont{and}
  \bibinfo{author}{\bibfnamefont{M.}~\bibnamefont{Wagner}},
  \bibinfo{journal}{Phys.\ Rev.\ B} \textbf{\bibinfo{volume}{43}},
  \bibinfo{pages}{7320} (\bibinfo{year}{1991}).

\bibitem[{\citenamefont{Pfannkuche et~al.}(1993)\citenamefont{Pfannkuche,
  Gudmundsson, and Maksym}}]{Pfannkuche93}
\bibinfo{author}{\bibfnamefont{D.}~\bibnamefont{Pfannkuche}},
  \bibinfo{author}{\bibfnamefont{V.}~\bibnamefont{Gudmundsson}},
  \bibnamefont{and} \bibinfo{author}{\bibfnamefont{P.~A.}
  \bibnamefont{Maksym}}, \bibinfo{journal}{Phys.\ Rev.\ B}
  \textbf{\bibinfo{volume}{47}}, \bibinfo{pages}{2244} (\bibinfo{year}{1993}).

\bibitem[{\citenamefont{Hawrylak}(1993)}]{Hawrylak93}
\bibinfo{author}{\bibfnamefont{P.}~\bibnamefont{Hawrylak}},
  \bibinfo{journal}{Phys.\ Rev.\ Lett.} \textbf{\bibinfo{volume}{71}},
  \bibinfo{pages}{3347} (\bibinfo{year}{1993}).

\bibitem[{\citenamefont{MacDonald and Johnson}(1993)}]{MacDonald93}
\bibinfo{author}{\bibfnamefont{A.~H.} \bibnamefont{MacDonald}}
  \bibnamefont{and} \bibinfo{author}{\bibfnamefont{M.~D.}
  \bibnamefont{Johnson}}, \bibinfo{journal}{Phys.\ Rev.\ Lett.}
  \textbf{\bibinfo{volume}{70}}, \bibinfo{pages}{3107} (\bibinfo{year}{1993}).

\bibitem[{\citenamefont{Yang et~al.}(1993)\citenamefont{Yang, MacDonald, and
  Johnson}}]{Yang93}
\bibinfo{author}{\bibfnamefont{S.-R.~E.} \bibnamefont{Yang}},
  \bibinfo{author}{\bibfnamefont{A.~H.} \bibnamefont{MacDonald}},
  \bibnamefont{and} \bibinfo{author}{\bibfnamefont{M.~D.}
  \bibnamefont{Johnson}}, \bibinfo{journal}{Phys.\ Rev.\ Lett.}
  \textbf{\bibinfo{volume}{71}}, \bibinfo{pages}{3194} (\bibinfo{year}{1993}).

\bibitem[{\citenamefont{Palacios et~al.}(1993)\citenamefont{Palacios,
  Mart\'{\i}n-Moreno, Chiappe, Louis, and Tejedor}}]{Palacios94}
\bibinfo{author}{\bibfnamefont{J.~J.} \bibnamefont{Palacios}},
  \bibinfo{author}{\bibfnamefont{L.}~\bibnamefont{Mart\'{\i}n-Moreno}},
  \bibinfo{author}{\bibfnamefont{G.}~\bibnamefont{Chiappe}},
  \bibinfo{author}{\bibfnamefont{E.}~\bibnamefont{Louis}}, \bibnamefont{and}
  \bibinfo{author}{\bibfnamefont{C.}~\bibnamefont{Tejedor}},
  \bibinfo{journal}{Phys.\ Rev.\ B} \textbf{\bibinfo{volume}{50}},
  \bibinfo{pages}{5760} (\bibinfo{year}{1993}).

\bibitem[{\citenamefont{Pfannkuche and Ulloa}(1995)}]{Pfannkuche95}
\bibinfo{author}{\bibfnamefont{D.}~\bibnamefont{Pfannkuche}} \bibnamefont{and}
  \bibinfo{author}{\bibfnamefont{S.~E.} \bibnamefont{Ulloa}},
  \bibinfo{journal}{Phys.\ Rev.\ Lett.} \textbf{\bibinfo{volume}{74}},
  \bibinfo{pages}{1194} (\bibinfo{year}{1995}).

\bibitem[{\citenamefont{Wojs and Hawrylak}(1996)}]{Wojs96}
\bibinfo{author}{\bibfnamefont{A.}~\bibnamefont{Wojs}} \bibnamefont{and}
  \bibinfo{author}{\bibfnamefont{P.}~\bibnamefont{Hawrylak}},
  \bibinfo{journal}{Phys.\ Rev.\ B} \textbf{\bibinfo{volume}{53}},
  \bibinfo{pages}{10841} (\bibinfo{year}{1996}).

\bibitem[{\citenamefont{Maksym}(1996)}]{Maksym96}
\bibinfo{author}{\bibfnamefont{P.~A.} \bibnamefont{Maksym}},
  \bibinfo{journal}{Phys.\ Rev.\ B} \textbf{\bibinfo{volume}{53}},
  \bibinfo{pages}{10871} (\bibinfo{year}{1996}).

\bibitem[{\citenamefont{Bruce and Maksym}(2000)}]{Bruce97}
\bibinfo{author}{\bibfnamefont{N.~A.} \bibnamefont{Bruce}} \bibnamefont{and}
  \bibinfo{author}{\bibfnamefont{P.~A.} \bibnamefont{Maksym}},
  \bibinfo{journal}{Phys.\ Rev.\ B} \textbf{\bibinfo{volume}{61}},
  \bibinfo{pages}{4718} (\bibinfo{year}{2000}).

\bibitem[{\citenamefont{Kouwenhoven et~al.}(1997)\citenamefont{Kouwenhoven,
  Oosterkamp, Danoesastro, Eto, Austing, Honda, and Tarucha}}]{Kouwenhoven97}
\bibinfo{author}{\bibfnamefont{L.~P.} \bibnamefont{Kouwenhoven}},
  \bibinfo{author}{\bibfnamefont{T.~H.} \bibnamefont{Oosterkamp}},
  \bibinfo{author}{\bibfnamefont{M.~W.~S.} \bibnamefont{Danoesastro}},
  \bibinfo{author}{\bibfnamefont{M.}~\bibnamefont{Eto}},
  \bibinfo{author}{\bibfnamefont{D.~G.} \bibnamefont{Austing}},
  \bibinfo{author}{\bibfnamefont{T.}~\bibnamefont{Honda}}, \bibnamefont{and}
  \bibinfo{author}{\bibfnamefont{S.}~\bibnamefont{Tarucha}},
  \bibinfo{journal}{Science} \textbf{\bibinfo{volume}{278}},
  \bibinfo{pages}{1788} (\bibinfo{year}{1997}).

\bibitem[{\citenamefont{Eto}(1997)}]{Eto97}
\bibinfo{author}{\bibfnamefont{M.}~\bibnamefont{Eto}},
  \bibinfo{journal}{Jpn.~J.~Appl.~Phys.} \textbf{\bibinfo{volume}{36}},
  \bibinfo{pages}{3924} (\bibinfo{year}{1997}).

\bibitem[{\citenamefont{Ezaki et~al.}(1997)\citenamefont{Ezaki, Mori, and
  Hamaguchi}}]{Ezaki97}
\bibinfo{author}{\bibfnamefont{T.}~\bibnamefont{Ezaki}},
  \bibinfo{author}{\bibfnamefont{N.}~\bibnamefont{Mori}}, \bibnamefont{and}
  \bibinfo{author}{\bibfnamefont{C.}~\bibnamefont{Hamaguchi}},
  \bibinfo{journal}{Phys.\ Rev.\ B} \textbf{\bibinfo{volume}{56}},
  \bibinfo{pages}{6428} (\bibinfo{year}{1997}).

\bibitem[{\citenamefont{Goldmann and Renn}(1999)}]{Goldmann99}
\bibinfo{author}{\bibfnamefont{E.}~\bibnamefont{Goldmann}} \bibnamefont{and}
  \bibinfo{author}{\bibfnamefont{S.~R.} \bibnamefont{Renn}},
  \bibinfo{journal}{Phys.\ Rev.\ B} \textbf{\bibinfo{volume}{60}},
  \bibinfo{pages}{16611} (\bibinfo{year}{1999}).

\bibitem[{\citenamefont{Creffield et~al.}(1999)\citenamefont{Creffield,
  H{\"a}usler, Jefferson, and Sarkar}}]{Creffield99}
\bibinfo{author}{\bibfnamefont{C.~E.} \bibnamefont{Creffield}},
  \bibinfo{author}{\bibfnamefont{W.}~\bibnamefont{H{\"a}usler}},
  \bibinfo{author}{\bibfnamefont{J.~H.} \bibnamefont{Jefferson}},
  \bibnamefont{and} \bibinfo{author}{\bibfnamefont{S.}~\bibnamefont{Sarkar}},
  \bibinfo{journal}{Phys.\ Rev.\ B} \textbf{\bibinfo{volume}{59}},
  \bibinfo{pages}{10719} (\bibinfo{year}{1999}).

\bibitem[{\citenamefont{Reimann et~al.}(2000)\citenamefont{Reimann, Koskinen,
  and Manninen}}]{ReimannPRB}
\bibinfo{author}{\bibfnamefont{S.~M.} \bibnamefont{Reimann}},
  \bibinfo{author}{\bibfnamefont{M.}~\bibnamefont{Koskinen}}, \bibnamefont{and}
  \bibinfo{author}{\bibfnamefont{M.}~\bibnamefont{Manninen}},
  \bibinfo{journal}{Phys.\ Rev.\ B} \textbf{\bibinfo{volume}{62}},
  \bibinfo{pages}{8108} (\bibinfo{year}{2000}).

\bibitem[{\citenamefont{Manninen et~al.}(2001)\citenamefont{Manninen, Koskinen,
  Reimann, and Mottelson}}]{Manninen01}
\bibinfo{author}{\bibfnamefont{M.}~\bibnamefont{Manninen}},
  \bibinfo{author}{\bibfnamefont{M.}~\bibnamefont{Koskinen}},
  \bibinfo{author}{\bibfnamefont{S.~M.} \bibnamefont{Reimann}},
  \bibnamefont{and}
  \bibinfo{author}{\bibfnamefont{B.}~\bibnamefont{Mottelson}},
  \bibinfo{journal}{Eur.~Phys.~J.~D} \textbf{\bibinfo{volume}{16}},
  \bibinfo{pages}{381} (\bibinfo{year}{2001}).

\bibitem[{\citenamefont{Mikhailov}(2002{\natexlab{a}})}]{Mikhailov02}
\bibinfo{author}{\bibfnamefont{S.~A.} \bibnamefont{Mikhailov}},
  \bibinfo{journal}{Phys.\ Rev.\ B} \textbf{\bibinfo{volume}{66}},
  \bibinfo{pages}{153313} (\bibinfo{year}{2002}{\natexlab{a}}).

\bibitem[{\citenamefont{Mikhailov}(2002{\natexlab{b}})}]{Mikhailov02bis}
\bibinfo{author}{\bibfnamefont{S.~A.} \bibnamefont{Mikhailov}},
  \bibinfo{journal}{Phys.\ Rev.\ B} \textbf{\bibinfo{volume}{65}},
  \bibinfo{pages}{115312} (\bibinfo{year}{2002}{\natexlab{b}}).

\bibitem[{\citenamefont{Yang and MacDonald}(2002)}]{Yang02}
\bibinfo{author}{\bibfnamefont{S.-R.~E.} \bibnamefont{Yang}} \bibnamefont{and}
  \bibinfo{author}{\bibfnamefont{A.~H.} \bibnamefont{MacDonald}},
  \bibinfo{journal}{Phys.\ Rev.\ B} \textbf{\bibinfo{volume}{66}},
  \bibinfo{pages}{R041304} (\bibinfo{year}{2002}).

\bibitem[{\citenamefont{G{\"u}\c{c}l{\"u}
  et~al.}(2002)\citenamefont{G{\"u}\c{c}l{\"u}, Sun, Guo, and
  Harris}}]{Guclu02}
\bibinfo{author}{\bibfnamefont{A.~D.} \bibnamefont{G{\"u}\c{c}l{\"u}}},
  \bibinfo{author}{\bibfnamefont{Q.~F.} \bibnamefont{Sun}},
  \bibinfo{author}{\bibfnamefont{H.}~\bibnamefont{Guo}}, \bibnamefont{and}
  \bibinfo{author}{\bibfnamefont{R.}~\bibnamefont{Harris}},
  \bibinfo{journal}{Phys.\ Rev.\ B} \textbf{\bibinfo{volume}{66}},
  \bibinfo{pages}{195327} (\bibinfo{year}{2002}).

\bibitem[{\citenamefont{G{\"u}\c{c}l{\"u}
  et~al.}(2003)\citenamefont{G{\"u}\c{c}l{\"u}, Wang, and Guo}}]{Guclu03}
\bibinfo{author}{\bibfnamefont{A.~D.} \bibnamefont{G{\"u}\c{c}l{\"u}}},
  \bibinfo{author}{\bibfnamefont{J.-S.} \bibnamefont{Wang}}, \bibnamefont{and}
  \bibinfo{author}{\bibfnamefont{H.}~\bibnamefont{Guo}},
  \bibinfo{journal}{Phys.\ Rev.\ B} \textbf{\bibinfo{volume}{68}},
  \bibinfo{pages}{035304} (\bibinfo{year}{2003}).

\bibitem[{\citenamefont{Cha and Yang}(2003)}]{Cha03}
\bibinfo{author}{\bibfnamefont{M.-C.} \bibnamefont{Cha}} \bibnamefont{and}
  \bibinfo{author}{\bibfnamefont{S.-R.~E.} \bibnamefont{Yang}},
  \bibinfo{journal}{Phys.\ Rev.\ B} \textbf{\bibinfo{volume}{67}},
  \bibinfo{pages}{205312} (\bibinfo{year}{2003}).

\bibitem[{\citenamefont{Tavernier et~al.}(2003)\citenamefont{Tavernier,
  Anisimovas, Peeters, Szafran, Adamowski, and Bednarek}}]{Tavernier03}
\bibinfo{author}{\bibfnamefont{M.~B.} \bibnamefont{Tavernier}},
  \bibinfo{author}{\bibfnamefont{E.}~\bibnamefont{Anisimovas}},
  \bibinfo{author}{\bibfnamefont{F.~M.} \bibnamefont{Peeters}},
  \bibinfo{author}{\bibfnamefont{B.}~\bibnamefont{Szafran}},
  \bibinfo{author}{\bibfnamefont{J.}~\bibnamefont{Adamowski}},
  \bibnamefont{and} \bibinfo{author}{\bibfnamefont{S.}~\bibnamefont{Bednarek}},
  \bibinfo{journal}{Phys.\ Rev.\ B} \textbf{\bibinfo{volume}{68}},
  \bibinfo{pages}{205305} (\bibinfo{year}{2003}).

\bibitem[{\citenamefont{Yannouleas and Landman}(2003)}]{Yannouleas03}
\bibinfo{author}{\bibfnamefont{C.}~\bibnamefont{Yannouleas}} \bibnamefont{and}
  \bibinfo{author}{\bibfnamefont{U.}~\bibnamefont{Landman}},
  \bibinfo{journal}{Phys.\ Rev.\ B} \textbf{\bibinfo{volume}{68}},
  \bibinfo{pages}{035326} (\bibinfo{year}{2003}).

\bibitem[{\citenamefont{Chung}(2004)}]{Chung04}
\bibinfo{author}{\bibfnamefont{M.-H.} \bibnamefont{Chung}},
  \bibinfo{journal}{Phys.\ Rev.\ B} \textbf{\bibinfo{volume}{70}},
  \bibinfo{pages}{113302} (\bibinfo{year}{2004}).

\bibitem[{\citenamefont{Anisimovas et~al.}(2004)\citenamefont{Anisimovas,
  Matulis, Tavernier, and Peeters}}]{Anisimovas04}
\bibinfo{author}{\bibfnamefont{E.}~\bibnamefont{Anisimovas}},
  \bibinfo{author}{\bibfnamefont{A.}~\bibnamefont{Matulis}},
  \bibinfo{author}{\bibfnamefont{M.~B.} \bibnamefont{Tavernier}},
  \bibnamefont{and} \bibinfo{author}{\bibfnamefont{F.~M.}
  \bibnamefont{Peeters}}, \bibinfo{journal}{Phys.\ Rev.\ B}
  \textbf{\bibinfo{volume}{69}}, \bibinfo{pages}{075305}
  (\bibinfo{year}{2004}).

\bibitem[{\citenamefont{Szafran
  et~al.}(2004{\natexlab{a}})\citenamefont{Szafran, Peeters, Bednarek, and
  Adamowski}}]{Szafran04a}
\bibinfo{author}{\bibfnamefont{B.}~\bibnamefont{Szafran}},
  \bibinfo{author}{\bibfnamefont{F.~M.} \bibnamefont{Peeters}},
  \bibinfo{author}{\bibfnamefont{S.}~\bibnamefont{Bednarek}}, \bibnamefont{and}
  \bibinfo{author}{\bibfnamefont{J.}~\bibnamefont{Adamowski}},
  \bibinfo{journal}{Phys.\ Rev.\ B} \textbf{\bibinfo{volume}{69}},
  \bibinfo{pages}{125344} (\bibinfo{year}{2004}{\natexlab{a}}).

\bibitem[{\citenamefont{Szafran
  et~al.}(2004{\natexlab{b}})\citenamefont{Szafran, Peeters, and
  Bednarek}}]{Szafran04b}
\bibinfo{author}{\bibfnamefont{B.}~\bibnamefont{Szafran}},
  \bibinfo{author}{\bibfnamefont{F.~M.} \bibnamefont{Peeters}},
  \bibnamefont{and} \bibinfo{author}{\bibfnamefont{S.}~\bibnamefont{Bednarek}},
  \bibinfo{journal}{Phys.\ Rev.\ B} \textbf{\bibinfo{volume}{70}},
  \bibinfo{pages}{205318} (\bibinfo{year}{2004}{\natexlab{b}}).

\bibitem[{\citenamefont{Korkusi\'nski et~al.}(2004)\citenamefont{Korkusi\'nski,
  Hawrylak, Ciorga, Pioro-Ladri\`ere, and Sachrajda}}]{Korkusinski04}
\bibinfo{author}{\bibfnamefont{M.}~\bibnamefont{Korkusi\'nski}},
  \bibinfo{author}{\bibfnamefont{P.}~\bibnamefont{Hawrylak}},
  \bibinfo{author}{\bibfnamefont{M.}~\bibnamefont{Ciorga}},
  \bibinfo{author}{\bibfnamefont{M.}~\bibnamefont{Pioro-Ladri\`ere}},
  \bibnamefont{and} \bibinfo{author}{\bibfnamefont{A.~S.}
  \bibnamefont{Sachrajda}}, \bibinfo{journal}{Phys.\ Rev.\ Lett.}
  \textbf{\bibinfo{volume}{93}}, \bibinfo{pages}{206806}
  (\bibinfo{year}{2004}).

\bibitem[{\citenamefont{Manninen et~al.}(2005)\citenamefont{Manninen, Reimann,
  Koskinen, Yu, and Toreblad}}]{Manninen05}
\bibinfo{author}{\bibfnamefont{M.}~\bibnamefont{Manninen}},
  \bibinfo{author}{\bibfnamefont{S.~M.} \bibnamefont{Reimann}},
  \bibinfo{author}{\bibfnamefont{M.}~\bibnamefont{Koskinen}},
  \bibinfo{author}{\bibfnamefont{Y.}~\bibnamefont{Yu}}, \bibnamefont{and}
  \bibinfo{author}{\bibfnamefont{M.}~\bibnamefont{Toreblad}},
  \bibinfo{journal}{Phys.\ Rev.\ Lett.} \textbf{\bibinfo{volume}{94}},
  \bibinfo{pages}{106405} (\bibinfo{year}{2005}).

\bibitem[{\citenamefont{Rontani et~al.}(2005)\citenamefont{Rontani, Cavazzoni,
  and Goldoni}}]{Rontani05a}
\bibinfo{author}{\bibfnamefont{M.}~\bibnamefont{Rontani}},
  \bibinfo{author}{\bibfnamefont{C.}~\bibnamefont{Cavazzoni}},
  \bibnamefont{and} \bibinfo{author}{\bibfnamefont{G.}~\bibnamefont{Goldoni}},
  \bibinfo{journal}{Computer Physics Commun.} \textbf{\bibinfo{volume}{169}},
  \bibinfo{pages}{430} (\bibinfo{year}{2005}).

\bibitem[{\citenamefont{Palacios and Hawrylak}(1995)}]{Palacios95}
\bibinfo{author}{\bibfnamefont{J.~J.} \bibnamefont{Palacios}} \bibnamefont{and}
  \bibinfo{author}{\bibfnamefont{P.}~\bibnamefont{Hawrylak}},
  \bibinfo{journal}{Phys.\ Rev.\ B} \textbf{\bibinfo{volume}{51}},
  \bibinfo{pages}{1769} (\bibinfo{year}{1995}).

\bibitem[{\citenamefont{Oh et~al.}(1996)\citenamefont{Oh, Chang, Ihm, and
  Lee}}]{Oh96}
\bibinfo{author}{\bibfnamefont{J.~H.} \bibnamefont{Oh}},
  \bibinfo{author}{\bibfnamefont{K.~J.} \bibnamefont{Chang}},
  \bibinfo{author}{\bibfnamefont{G.}~\bibnamefont{Ihm}}, \bibnamefont{and}
  \bibinfo{author}{\bibfnamefont{S.~J.} \bibnamefont{Lee}},
  \bibinfo{journal}{Phys.\ Rev.\ B} \textbf{\bibinfo{volume}{53}},
  \bibinfo{pages}{R13264} (\bibinfo{year}{1996}).

\bibitem[{\citenamefont{Imamura et~al.}(1996)\citenamefont{Imamura, Maksym, and
  Aoki}}]{Imamura96}
\bibinfo{author}{\bibfnamefont{H.}~\bibnamefont{Imamura}},
  \bibinfo{author}{\bibfnamefont{P.~A.} \bibnamefont{Maksym}},
  \bibnamefont{and} \bibinfo{author}{\bibfnamefont{H.}~\bibnamefont{Aoki}},
  \bibinfo{journal}{Phys.\ Rev.\ B} \textbf{\bibinfo{volume}{53}},
  \bibinfo{pages}{12613} (\bibinfo{year}{1996}).

\bibitem[{\citenamefont{Imamura et~al.}(1998)\citenamefont{Imamura, Aoki, and
  Maksym}}]{Imamura98}
\bibinfo{author}{\bibfnamefont{H.}~\bibnamefont{Imamura}},
  \bibinfo{author}{\bibfnamefont{H.}~\bibnamefont{Aoki}}, \bibnamefont{and}
  \bibinfo{author}{\bibfnamefont{P.~A.} \bibnamefont{Maksym}},
  \bibinfo{journal}{Phys.\ Rev.\ B} \textbf{\bibinfo{volume}{57}},
  \bibinfo{pages}{R4257} (\bibinfo{year}{1998}).

\bibitem[{\citenamefont{Imamura et~al.}(1999)\citenamefont{Imamura, Maksym, and
  Aoki}}]{Imamura99}
\bibinfo{author}{\bibfnamefont{H.}~\bibnamefont{Imamura}},
  \bibinfo{author}{\bibfnamefont{P.~A.} \bibnamefont{Maksym}},
  \bibnamefont{and} \bibinfo{author}{\bibfnamefont{H.}~\bibnamefont{Aoki}},
  \bibinfo{journal}{Phys.\ Rev.\ B} \textbf{\bibinfo{volume}{59}},
  \bibinfo{pages}{5817} (\bibinfo{year}{1999}).

\bibitem[{\citenamefont{Tokura et~al.}(1999)\citenamefont{Tokura, Austing, and
  Tarucha}}]{Tokura99}
\bibinfo{author}{\bibfnamefont{Y.}~\bibnamefont{Tokura}},
  \bibinfo{author}{\bibfnamefont{D.~G.} \bibnamefont{Austing}},
  \bibnamefont{and} \bibinfo{author}{\bibfnamefont{S.}~\bibnamefont{Tarucha}},
  \bibinfo{journal}{J.\ Phys.: Condens.~Matter} \textbf{\bibinfo{volume}{11}},
  \bibinfo{pages}{6023} (\bibinfo{year}{1999}).

\bibitem[{\citenamefont{Tokura et~al.}(2000)\citenamefont{Tokura, Sasaki,
  Austing, and Tarucha}}]{Tokura00}
\bibinfo{author}{\bibfnamefont{Y.}~\bibnamefont{Tokura}},
  \bibinfo{author}{\bibfnamefont{S.}~\bibnamefont{Sasaki}},
  \bibinfo{author}{\bibfnamefont{D.~G.} \bibnamefont{Austing}},
  \bibnamefont{and} \bibinfo{author}{\bibfnamefont{S.}~\bibnamefont{Tarucha}},
  \bibinfo{journal}{Physica E} \textbf{\bibinfo{volume}{6}},
  \bibinfo{pages}{676} (\bibinfo{year}{2000}).

\bibitem[{\citenamefont{Mart\'{\i}n-Moreno
  et~al.}(2000)\citenamefont{Mart\'{\i}n-Moreno, Brey, and Tejedor}}]{Moreno00}
\bibinfo{author}{\bibfnamefont{L.}~\bibnamefont{Mart\'{\i}n-Moreno}},
  \bibinfo{author}{\bibfnamefont{L.}~\bibnamefont{Brey}}, \bibnamefont{and}
  \bibinfo{author}{\bibfnamefont{C.}~\bibnamefont{Tejedor}},
  \bibinfo{journal}{Phys.\ Rev.\ B} \textbf{\bibinfo{volume}{62}},
  \bibinfo{pages}{R10633} (\bibinfo{year}{2000}).

\bibitem[{\citenamefont{Hu and Das~Sarma}(2001)}]{Hu01}
\bibinfo{author}{\bibfnamefont{H.}~\bibnamefont{Hu}} \bibnamefont{and}
  \bibinfo{author}{\bibfnamefont{S.}~\bibnamefont{Das~Sarma}},
  \bibinfo{journal}{Physical Review A} \textbf{\bibinfo{volume}{64}},
  \bibinfo{pages}{042312} (\bibinfo{year}{2001}).

\bibitem[{\citenamefont{Harju et~al.}(2002)\citenamefont{Harju, Siljam{\"a}ki,
  and Nieminen}}]{Harju02}
\bibinfo{author}{\bibfnamefont{A.}~\bibnamefont{Harju}},
  \bibinfo{author}{\bibfnamefont{S.}~\bibnamefont{Siljam{\"a}ki}},
  \bibnamefont{and} \bibinfo{author}{\bibfnamefont{R.~M.}
  \bibnamefont{Nieminen}}, \bibinfo{journal}{Phys.\ Rev.\ Lett.}
  \textbf{\bibinfo{volume}{88}}, \bibinfo{pages}{226804}
  (\bibinfo{year}{2002}).

\bibitem[{\citenamefont{Jacob et~al.}(2004)\citenamefont{Jacob, Wunsch, and
  Pfannkuche}}]{Jacob04}
\bibinfo{author}{\bibfnamefont{D.}~\bibnamefont{Jacob}},
  \bibinfo{author}{\bibfnamefont{B.}~\bibnamefont{Wunsch}}, \bibnamefont{and}
  \bibinfo{author}{\bibfnamefont{D.}~\bibnamefont{Pfannkuche}},
  \bibinfo{journal}{Phys.\ Rev.\ B} \textbf{\bibinfo{volume}{70}},
  \bibinfo{pages}{R081314} (\bibinfo{year}{2004}).

\bibitem[{\citenamefont{Wunsch et~al.}(2005)\citenamefont{Wunsch, Jacob, and
  Pfannkuche}}]{Wunsch05}
\bibinfo{author}{\bibfnamefont{B.}~\bibnamefont{Wunsch}},
  \bibinfo{author}{\bibfnamefont{D.}~\bibnamefont{Jacob}}, \bibnamefont{and}
  \bibinfo{author}{\bibfnamefont{D.}~\bibnamefont{Pfannkuche}},
  \bibinfo{journal}{Physica E} \textbf{\bibinfo{volume}{26}},
  \bibinfo{pages}{464} (\bibinfo{year}{2005}).

\bibitem[{\citenamefont{Ota et~al.}(2005)\citenamefont{Ota, Rontani, Tarucha,
  Nakata, Song, Miyazawa, Usuki, Takatsu, and Yokoyama}}]{Ota05}
\bibinfo{author}{\bibfnamefont{T.}~\bibnamefont{Ota}},
  \bibinfo{author}{\bibfnamefont{M.}~\bibnamefont{Rontani}},
  \bibinfo{author}{\bibfnamefont{S.}~\bibnamefont{Tarucha}},
  \bibinfo{author}{\bibfnamefont{Y.}~\bibnamefont{Nakata}},
  \bibinfo{author}{\bibfnamefont{H.~Z.} \bibnamefont{Song}},
  \bibinfo{author}{\bibfnamefont{T.}~\bibnamefont{Miyazawa}},
  \bibinfo{author}{\bibfnamefont{T.}~\bibnamefont{Usuki}},
  \bibinfo{author}{\bibfnamefont{M.}~\bibnamefont{Takatsu}}, \bibnamefont{and}
  \bibinfo{author}{\bibfnamefont{N.}~\bibnamefont{Yokoyama}},
  \bibinfo{journal}{Phys.\ Rev.\ Lett.} \textbf{\bibinfo{volume}{95}},
  \bibinfo{pages}{236801} (\bibinfo{year}{2005}).

\bibitem[{\citenamefont{Yannouleas and Landman}(1999)}]{Yannouleas99}
\bibinfo{author}{\bibfnamefont{C.}~\bibnamefont{Yannouleas}} \bibnamefont{and}
  \bibinfo{author}{\bibfnamefont{U.}~\bibnamefont{Landman}},
  \bibinfo{journal}{Phys.\ Rev.\ Lett.} \textbf{\bibinfo{volume}{82}},
  \bibinfo{pages}{5325} (\bibinfo{year}{1999}).

\bibitem[{\citenamefont{Pederiva et~al.}(2000)\citenamefont{Pederiva, Umrigar,
  and Lipparini}}]{Pederiva00}
\bibinfo{author}{\bibfnamefont{F.}~\bibnamefont{Pederiva}},
  \bibinfo{author}{\bibfnamefont{C.~J.} \bibnamefont{Umrigar}},
  \bibnamefont{and}
  \bibinfo{author}{\bibfnamefont{E.}~\bibnamefont{Lipparini}},
  \bibinfo{journal}{Phys.\ Rev.\ B} \textbf{\bibinfo{volume}{62}},
  \bibinfo{pages}{8120} (\bibinfo{year}{2000}).

\bibitem[{\citenamefont{Rontani et~al.}(1999)\citenamefont{Rontani, Rossi,
  Manghi, and Molinari}}]{Rontani99}
\bibinfo{author}{\bibfnamefont{M.}~\bibnamefont{Rontani}},
  \bibinfo{author}{\bibfnamefont{F.}~\bibnamefont{Rossi}},
  \bibinfo{author}{\bibfnamefont{F.}~\bibnamefont{Manghi}}, \bibnamefont{and}
  \bibinfo{author}{\bibfnamefont{E.}~\bibnamefont{Molinari}},
  \bibinfo{journal}{Phys.\ Rev.\ B} \textbf{\bibinfo{volume}{59}},
  \bibinfo{pages}{10165} (\bibinfo{year}{1999}).

\bibitem[{not()}]{notaReimann}
\bibinfo{note}{Indeed, authors of Ref.~\onlinecite{ReimannPRB} were not able to
  investigate the full $\lambda$ range since they used a serial CI code with
  strong power and memory limitations.}

\bibitem[{\citenamefont{Koskinen et~al.}(2001)\citenamefont{Koskinen, Manninen,
  Mottelson, and Reimann}}]{Koskinen01}
\bibinfo{author}{\bibfnamefont{M.}~\bibnamefont{Koskinen}},
  \bibinfo{author}{\bibfnamefont{M.}~\bibnamefont{Manninen}},
  \bibinfo{author}{\bibfnamefont{B.}~\bibnamefont{Mottelson}},
  \bibnamefont{and} \bibinfo{author}{\bibfnamefont{S.~M.}
  \bibnamefont{Reimann}}, \bibinfo{journal}{Phys.\ Rev.\ B}
  \textbf{\bibinfo{volume}{63}}, \bibinfo{pages}{205323}
  (\bibinfo{year}{2001}).

\bibitem[{\citenamefont{Ruan et~al.}(1995)\citenamefont{Ruan, Liu, Bao, and
  Zhang}}]{Ruan95}
\bibinfo{author}{\bibfnamefont{W.~Y.} \bibnamefont{Ruan}},
  \bibinfo{author}{\bibfnamefont{Y.~Y.} \bibnamefont{Liu}},
  \bibinfo{author}{\bibfnamefont{C.~G.} \bibnamefont{Bao}}, \bibnamefont{and}
  \bibinfo{author}{\bibfnamefont{Z.~Q.} \bibnamefont{Zhang}},
  \bibinfo{journal}{Phys.\ Rev.\ B} \textbf{\bibinfo{volume}{51}},
  \bibinfo{pages}{7942} (\bibinfo{year}{1995}).

\bibitem[{\citenamefont{Bolton and R{\"o}ssler}(1993)}]{Bolton93}
\bibinfo{author}{\bibfnamefont{F.}~\bibnamefont{Bolton}} \bibnamefont{and}
  \bibinfo{author}{\bibfnamefont{U.}~\bibnamefont{R{\"o}ssler}},
  \bibinfo{journal}{Superlatt.~Microstruct.} \textbf{\bibinfo{volume}{13}},
  \bibinfo{pages}{139} (\bibinfo{year}{1993}).

\bibitem[{\citenamefont{Bedanov and Peeters}(1994)}]{Bedanov94}
\bibinfo{author}{\bibfnamefont{V.~M.} \bibnamefont{Bedanov}} \bibnamefont{and}
  \bibinfo{author}{\bibfnamefont{F.~M.} \bibnamefont{Peeters}},
  \bibinfo{journal}{Phys.\ Rev.\ B} \textbf{\bibinfo{volume}{49}},
  \bibinfo{pages}{2667} (\bibinfo{year}{1994}).

\bibitem[{\citenamefont{Anisimovas and Matulis}(1998)}]{Anisimovas98}
\bibinfo{author}{\bibfnamefont{E.}~\bibnamefont{Anisimovas}} \bibnamefont{and}
  \bibinfo{author}{\bibfnamefont{A.}~\bibnamefont{Matulis}},
  \bibinfo{journal}{J.\ Phys.: Condens.~Matter} \textbf{\bibinfo{volume}{10}},
  \bibinfo{pages}{601} (\bibinfo{year}{1998}).

\bibitem[{\citenamefont{Rontani}(1999)}]{RontaniPhD}
\bibinfo{author}{\bibfnamefont{M.}~\bibnamefont{Rontani}}, Ph.D. thesis,
  \bibinfo{school}{University of Modena and Reggio Emilia},
  \bibinfo{address}{Via Campi 213A 41100 Modena Italy} (\bibinfo{year}{1999}),
  \bibinfo{note}{available at {\tt http://www.nanoscience.unimo.it}}.

\bibitem[{\citenamefont{Wensauer et~al.}(2004)\citenamefont{Wensauer,
  Korkusi\'nski, and Hawrylak}}]{Wensauer04}
\bibinfo{author}{\bibfnamefont{A.}~\bibnamefont{Wensauer}},
  \bibinfo{author}{\bibfnamefont{M.}~\bibnamefont{Korkusi\'nski}},
  \bibnamefont{and} \bibinfo{author}{\bibfnamefont{P.}~\bibnamefont{Hawrylak}},
  \bibinfo{journal}{Solid State Commun.} \textbf{\bibinfo{volume}{130}},
  \bibinfo{pages}{115} (\bibinfo{year}{2004}).

\end{thebibliography}
%

\end{document}